\pdfoutput=1
\RequirePackage{ifpdf}
\ifpdf
\documentclass[pdftex]{sigma}
\else
\documentclass{sigma}
\fi

\usepackage{cite}

\numberwithin{equation}{section}

\newcommand{\R}{\mathbb{R}}
\newcommand{\C}{\mathbb{C}}

\newcommand{\cF}{{\mathcal{F}}}

\newcommand{\cC}{{\cal C}}

\def\kk{{\cal K}}
\def\mm{{\cal M}}
\def\uu{{\cal U}}
\newcommand{\cL}{{\cal L}}
\newcommand{\cM}{{\cal M}}
\newcommand{\cP}{{\cal P}}
\newcommand{\cT}{{\cal T}}

\newcommand{\mone}{^{-1}}
\newcommand{\smp}{{\kappa}}

\def\poi#1{\{ #1 \}}
\def\com#1{[ #1 ]}

\def\act{\rhd}
\def\mn{{\mu\nu}}
\def\vphi{{\varphi}}

\def\e{\mbox{e}}

\def\dr{{\rightarrow}}
\newcommand{\one}{\mbox{$1\hspace{-1.0mm}{\bf l}$}}
\def\cop{{\bigtriangleup}}
\def\hphi{{\hat \phi}}
\def\hpsi{{\hat \psi}}
\newcommand{\so}{\mathfrak{so}}
\newcommand{\su}{\mathfrak{su}}
\newcommand{\SU}{\mathrm{SU}}
\newcommand{\AN}{\mathrm{AN}}
\newcommand{\SO}{\mathrm{SO}}
\newcommand{\SL}{\mathrm{SL}}

\newcommand{\bra}[1]{\langle #1 \! \mid}
\newcommand{\ket}[1]{\mid \! #1 \rangle}
\newcommand{\braket}[2]{\langle #1 \! \mid #2 \! \rangle}
\newcommand{\ev}[1]{\left\langle #1 \right\rangle}

\begin{document}

\allowdisplaybreaks

\renewcommand{\thefootnote}{$\star$}

\renewcommand{\PaperNumber}{098}

\FirstPageHeading

\ShortArticleName{Loop Quantum Gravity Phenomenology}

\ArticleName{Loop Quantum Gravity Phenomenology:\\ Linking Loops to Observational Physics\footnote{This
paper is a~contribution to the Special Issue ``Loop Quantum Gravity and Cosmology''.
The full collection is available at \href{http://www.emis.de/journals/SIGMA/LQGC.html}{http://www.emis.de/journals/SIGMA/LQGC.html}}}

\Author{Florian GIRELLI~$^{\dag^1\dag^2}$, Franz HINTERLEITNER~$^{\dag^3}$ and Seth A.~MAJOR~$^{\dag^4}$}

\AuthorNameForHeading{F.~Girelli, F.~Hinterleitner and S.A.~Major}

\Address{$^{\dag^1}$~Department of Applied Mathematics, University of Waterloo, Waterloo, Ontario, Canada}
\EmailDD{\href{mailto:fgirelli@uwaterloo.ca}{fgirelli@uwaterloo.ca}}
\Address{$^{\dag^2}$~University Erlangen-Nuremberg, Institute for Theoretical Physics III, Erlangen, Germany}

\Address{$^{\dag^3}$~Department of Theoretical Physics and Astrophysics,
Faculty of Science \\
\hphantom{$^{\dag^3}$}~of the Masaryk University, Kotl\'{a}\v{r}sk\'{a}
2, 611\,37 Brno, Czech Republic}
\EmailDD{\href{mailto:franz@physics.muni.cz}{franz@physics.muni.cz}}

\Address{$^{\dag^4}$~Department of Physics, Hamilton College, Clinton NY 13323, USA}
\EmailDD{\href{mailto:smajor@hamilton.edu}{smajor@hamilton.edu}}

\ArticleDates{Received May 30, 2012, in f\/inal form December 03, 2012; Published online December 13, 2012}

\Abstract{Research during the last decade demonstrates that ef\/fects
originating on the Planck scale are currently being tested in
multiple observational contexts.
In this review we discuss quantum gravity
phenomenological models and their possible links to loop quantum
gravity.
Particle frameworks, including kinematic models, broken
and deformed Poincar\'e symmetry, non-commutative geometry, relative
locality and generalized uncertainty principle, and f\/ield theory
frameworks, including Lorentz violating operators in ef\/fective f\/ield
theory and non-commutative f\/ield theory, are discussed.
The arguments
relating loop quantum gravity to models with modif\/ied dispersion relations are reviewed,
as well as, arguments supporting the preservation of
local Lorentz invariance.
The phenomenology
related to loop quantum cosmology is brief\/ly reviewed, with a~focus
on possible ef\/fects that might be tested in the near future.
As the
discussion makes clear, there remains much interesting work to do in
establishing the connection between the
fundamental theory of loop quantum gravity and these specif\/ic
phenomenological models, in determining observational consequences of the characteristic aspects of loop quantum gravity, and in further ref\/ining current observations.
Open problems related to these developments
are highlighted.}


\Classification{83-02; 83B05; 83C45; 83C47; 83C65}

\tableofcontents

\renewcommand{\thefootnote}{\arabic{footnote}}
\setcounter{footnote}{0}

\section{Introduction}

Twenty f\/ive years ago Ashtekar, building on earlier work by Sen,
laid the foundations of Loop Quantum Gravity (LQG) by reformulating
general relativity (GR) in terms of canonical connection and
triad variables~-- the ``new variables''.
The completion of the kinematics~-- the quantum theory of spatial
geometry~-- led to the prediction of a~granular structure of space,
described by specif\/ic discrete spectra of geometric operators, area
\cite{RS}, volume~\cite{RS,Lo,Le}, length \cite{length,Blength,MSY} and
angle~\cite{angle} operators.
The discreteness of area led to
an explanation of black hole entropy~\cite{R2,Kbh,K1,Bnova} (see \cite{AlC} for a~recent review).
Although granularity in spatial geometry is predicted not only by
LQG, but also by some string theory and non-commutative
geometry mo\-dels, the specif\/ic predictions for the spectra of geometry
operators bears the unique stamp of LQG.\looseness=-1

\looseness=1
Quantum ef\/fects of gravity are expected to be directly perceptible
at distances of the order of the Planck length, about $10^{-35}$ m, in particle processes
at the Planck energy, about $10^{28}$~eV $(c=1)$, and at a~Planck
scale density.
With the typical energy $M_{\rm QG}$ for quantum
gravity (QG) assumed to be of the order of the Planck energy, there
are f\/ifteen orders of magnitude between this energy scale and the
highest attainable center-of-mass energies in accelerators and, in the Earth's frame,
eight orders of magnitude above the highest energy cosmic rays.
So fundamental quantum theories of gravity and the realm of particle
physics appear like continents separated by a~wide ocean.
(Although, if the
world has large extra dimensions, the typical energy scale of
quantum gravity may be signif\/icantly lower.) The
situation is worsened by the fact that none of the tentative QG
theories has attained such a~degree of maturity that would allow to
derive reliable predictions of such a~kind that could be
extrapolated to our ``low-energy'' reality.
It would appear that there is
little hope in directly accessing the deep quantum gravity
regime via experiment.
One can hope, however, to probe the quantum
gravity semi-classical regime, using particle, astrophysical and cosmological
phenomena to enhance the observability of the ef\/fects.

In spite of this discouraging perspective, over one decade
ago a~striking paper by Amelino--Camelia et al.~\cite{A-CEMNS}
on {\em quantum gravity phenomenology} appeared.
The paper was based on a~plausibility
argument: The strong gravity regime is inaccessible but
quantum gravity, as modeled in certain models of string theory and, perhaps, in the quantum geometry of LQG, has a~notion of discreteness in its very core.
This discreteness is understood to be a
genuine property of space, independent of the strength of the actual
gravitational f\/ield at any given location.
Thus it may be possible
to observe QG ef\/fects even without strong gravitational f\/ield, in
the f\/lat space limit.
In~\cite{A-CEMNS} the authors proposed that granularity of space
inf\/luences the propagation of particles, when their energy is
comparable with the QG energy scale.
Further, the assumed invariance
of this energy scale, or the length scale, respectively, are in apparent
contradiction with special relativity (SR).
So it is expected that
the energy-momentum dispersion relation could be modif\/ied to include
dependence on the ratio of the particle's energy and the QG energy.
At
lowest order
\begin{gather}\label{E}
E\simeq p+\frac{m^2}{2p}\pm\xi\frac{E^2}{M_{\rm QG}}
\end{gather}
with the parameter $\xi>0$ of order unity.
Relations like~\eqref{E}
violate, or modify, local Lorentz invariance (LLI).
According
to the sign in~\eqref{E}, the group velocity of high-energy photons could
be sub- or super-luminal, when def\/ined in the usual way by $\partial
E/\partial p$.
Like with all QG ef\/fects, the suppression of Lorentz
invariance violation by the ratio of the particle energy to the QG energy may appear
discouraging at f\/irst sight.
To have a~chance to detect an ef\/fect of the
above modif\/ication, we need an amplif\/ication mechanism, or
``lever arm''.

The authors of~\cite{A-CEMNS} showed that if the tiny ef\/fect on the
speed of light accumulates as high energy photons travel cosmic
distances, the spectra of $\gamma$ ray bursts (GRB) would reveal an
energy-dependent speed of light through a~measurable dif\/ference of
the time of arrival of high and low energy photons.
Due to dif\/ferent group velocities $v=
\partial\omega/\partial k\simeq1+\xi k/M_{\rm QG}$, photons
emitted at dif\/ferent momenta, $k_1$ and $k_2$, would arrive at a
distant observer (at distance $D$) at times separated by the
interval $\Delta t\simeq\xi(k_2-k_1)D/M_{\rm QG}$.
Distant
sources of $\gamma$-ray photons are the best for this test.
Despite
the uncertainties concerning the physics of the psroduction of such
$\gamma$-rays, one can place limits on the parameter $\xi_0$.
The
current strongest limit is $\xi\lesssim0.8$ reported by the
Fermi Collaboration using data from the $\gamma$-ray burst GRB
090510~\cite{FERMI}.
This is discussed further in Section~\ref{eftconstraints}.

In the years following this work, the nascent f\/ield of QG
phenomenology developed~\cite{A-Crev} from ad hoc ef\/fective theories,
like isolated isles lying between the developing QG theories and reality, linked to
the former ones loosely by plausibility arguments.
Today
the main ef\/forts of QG phenomenology go in two
directions: to establish a~bridge between the intermediate ef\/fective
theories and the fundamental QG theory and the ref\/inement
of observational methods, through new ef\/fective theories and
experiments that could shed new light on QG ef\/fects.
These are
exceptionally healthy developments for the f\/ield.
The development
of physical theory relies on the link between theory and experiment.
Now these links between current observation and quantum gravity
theory are possible and under active development.

The purpose of this review on quantum gravity phenomenology is three-fold.
First, we wish to
provide a~summary of the state of the art in LQG phenomenology and
closely related f\/ields with particular attention to theoretical structures related to LQG
and to possible observations that hold near-term promise.
Second, we wish to provide a~road map for
those who wish to know which physical ef\/fects have been studied and
where to f\/ind more information on them.
Third, we wish to highlight
open problems.

Before describing in more details what is contained in this review, we
remind the reader that, of course, the LQG dynamics
remains open.
Whether a~fully discrete space-time follows from
the discreteness of spatial geometry is a~question for the solution
to LQG dynamics generated by the master constraint, the Hamiltonian
constraint operator, and/or spin foam models~\cite{T1,R1}.
Nevertheless
if the granular spatial geometry of LQG is physically correct then
it must manifest itself in observable ways.
This review concerns
the various avenues in which such phenomenology is explored.

The contents of our review is organized in the the following four
sections:
\begin{description}\itemsep=0pt
\item{Section~\ref{section2}:} A brief introduction to the {\em geometric operators} of LQG,
area, volume, length and angle, where discreteness shows up.
\item{Section~\ref{section3}:} An overview of {\em particle} ef\/fective theories of the type
introduced above.
In this section we review particle kinematics, discuss arguments in LQG that lead
to modif\/ied dispersion relations (MDR) like the kind~\eqref{E} and discuss models of symmetry deformation.
The variety of models underlines their loose relation to fundamental theories such as LQG.
\item{Section~\ref{section4}:} A brief review of {\em field theories} leading
to phenomenology including ef\/fective f\/ield theory with Lorentz symmetry violation and non-commutative f\/ield theory.
The ef\/fective f\/ield theories incorporate MDR and contain explicit Lorentz symmetry violation.
A model with LLI is discussed and, in the f\/inal part, actions for f\/ield theories
over non-commutative geometries are discussed.
\item{Section~\ref{section5}:} A brief discussion of {\em loop quantum cosmology}
and possible observational windows.
Cosmology is a~promising observational window and a~chance to bridge the gap
between QG and reality directly, without intermediate ef\/fective theories:
The cosmic microwave and gravitational wave background f\/luctuations allow a~glimpse into the far past, closer to the conditions when the discreteness of space would play a~more dominant role.
\end{description}

The Planck length is given when the Compton length is equivalent to the Schwarzschild length,
$\ell_{\rm P}:=\sqrt{\hbar G/c^3}\simeq1.6\times10^{-35}$~m.
Similarly, the Planck mass is given when the Compton mass is equivalent to the Schwarzschild mass,
$M_{\rm P}:=\sqrt{\hbar c/G}\simeq1.2\times10^{28}\text{~eV/c}^2$.
These conditions mean that at these scales, quantum ef\/fects are comparable to gravitational ef\/fects.
The usual physical argument, which~\cite{Doplicher:1994tu} made more rigorous,
is that to make a~very precise measurement of a~distance, we use a~photon with very high energy.
The higher the precision, the higher the energy will be in a small volume of space,
so that gravitational ef\/fects will kick in due to a~large energy density.
When the volume is small enough, i.e.\ the precision very high, the energy density is
so large that a black hole is created and the photon can not come back to the observer.
Hence there is a maximum precision and a notion of minimum length $\ell_{\rm P}$.
This argument goe
s beyond the simple application of dimensional analysis of the fundamental scales
of the quantum gravitational problem, $c$, $G$, $\hbar$, and $\Lambda$, the cosmological constant.
In the remainder of this review, except where it could lead to confusion, we set $c=1$
and {\em denote the Planck scale mass by $\kappa$ so that the Planck scale
$\smp=M_{\rm P}=\frac{1}{\ell_{\rm P}}$} can be interpreted
as Planck momentum $\smp=M_{\rm P}c$, Planck energy $\smp=M_{\rm P}c^2$ or Planck rest mass $\smp=M_P$.

\section{Discreteness of LQG geometric operators}\label{section2}

Loop quantum gravity hews close to the classical theory of general
relativity, taking the notion of background independence and
apparent four dimensionality of space-time seriously.
The
quantization has been approached in stages, with work on kinematics,
the quantization of spatial geometry, preceding the dynamics, the
full description of space-time.
The kinematics, all but unique,
reveals a~picture of quantized space.
This quantization, this
granularity, inspired the phenomenological models in this review
and is the subject of this section.
We focus on the geometric
observables here.
For a~brief review of the elements of LQG and the new variables see Appendix~\ref{newvariables}.

In LQG the operators representing area, angle, length, and volume
have discrete spectra, so discreteness is naturally incorporated
into LQG. This fundamental discreteness predicted by LQG, must be
at some level physically manifest.
Much of the work in
phenomenology related to LQG has been an ef\/fort to link up this
predicted discreteness with possible observational contexts.
In
later sections we will introduce a~fundamental length or energy
scale, adding terms to the ef\/fective action for particles, exploring
ef\/fects of an minimum area on cosmological models, and studying the
af\/fects of underlying combinatorics on geometric quantities.

\subsection{Area}
Classically the area of a~two-dimensional surface is the integral
over the square root of the determinant of the induced
two-dimensional metric.
Thus,
\begin{gather*}%\label{26}
A({\cal S})=\int_{\cal S}\sqrt{n_a {E^{ai}} n_b {E^b}_i}\,{\rm d}^2\sigma.
\end{gather*}
(For details see~\cite{R1,AL}.) The operators related to ${E^a}_i$,
namely the f\/lux operators $E_i({\cal S})$, associated to $\cal S$,
have a~Lie algebra index and so are not gauge invariant.
Nor does
its ``square'' $E^2({\cal S}):=E^i({\cal S})E_i({\cal S})$ give
rise to a~gauge-invariant operator in general, because the
integration over $\cal S$ complicates the transformation properties,
when there are more than one intersection of a~spin network (SNW)
graph $\gamma$ with $\cal S$.
Its action on a~single link
intersecting $\cal S$, however, is simple: Each $E_i$ inserts an
$\su(2)$ generator $\tau_i^{(j)}$ into the corresponding holonomy,
which results in the Casimir operator of $\SU(2)$ in the action of
$E^2({\cal S})$, namely
\begin{gather*}%\label{28}
\sum_i\tau_i^{(j)}\tau_i^{(j)}=j(j+1)\cdot \one.
\end{gather*}
To make use of this simple result in the case of extended graphs
intersecting $\cal S$, one partitions the surface into $n$ small
surfaces ${\cal S}_i$, such that each of them contains not more than
one intersection point with the given graph, and then takes the sum
over the small sub-surfaces,
\begin{gather*}%\label{29}
A({\cal S}):=\lim_{n\rightarrow\infty}\sum_{i=1}^n\sqrt{E^2({\cal
S}_i)}.
\end{gather*}
This def\/ines the area operator.
Area can be equally well def\/ined in
a combinatorial framework as discussed in Appendix~\ref{newvariables}.

The action on a~SNW function $\Psi_\Gamma$ is
\begin{gather}\label{30}
A({\cal S})|\Psi_\Gamma\rangle=\frac{8\pi\gamma}{\kappa^2}\sum_{p\in\Gamma\cap\cal
S}\sqrt{j_p(j_p+1)}\,|\Psi_\Gamma\rangle,
\end{gather}
where $j_p$ is the spin, or color, of the link that intersects $\cal S$ at
$p$ and $\gamma$ is the Barbero--Immirizi parameter.
The area operator acts only on the intersection points of the
surface with the SNW graph, $\gamma\cap\cal S$ and so gives a~f\/inite number of
contributions.
SNW functions are eigenfunctions.
The eigenvalues are
obviously discrete.
The quanta of area live on the edges of the graph and are the simplest elements of quantum geometry.
There is a~minimal eigenvalue, the so-called
area gap, which is the area when a~single edge with $j=1/2$
intersects $\cal S$,
\begin{gather*}%\label{31}
\Delta A=4\sqrt{3}\pi\hbar G c^{-3}\sim10^{-70}~{\rm m}^2.
\end{gather*}
This is the minimal quantum of area, which can be carried by a~link.

The eigenvalues~\eqref{30} form only the main
sequence of the spectrum of the area operator.
When nodes of the SNW lie on $\cal S$ and some links are
tangent to it the relation is modif\/ied, see~\cite{AL,R1}.
The important fact, independent of
these details, is that discreteness of area with the SNW links,
carrying its quanta, comes out in a~natural way.
The interpretation
of discrete geometric eigenvalues as observable quantities goes back
to early work in~\cite{GCQ}.
This discreteness made the calculation of black hole
entropy possible by counting the number of microstates of the
gravitational f\/ield that lead to a~given area of the horizon within
some small interval.

Intriguingly, area operators acting on surfaces that intersect in a~line fail to commute,
when SNW nodes line in that intersection~\cite{Ashtekar:1998ak}.
One may see this as resulting from the commutation relations among angular momentum operators in the two area operators.
Recently additional insight into this non-commutativity comes from the formulation of discrete
classical phase space of loop gravity, in which the f\/lux operators also depend on the connection \cite{FGZdiscrete}.

Another, inequivalent, form of the area operator was proposed in~\cite{krasnov_area}. This operator,
$\tilde{A}_{\cal S}$, is based on a~non gauge-invariant expression of the surface metric.
Fix a unit vector is the Lie algebra, $r^i$ then the classical area may be expressed as the maximum value of
\begin{gather*}
\tilde{A}_{\cal S}=\int_{\cal S}\sqrt{n_a {E^{ai}} r_i}\,{\rm
d}^2\sigma,
\end{gather*}
where the maximum is obtained by gauge rotating the triad.
On the quantum mechanical side this value is the maximum magnetic quantum number, simply $j$ so the spectrum is simply
\begin{gather*}
\tilde{A}({\cal S})|\Psi_\Gamma\rangle=\frac{8\pi\gamma}{\kappa^2}\sum_{p\in\Gamma\cap\cal
S}j_p\, |\Psi_\Gamma\rangle.
\end{gather*}
This operator, frequently used in the spin foam context, is particularly useful
in systems with boundary such as where gauge invariance might be (partially) f\/ixed.

\subsection{Volume}
\label{volume}

Like the area of a~surface, the volume of a region $\cal
R$ in three-dimensional space, the integral of the square root of
the determinant of the metric, can be expressed in terms of
densitized triads,
\begin{gather}\label{32}
V({\cal R})=\int_{\cal R}{\rm
d}^3x\sqrt{\frac{1}{3!}\left|\epsilon_{abc} \epsilon^{ijk} {E^a}_i(x) {E^b}_j(x) {E^c}_k(x)\right|}.
\end{gather}
Regularizations of this expression consist of
partitioning the region under consideration into cubic cells in some
auxiliary coordinates and constructing an operator for each cell.
The cells are shrunk to zero coordinate volume.
This continuum limit
is well-def\/ined, thanks to discreteness
reached when the cells are suf\/f\/iciently small, but f\/inite.
Readers
interested in precisely how this is done should consult~\cite{RS,vo}.

There are, primarily, two def\/initions of the operator, one due to Rovelli and Smolin
(RS)~\cite{RS} and the other due to Ashtekar and Lewandowski (AL)~\cite{vo}.
Here we present the AL volume operator of~\cite{vo}, presented also in \cite{T1}.
For a~given
SNW function based on a~graph $\Gamma$, the operator $\hat V_{{\cal
R},\Gamma}$ of the volume of a~region $\cal R$ acts nontrivially
only on (at least four-valent~\cite{Lo}) vertices in $\cal R$.
According to the three triad components in~\eqref{32}, which become
derivatives upon quantization, in the volume operator three
derivative operators $\hat X^i_{v,e_I}$ act at every node or vertex $v$ on
each triple of adjacent edges $e_I$,
\begin{gather}
\label{vol_op}
\hat V_{{\cal R},\Gamma}
=\left(\frac{\ell_{\rm P}}{2}\right)^3
\sum_{v\in\cal R}\sqrt{\Bigg|\frac{i}{3!\cdot 8}\sum_{I,J,K}s(e_I,e_J,e_K)
\epsilon_{ijk} \hat X^i_{v,e_I}\hat X^j_{v,e_J}\hat X^k_{v,e_K}\Bigg|}.
\end{gather}
Dependence on the tangent space structure of the embedding is
manifest in $s(e_I,e_J,e_K)$.
This is $+1$ $(-1)$, when $e_I$,
$e_J$, and $e_K$ are positive (negative) oriented, and is zero when
the edges are coplanar.
The action of the operators $\hat
X^i_{v,e_I}$ on a~SNW function
$\Psi_\Gamma=\psi(h_{e_1}(A),\ldots,h_{e_N}(A))$ based on the graph
$\Gamma$ is
\begin{gather*}
\hat X^i_{v,e_I} \Psi_\Gamma(A)=i\operatorname{tr}\left(h_{e_I}(A)\tau^i \frac{\partial\psi}{\partial
h_{e_I}(A)}\right),
\end{gather*}
when $e_I$ is outgoing at $v$.
This is the action of the
left-invariant vector f\/ield on $\SU(2)$ in the direction of $\tau^i$;
for ingoing edges it would be the right-invariant vector f\/ield.

Given the ``triple-product'' action of the operator~\eqref{vol_op}, vertices carry discrete quanta of volume.
The
volume operator of a~small region containing a node does not
change the graph, nor the colors of the adjacent edges, it acts in
the form of a~linear transformation in the space of intertwiners at the
vertex for given colors of the adjacent edges.
It is then this space of intertwiners that
forms the ``atoms of quantum geometry''.

The complete spectrum is not known, but it has been investigated
\cite{Pi,TT,Meissner,BruThi,BruRid,BruRid2}.
In the thorough
analysis of~\cite{BruRid,BruRid2}, Brunnemann and Rideout showed
that the volume gap, i.e.\
the lower boundary for the smallest
non-zero eigenvalue, depends on the geometry of the graph and
doesn't in general exist.
In the simplest nontrivial case, for a
four-valent vertex, the existence of a~volume gap is demonstrated
analytically.

The RS volume operator~\cite{RS} (see
also~\cite{R1}) dif\/fers from the AL operator outlined
above.
In this def\/inition the densitized triad operators are
integrated over surfaces bounding each cell with the results that
the square root is inside the sum over $I$, $J$, $K$ and the orientation
factor $s(e_I,e_J,e_K)$ is absent.
Due to the orientation factor the
volume of a~node with coplanar tangent vectors of the adjacent links
is zero, when calculated with the AL operator, whereas the RS
operator does not distinguish between coplanar and non-coplanar
links.

The two volume operators are inequivalent, yielding dif\/ferent
spectra.
While the details of the spectra of the Rovelli--Smolin and
the Ashtekar--Lewnadowski def\/initions of the volume operator dif\/fer,
they do share the property that the volume operator vanishes on all
gauge invariant trivalent vertices~\cite{Lo,Lo2}.

According to an analysis in~\cite{GT,GT2} the AL operator is
compatible with the f\/lux operators, on which it is based, and the RS
operator is not.
On the other hand, thanks to its topological
structure the RS volume does not depend on tangent space structure;
the operator is `topological' in that is invariant under spatial
homeomorphisms.
It is also covariant also under ``extended
dif\/feomorphisms'', which are everywhere continuous mappings that are
invertible everywhere except at a~f\/inite number of isolated points;
the AL operator is invariant under dif\/feomorphisms.
For more on the
comparison see~\cite{T1,R1}.

Physically, the distinction between the two operators is the role of
the tangent space structure at SNW nodes.
There is some
tension in the community over the role of this structure.
Recent developments in twisted discrete geometries~\cite{twist} and the polyhedral point of view
\cite{polyhedral} may help resolve these issues.
It would be valuable to investigate ways in which the
tangent space structure, and associated moduli~\cite{moduli}, could be observationally manifest.

In~\cite{BiaHag} Bianchi and Haggard show that the volume
spectrum of the 4-valent node may be obtained by direct
Bohr--Sommerfeld quantization of geometry.
The description of the
geometry goes all the way back to Minkowski, who showed that the
shapes of convex polyhedra are determined from the areas and unit
normals of the faces.
Kapovich and Millson showed that this space
of shapes is a~phase space, and it is this phase space~-- the same
as the phase space of intertwiners~-- that Bianchi and Haggard used
for the Bohr--Sommerfeld quantization.
The agreement between the
spectra of the Bohr--Sommerfeld and LQG volume is quite good~\cite{BiaHag}.

\subsection{Length}

In constructing the length operator one faces with the challenges of
constructing a~one-di\-men\-sio\-nal operator in terms of f\/luxes and of
constructing the inverse volume operator.
There are three versions
of the length operator.
One~\cite{length} requires the same trick,
due to Thiemann~\cite{trick}, that made the construction of the
inverse volume operator in cosmology and the Hamiltonian constraint
operator in the real connection representation possible.
The second
operator~\cite{Blength}, due to Bianchi, uses instead a
regularization guided by the dual picture in LQG, where one
considers (quantum) convex polyhedral geometries dual to SNW nodes,
the atoms of quantum geometry.
For more
discussion on the comparison between these two operators, see~\cite{Blength}.
The third operator can be seen to be an average of a
formula for length based on area, volume and f\/lux operators~\cite{MSY}.
To give a~f\/lavor of the construction we will review the
f\/irst def\/inition based on~\cite{length}.

Classically the length of a~(piecewise smooth) curve
$c:[0,1]\rightarrow\Sigma$ in the spatial 3-mani\-fold~$\Sigma$ with
background metric $q_{ab}$ is given by
\begin{gather*}
L=\int_0^1{\rm d}t\sqrt{q_{ab}(c(t)) \dot c^a(t) \dot c^b(t)}.
\end{gather*}
In LQG the metric is not a~background structure, but can be given in
terms of the inverse fundamental triad variables,
\begin{gather*}
q_{ab}=\det({E^a}_i) {E_a}^i{E_b}^i,
\end{gather*}
that is
\begin{gather*}
q_{ab}=\epsilon_{acd} \epsilon_{bef} \epsilon^{ijk} \epsilon^{imn}
\frac{{E^c}_j {E^d}_k {E^e}_m {E^f}_n}{4\det({E^g}_l)}.
\end{gather*}
The problem is to f\/ind an operator equivalent to this complicated
non-polynomial expression: any operator version of the denominator
would have a~huge kernel in the Hilbert space, so that the above
expression cannot become a~densely def\/ined operator.

Fortunately $q_{ab}$ can be expressed in terms of Poisson brackets
of the connection $A_a:={A_a}^i\tau_i$ $(\tau_i\in \su(2))$ with the
volume
\begin{gather*}
q_{ab}=-\frac{1}{8\pi^2G^2}\operatorname{tr} \big(\{A_a,V\}\{A_b,V\}\big).
\end{gather*}
$V$ can be formulated as a~well-def\/ined operator.
The connection
$A_a$, on the other hand, can be replaced by its holonomy, when the
curve is partitioned into small pieces, so that the exponent $\int
A_a\dot c^a$ of the holonomy is small and higher powers can be
neglected in f\/irst approximation.
The zeroth-order term (which is
the unity operator) does not contribute to the Poisson brackets.

The length operator is constructed as a~Riemann sum over $n$ pieces
of the curve and by inserting the volume operator $\hat V$ and
replacing the Poisson brackets by $1/i\hbar$ times the commutators,
\begin{gather*}
\hat L_n(c)=\ell_{\rm P}\sum_{i=1}^n\sqrt{-8\operatorname{tr}\big([h_c(t_{i-1},t_i),\hat V][h_c(t_{i-1},t_i)^{-1},\hat
V]\big)}.
\end{gather*}
In the limit $n\rightarrow\infty$ the approximation of $A$ by its
holonomy becomes exact.

In~\cite{length} it is shown that this is indeed a~well-def\/ined
operator on cylindrical functions and, due to the occurrence of the
volume operator, its action on SNW functions gives rise to nonzero
contributions only when the curve contains SNW vertices.
As soon as
the partition is f\/ine enough for each piece to contain not more than
one vertex, the result of $\hat L_n\Psi$ remains unchanged when the
partition is further ref\/ined.
So the continuum limit is reached for a f\/inite partition.\looseness=1

However this action on SNWs raises a~problem.
For any given generic
SNW a~curve $c$ will rarely meet a vertex, so that for macroscopic
regions lengths will always be predicted too short in relation to
volume and surface areas: $c$ is ``too thin''.
To obtain reasonable
results in the classical limit, one combines curves together to
tubes, that is two-dimensional congruences of curves with $c$ in the
center and with cross-sections of the order of $\ell_{\rm P}^2$.
The
spectra of the so-constructed tube-operators are purely discrete.

None of the phenomenological models discussed in this review depend on
the specif\/ic form of the length operator.
These have already been compared
from the geometric point of view~\cite{Blength}.
As with the volume operators
it would be interesting to develop phenomenological models that observationally
distinguish the dif\/ferent operators.

\subsection{Angle}
\label{angle}

The angle operator is def\/ined using a~partition of the closed dual surface around a single SNW node
into three surfaces, ${\cal S}_1$, ${\cal S}_2$, ${\cal
S}_3$, the angle operator is def\/ined in terms of the associated f\/lux
variables $E^i({\cal S}_I)$~\cite{angle}
\begin{gather}\label{angledef}
\theta^{(12)}_{n}:=\arccos
\frac{E^{i}({\cal S}_{1}) E_{i}({\cal S}_{2})}
{A({\cal S}_{1})A({\cal S}_{2})}.
\end{gather}
As is immediately clear from the form of the operator (and dimensional analysis), there is no scale associated to the angle operator.
It is determined purely by the state of the intertwiner, the atom of quantum geometry.
Deriving the spectrum of the angle operator of equation~\eqref{angledef} is a~simple exercise in angular momentum algebra~\cite{angle}.
Dropping all labels on the intertwiner except those that label the spins originating from one of the three partitions, we have
\begin{gather*}
\hat{\theta}_{(12)}\ket{j_1 j_2 j_3}
=\theta_{(12)}\ket{j_1 j_2 j_3}\qquad\text{with}\nonumber\\
\theta_{(12)}
=\arccos \left(\frac{j_{3}(j_{3}+1)-
j_{1}(j_{1}+1)-j_{2}(j_{2}+1)}{2\left[j_{1}(j_{1}+1)
j_{2}(j_{2}+1)\right]^{1/2}}\right).%\label{angle_spectrum}
\end{gather*}
where the $j_i$ are the spins on the internal graph labeling the intertwiner.
As such they can be seen to label ``internal faces'' of a~polyhedral decomposition of the node.
For a~single partition of the dual surface the angle operators commute.
But, ref\/lecting the quantum nature of the atom of geometry and the same non-commutativity as for area operators for intersecting surfaces, the angle operators for dif\/ferent partitions do not commute.

As is clear from a~glance at the spectrum there are two aspects of
the continuum angular spatial geometry that are hard to model with
low spin.
First, small angles are sparse.
Second, the distribution
of values is asymmetric and weighted toward large angles.
As
discussed in~\cite{mikes,Seife} the asymmetry persists even when the
spins are very large.

\subsection{Physicality of discreteness}

A characteristic feature of the above geometric operators is their discrete spectra.
It is natural to ask whether it is physical.
Can it be used as a~basis for the phenomenology of quantum
geometry? Using examples, Dittrich and Thiemann~\cite{BD} argue that
the discreteness of the geometric operators, being gauge
non-invariant, may not survive implementation in the full dynamics of LQG.
Ceding the point in general, Rovelli~\cite{Com}
argues in favor of the reasonableness of physical geometric
discreteness, showing in one case that the preservation of
discreteness in the generally covariant context is immediate.
In
phenomenology this discreteness has been a~source of inspiration for
models.
Nonetheless as the discussion of these operators makes
clear, there are subtleties that wait to be resolved, either through
further completion of the theory or, perhaps, through observational
constraints on phenomenological models.

\subsection{Local Lorentz invariance and LQG} \label{LLILQG}

It may seem that discreteness immediately gives rise to
compatibility problems with LLI. For instance, the length derived
from the minimum area eigenvalue may appear to be a~new fundamental
length.
However, as SR does not contain an invariant length, must
such a~theory with a~distinguished characteristic length be in
contradiction with SR? That this is not necessarily the case has
been known since the 1947 work of Snyder~\cite{Sn} (see \cite{E}
for a~recent review).

In~\cite{SS1} Rovelli and Speziale explain that a~discrete spectrum
of the area operator with a~minimal non-vanishing eigenvalue can be
compatible with the usual form of Lorentz symmetry.
To show this, it
is not suf\/f\/icient to set discrete eigenvalues in relation to Lorentz
transformations, rather, one must consider what an observer is able
to measure.
The main argument of~\cite{SS1} is that in quantum theory
the spectra of geometric variables are observer invariant, but
expectation values are not.
The authors explain this idea by means
of the area of a~surface.
Assume an observer~$\cal O$ measures the
area of a~small two-dimensional surface to be~$A$, a second observer~${\cal O}'$, who moves at a~velocity $v$ tangential to the surface,
measures~$A'$.
In classical SR, when~$\cal O$ is at rest with
respect to the surface in f\/lat space, the two areas are related as
$A'=\sqrt{1-v^2} A$.
If $A$ is suf\/f\/iciently small, this holds also
in GR.

However this relation, which allows for arbitrarily small values of $A'$,
cannot be simply taken over as a~relation between the area operators
$\hat A$ and~${\hat A}'$ in LQG. The above form suggests that~$A'$
is a~simple function of~$A$ and so~$\hat A$ and~${\hat A}'$ should
commute.
This is not the case.
The velocity~$v$, as the
physical relative velocity between $\cal O$ and ${\cal O}'$, depends
on the metric, which of course is an operator, too.
Rovelli and
Speziale show that $\hat v$ does not commute with $\hat A$, and so
$[\hat A,{\hat A}']\neq0$.
This means that the measurements of the
area and the velocity of a~surface are incompatible.

The apparent conf\/lict between discreteness and Lorentz contraction
is resolved in the follo\-wing way: The velocity of an observer, who
measures the area of a~surface sharply, is completely undetermined
with respect to this surface and has vanishing expectation value.
The indeterminacy of the velocity means that an observer who
measures the area~$A$ precisely cannot be at rest with respect to
the surface.
On the other hand, an observer with a~nonzero
expectation value of velocity relative to the surface cannot measure
the area exactly.
For this observer the {\em expectation value} is
Lorentz-contracted, whereas the {\em spectrum} of the area operator
is the same.\looseness=1

More recent considerations in the spin-foam framework can be found in
\cite{SS2}, where the Hilbert space of functions on $\SU(2)$ is mapped to a~set
$\cal K$ of functions on $\SL(2,\C)$ by the Dupuis--Livine map~\cite{Dup}.
In this way $\SU(2)$ SNW functions are mapped to $\SL(2,\C)$ functions
that are manifestly Lorentz covariant.
Furthermore these functions
are completely determined by their projections on $\SU(2)$, so $\cal
K$ is linearly isomorphic to a~space of functions on $\SU(2)$.
It is
shown in~\cite{SS2} that the transition amplitudes are invariant
under $\SL(2,\C)$ gauge transformations in the bulk and manifestly
satisfy LLI.

While these papers suggest strongly that LLI is part of LQG~-- just
as might be expected from a~quantization of GR~-- other researchers
have explored the possibility that the discreteness spoils or
deforms LLI through the modif\/ication of dispersion relations and interaction terms.

From a~fundamental theory point of view, the symmetry group associated to the f\/ield theory
of the continuum approximation, from which particles acquire their properties through irreducible representations,
will be dynamically determined by quantum gravity theory and the associated ground state.
Originating in work by Kodama, a~line work work contains hints that this group may be deformed.

Found in the late 80's~\cite{kodama,kodama2,Leereview}, the Kodama
state,
\begin{gather*}
\Psi_{\rm K}[A]=e^{\frac{3}{2\Lambda G\hbar} S_{\rm CS}[A]},
\end{gather*}
was a~source of hope that one could model the ground state
(and maybe excited states) of QG with a cosmological constant~$\Lambda$.
The phase, $S_{\rm CS}[A]$, is the Chern--Simons action
for complex Ashtekar connections, with the same symmetry group
as deSitter or anti deSitter space, according to the sign of $\Lambda$.
The wavefunction $\Psi_{\rm K}$ is (locally) gauge invariant,
spatially dif\/feomorphism invariant, and a~solution to the Hamiltonian constraint
of LQG for a~(triads-on-left) factor ordering in complex Ashtekar variables.
When $\Psi_{\rm K}[A]$ is multiplied by SNW functions of the connections a~picture emerges
in which states of quantum gravity are labeled by knot (or more accurately, graph)
classes of framed spin networks~\cite{qloops,majordis,Leereview}.
The space-time has DeSitter as a semiclassical limit~\cite{Leereview}.
There is also an intriguing link between the cosmological constant and particle statistics~\cite{majordis}.

It is well-known that for space-time with boundary, boundary terms and/or conditions
must be added to the Einstein--Hilbert action to ensure that the variational principle
is well def\/ined and Einstein's equations are recovered in the bulk.
(Possible boundary conditions and boundary terms for real Ashtekar variables were worked out
in~\cite{boundary, majordis}.)

However, there are severe dif\/f\/iculties with this choice of complex-valued self-dual connection variables
and the Kodama state: The kinematic state space of complex-valued connections is not yet rigorously
constructed~-- we lack a~uniform measure.
The state itself is both not normalizable in the linearized theory,
violates CPT and is not invariant under f\/inite gauge transformations (see~\cite{T1} for discussion).
An analysis of perturbations around the Kodama state shows that the perturbations of the Kodama state
mix positive-frequency right-handed gravitons with negative-frequency left-handed gravitons~\cite{BM2}.
The graph transform of the Kodama states, def\/ined through variational methods, acquires a~sensitivity
to tangent space structure at vertices~\cite{kodama_embed}.
Finally, the original $q$-deformation of the loop algebra suggested in~\cite{qloops,majordis}
is inconsistent~\cite{loop_algebra}.
These dif\/f\/iculties have made further progress in this area challenging, although there is work
on generalizing the Kodama state to real Ashtekar variables, where some of these issues are addressed~\cite{Ran}.

Following the lead of developments in 3D gravity coupled to point particles, where particle kinematics
is deformed when the topological degrees of freedom are integrated out, one may wonder whether
a~similar situation holds in $3+1$ when the local gravitational ef\/fects are integrated out~\cite{KGS}.
In~\cite{KGS} the authors showed that, for BF theory with a~symmetry breaking term controlled
by a parameter~\cite{S,SS3,FS}, (point) particles enjoy the usual dispersion relations and
any deformation appears only in interaction terms.

In the next section we review frameworks in which the symmetry groups are deformed or broken.

\section{Quantum particle frameworks}\label{section3}

\subsection{Relativistic particles and plane-waves}\label{sec:particle}
We start by recalling the fundamental structures associated with the
physics of free particles in the phase space picture.
Constructing a
phenomenological model to incorporate the Planck scale consists in generalizing or modifying this structure.

A relativistic particle (with no spin) propagating in Minkowski spacetime is described in the Hamiltonian formalism
by the following structures.
\begin{itemize}\itemsep=0mm
\item \textit{A phase space} $\cP\sim T^*\R^4\sim\R^4\times\R^4$,
the cotangent bundle of the f\/lat manifold $\R^4$.
It is
parameterized by the conf\/iguration coordinates $x^\mu\in\R^4$ and
the momentum coordinates $p_\mu\in\R^4$.
These coordinates have a
physical meaning, i.e.\ they are associated with outcome of
measurements (e.g.\
using rods, clocks, calorimeters, etc.).
$\cP$ is equipped
with a~Poisson bracket, that is, the algebra of (dif\/ferentiable)
functions over the phase space $\cC(\cP)$ is equipped with a~map
$\lbrace,\rbrace:\,\cC(\cP)\times\cC(\cP)\dr\cC(\cP)$ which
satisf\/ies the Jacobi identity.
For the coordinate functions, the standard Poisson bracket is given by
\begin{gather*}%\label{poisson}
\{x^\mu,x^\nu\}=0,\qquad\{x^\mu,p_\nu\}=\delta^\mu_\nu,\qquad\{p_\mu,p_\nu\}=0.
\end{gather*}

\item \textit{Symmetries} given by the Poincar\'e group $\cP\sim\SO(3,1)\ltimes
\cT$, given in terms of the semi-direct product of the Lorentz
group
$\SO(3,1)$ and the translation group $\cT$.
So that there exists an
action of the Lorentz group on the translation, which we note
$\Lambda\act h$, $\forall\, (\Lambda,h)\in\cP$.
The product of group
elements is hence given by
\begin{gather*}%\label{cross product}
(\Lambda_1,
h_1)(\Lambda_2,h_2)=(\Lambda_1\Lambda_2,h_1(\Lambda_1\act h_2)).
\end{gather*} The Lie algebra $\mathfrak{P}$ of $\cP$ is generated by the
inf\/initesimal Lorentz transformations $J_{\mu\nu}$ and translations
$T_\mu$ which satisfy
\begin{gather*}
\com{J_{\mu\nu},
J_{\alpha\beta}}=\eta_{\mu\beta}J_{\nu\alpha}+\eta_{\nu\alpha}J_{\mu\beta}
-\eta_{\mu\alpha}J_{\nu\beta}-\eta_{\nu\beta}J_{\mu\alpha},
\nonumber\\
\com{T_\alpha,T_\beta}=0,\qquad \com{J_{\mu\nu},
T_\alpha}=\eta_{\nu\alpha}T_\mu-\eta_{\mu\alpha}T_\nu.%\label{poinc lie algebra}
\end{gather*}
The action of $\mathfrak{P}$ is
given on the phase space coordinates by
\begin{gather*}
J_{\mu\nu}\act x^\alpha=\eta_\nu^\alpha x_\mu-\eta_\mu
^\alpha x_\nu,
\qquad
J_{\mu\nu}\act p_\alpha=\eta_{\nu\alpha}
p_\mu-\eta_{\mu\alpha}p_\nu,
\nonumber\\
T_\mu\act x^\nu=\delta_{\mu}^{\nu},
\qquad
T_\mu\act p_\nu=0.%\label{poinc lie action}
\end{gather*}
This is extended naturally to the functions on phase space.

\item \textit{Particle dynamics} given by the mass-shell or dispersion relation\footnote{We use
the metric $\eta^{\mu\nu}=\operatorname{diag}(+,-,-,-)$.}
$p^2=p_\mu\eta^{\mu\nu}p_\nu=m^2$.
This is a~constraint on phase space which implements the time reparameterization invariance of the following action
\begin{gather*}
\text{\ss}=\int{\rm d}\tau \left(\dot x^\mu p_\mu-\lambda\left(
p^2-m^2\right) \right).
\end{gather*}
 $\lambda$ is the Lagrange multiplier
implementing the constraint $p^2-m^2=0$.
This action contains the
information about the phase space structure and the dynamics.
We
can perform a~\textit{Legendre transform} in the massive case (or a
\textit{Gauss transform} in the massless case) to express this
action in the tangent bundle $T\R^4$,
\begin{gather*} %\label{action particle}
\text{\ss}= m\int{\rm d}\tau \sqrt{\dot x^\mu \dot x^\nu \eta_{\mu\nu}(x)},
\qquad \dot x^\mu= \frac{dx^\mu}{d\tau}.
\end{gather*}
With this description, we recover the familiar fact that the relativistic particle worldline given by a
geodesic of the metric.
\end{itemize}
When we require the Poincar\'e symmetries to be consistent with all these phase space and particle
dynamics structures, these pieces f\/it together very tightly.

\begin{itemize}\itemsep=0mm
\item The Poincar\'e symmetries should be compatible with the Poisson bracket.
If we def\/ine our theory in a~given inertial frame, physics will not change if we use a dif\/ferent inertial frame,
related to the initial one by a Poincar\'e transformation~$t$,
\begin{gather*}
\poi{f_1(x,p),f_2(x,p)}=f_3(x,p)\nonumber\\
\qquad\Leftrightarrow\quad
\poi{f_1(t\act x,t\act p),f_2(t\act x,t\act p)}=f_3(t\act x,t\act p),%\label{poinpoi}
\qquad f_i\in\cC(P).
\end{gather*}

\item The mass-shell condition/dispersion relation $p^2=m^2$ encodes the mass Casimir of the Poincar\'e group.
As such this mass-shell condition is invariant under Lorentz transformations.
\end{itemize}

When dealing with f\/ields or multi-particles states, we have also the following important structures.
\begin{itemize}\itemsep=0mm
\item \textit{The total momentum } of many particles is obtained using a~group law for the momentum,
adding extra structure to the phase space.
We are using~$\R^4$, which is naturally equipped with an Abelian group
structure\footnote{$\R^4$ is even a~vector space but for the
generalizations we shall consider, it is only the group structure
that is relevant.}.
From this perspective, one can consider
the phase space as a~cotangent bundle over the group~$\R^4$.
This picture will be at the root at the generalization to the
non-commutative case.

\item \textit{Plane-waves} $e^{ix^\mu k_\mu}$, where $k_\mu$ is the
wave-covector, are an important ingredient when we deal with f\/ield
theories.
The plane-wave is usually seen as the eigenfunction of the
dif\/ferential operators encoding the inf\/initesimal translations on
momentum or con\-f\/i\-gu\-ra\-tion space
\begin{gather}\label{planewave translation}
\partial_{x^\mu}e^{ix^\nu k_\nu}=ik_\mu e^{ix^\nu k_\nu},\qquad \partial_{k_\mu}e^{ix^\nu k_\nu}=ix^\mu e^{ix^\nu k_\nu}.
\end{gather}
Since the momentum operator $P_\mu$ is usually represented as
$-i\partial_{x^\mu}$, it is natural to identify the wave-covector to
the momentum $k_\mu=p_\mu$.
When this identif\/ication is implemented, the product of plane-waves is intimately related to the addition of momenta,
hence the group structure of momentum space.
\begin{gather*}
e^{ix\cdot p_1}e^{ix\cdot p_2}=e^{ix\cdot(p_1+p_2)}.
\end{gather*}

\item The inf\/initesimal translation $T_\mu$ is represented as $\partial_{x^\mu}$ therefore it
can be related to the momentum operator from~\eqref{planewave translation}.
Modi\-fying momentum space is then synonymous to modi\-fying the translations.
\end{itemize}

As we are going to see in the next sections, introducing QG ef\/fects
in an ef\/fective framework will consist in modifying some of the
above structures, either by brute force by breaking some
symmetries or, in a~smoother way, by deforming these symmetries.

\subsection{Introducing Planck scales into the game: modif\/ied dispersion relations}
\label{MDRmodels}

Light, or the electromagnetic f\/ield, is a~key object to explore the structure of spacetime.
In
1905, light performed a~preferred role in understanding Special Relativity.
In 1919, Eddington measured the bending of light induced by the curvature of spacetime.
As a consequence, these
results pointed to the fact that a~Lorentzian metric is the right structure to describe a classical spacetime.

In the same spirit, a~common idea behind QG phenomenology is that a semi-classical spacetime
should leave imprints on the propagation of the electromagnetic f\/ield such as in~\cite{A-CEMNS} discussed in the Introduction.
In this example the lever arm that raises possible QG ef\/fects into view is the proposed \textit{cumulative} ef\/fects over great distances.

The concept of a~modif\/ied dispersion relation (MDR) is at the root
of most QG phenomenology ef\/fective theories.
Depending on the approach one
follows, there can also be some modif\/ications at the level of the
multiparticle states, i.e.\
how momenta are added.
One can readily see
that such a~modif\/ied dispersion relation is not consistent with the
Lorentz symmetries, so that they have to be broken or deformed.
We
shall discuss both possibilities below.

There is nevertheless a~semi-classical regime where the Planck scale is relevant and possible non trivial ef\/fects regarding symmetries could appear.
Indeed, the natural f\/lat semi-classical limit in the QG regime is given
by\footnote{$\Lambda$ is the cosmological constant, $G$ the Newton
constant and $\hbar$ the Planck constant.} $\Lambda,G,\hbar\dr0$.
There are a~number of possibilities to implement these limits~\cite{Freidel:2005bb}.
An interesting f\/lat semi-classical limit is
when $\Lambda=0$ and $\frac{\hbar}{G}=\smp^2$ is kept constant in the
limit $G,\hbar\dr0$.
This regime is therefore characterized by a~new constant
$\smp$, which has dimension either energy, momentum or mass.
Note
that in this regime the Planck length $\ell_{\rm P}^2={\hbar
G}$ naturally goes to zero, hence there is no minimum length
from a~dimensional argument.
The key question is how to implement this momentum scale $\smp$,
that is to identify the physical motivations which will dictate how to encode this scale in the theory.

Following the paper by Amelino-Camelia~et al.~\cite{A-CEMNS}, many modeled
potentially observable QG ef\/fects with ``semi-classical'' ef\/fective theories.
In some cases discreteness was put in ``by hand''.
In others deviations from Special Relativity, suppressed by the ratio (particle energy)/(QG
scale) or some power of it, were modeled.
This is the approach followed when considering Lorentz symmetry violation discussed in Section~\ref{LV}.
Another approach taken was to introduce the Planck length in the game as a~minimum length and investigate possible consequences.
For a~recent review on this notion and implications of minimum length see~\cite{sabine:mini-length}.
Alternatively the Planck energy, or the Planck momentum, was set as the maximum energy~\cite{Bruno:2001mw} (or maximum momentum) that a~fundamental particle could obtain.
Implementing this feature can also generate a~modif\/ied dispersion relation.
This is the approach which is often considered in the deformed symmetries approach.

Both of these later proposals af\/fect dispersion relations and hence the Poincar\'e symmetries.
Therefore in the regime $\lim\limits_{{\hbar},{G}\dr0}\frac{\hbar}{G}=\smp^2$,
it is not clear that the symmetries must be preserved and some non-trivial ef\/fects can appear.

In general, the idea is to cook up more or less rigorously an ef\/fective model and then try to relate it to a
given QG model (bottom-top approach).
The models which are (the most) well def\/ined mathematically
are, to our knowledge, given by the non-commutative approach and the Finsler geometry approach.
Among these two, Finsler geometry is the easiest to make sense at the physical level.

There are fewer attempts to derive semi-classical ef\/fects from QG models (top-down approach).
Most of the time, these attempts to relate the deep QG regime and the semi-classical are heuristic: there is no real complete QG theory at this time and the semi-classical limit is often problematic.
We shall review some of them when presenting the dif\/ferent QG phenomenological models.
Even though these attempts were few and heuristic, they were inf\/luential, promoting the idea that it is possible to measure ef\/fects originating at the Planck scale.

Currently, QG phenomenology is therefore not f\/irmly tied to a~particular quantum theory of gravity.
For a~brief, general review over quantum gravity phenomenology, independent of a~fundamental theory,
see~\cite{LM}.
Contemporary observational data are not suf\/f\/icient to rule out QG theories, not only
because of the lack of stringent data, but particularly because the link between fundamental theories
and QG phenomenology is loose.
Nevertheless, present observational data restrict parameters in some models,
ef\/fectively ruling out certain modif\/ications, such as cubic modif\/ications to dispersion relations in the ef\/fective f\/ield theory
(EFT) context.
We shall review this in Section~\ref{LV}.

In the following we are going to present the main candidates to encode some
QG ef\/fective semi-classical ef\/fects.
When available we shall also recall the
arguments relating them to LQG. As a~starter, we now recall dif\/ferent arguments
which attempt to justify a~MDR from the LQG perspective.

\subsection{Arguments linking modif\/ied dispersion relations and LQG}
\label{MDR}

We present three quite dif\/ferent strategies to
establish a~f\/irmer tie between LQG and modif\/ied dispersion relations.
The f\/irst one introduces a~heuristic set of weave
states, f\/lat and continuous above a~characteristic scale~$L$, and
then expands the f\/ields around this scale.
The second strategy
starts from full LQG and aims at constructing quantum f\/ield theory
(QFT) on curved space-time which is an adaptation of conventional QFT
to a~regime of non-negligible, but not too strong gravitational
f\/ield.
In this construction coherent states of LQG are employed,
which are quantum counterparts of classical f\/lat space.
Due to the
enormous complications, this venture must resort to many
approximations.
The third strategy deals in a~very general way with
quantum f\/luctuations around classical solutions of GR. This approach
is rather sketchy and less worked-out in details.
Given the
preliminary stage of development of LQG all of the derivations
employ additional assumptions.
Nevertheless they provide a~starting
point for exploring the possible ef\/fects of the discreteness of LQG.

Departures from the standard quadratic energy-momentum relations and
from the standard form of Lorentz transformations can of course
originate from the existence of a~preferred reference frame in the
limit of a~vanishing gravitational f\/ield, i.e.\
a breaking of Lorentz
invariance at high energies.
Nevertheless, this need not necessarily
be the case.
The relativity principle can be valid also under the
conditions of modif\/ied dispersion relations and Lorentz
transformations.
In~\cite{Rel} the compatibility of a~second
invariant quantity in addition to the speed of light, a~length of
the order of the Planck length, with the relativity principle was
shown.
The product of this length with a~particle energy is a
measure for the modif\/ication of the dispersion relation.
Frameworks
with two invariant scales, where the second one may also be an
energy or a~momentum, were dubbed ``doubly special relativity
theories'' (DSR).
As an outcome of the theory's development, it was
found that ``DSR'' may also be an acronym for ``deformed special
relativity'' in that Poincar\'{e} Lie algebra of symmetry generators,
namely the energy and momentum operators, may be deformed or
embedded into a~Hopf algebra~\cite{majid}, whereas in the doubly
special relativity framework the representation of the Poincar\'{e}
group, i.e.\
the action on space-time or momentum space, is
nonlinearly deformed.
Deformed algebras are used in the
$\kappa$-Minkowski and in the $\kappa$-Poincar\'{e} approach
\cite{kappa,kappa2}.
For relations to doubly special relativity see
\cite{dsr}.

\subsubsection{MDR from weave states}\label{GP}

Following the f\/irst strategy of introducing an heuristic state
Gambini and Pullin~\cite{GP} modeled a~low energy semi-classical kinematic state with a~``weave'',
a discrete approximation of smooth f\/lat geometry,
characterized by a~scale $L$.
In an inertial frame, the spatial
geometry reveals its atomic nature below the characteristic length
scale.
Above this length scale~$L$, space appears f\/lat and
continuous.
In this preferred frame the expectation value of the
metric is of the form
\begin{gather*}
\ev{q_{ab}}=\delta_{ab}+O\left(\frac{\ell_{\rm P}}{L}\right).
\end{gather*}
To see the leading order ef\/fect for photons Gambini and Pullin
analyzed the Maxwell Hamiltonian,
\begin{gather*}
H=\frac{1}{2}\int{\rm d}^3x\frac{q_{ab}}{\sqrt{q}}\big(E^a E^b+B^a B^b\big)
\end{gather*}
importing one key idea from LQG.
The densitized metric operator $q_{ab}/\sqrt{q}$ is expressed as a~pro\-duct of two operators $\hat{w}_a(v_i)$,
which are commutators of the connection and the volume operator.
These operators are f\/inite and take non-vanishing values only at
vertices $v_i$ of the graph.
Re\-gu\-la\-ting the Hamiltonian with point
splitting, the authors took the expectation value of the
Hamiltonian in the weave state, averaging over a~cell of size~$L$.
They expanded the f\/ields around the center of the cell~$P$ and found that the leading order term
\begin{gather*}
\ev{\hat{w}_a(v_i)\hat{w}_b(v_j)}(v_i-P)_c
\end{gather*}
is a~tensor with three indices.
Assuming rotational symmetry, this term is
proportional to $\epsilon_{abc}\ell_{\rm P}/L$, thus modif\/ing Maxwell's equations.
The correction is parity violating.
The resulting dispersion relations enjoy cubic modif\/ications, taking the form
\begin{gather*}%\label{photonMDRGP}
\omega_\pm^2=k^2\mp4\chi\frac{k^3}{\smp}
\end{gather*}
in the helicity basis.
The constant
$\chi$ was assumed to be order~1.
Hence the weave states led to
birefringence.
As discussed in Section~\ref{biref} these ef\/fects may
be constrained by observation.
Furthermore some theoretical arguments can also be proposed against the validity of such proposal as we shall see in Section~\ref{sec:finsler}.

Taking a~similar approach and specifying general properties of a
semi-classical state, Alfaro  et~al.\ found that, in an
analysis of particle propagation, photon~\cite{AM-TUphot} and
fermion~\cite{AM-TUneut,AM-TUfermions} dispersion relations are
modif\/ied.
They f\/ind these by applying LQG techniques on the
appropriate quantum Hamiltonian acting on their states.
Following
similar steps to Gambini and Pullin, Alfaro~et~al.\ expand
the expectation value of the matter Hamiltonian operators in these
states.

To determine the action of the Hamiltonian operator of the f\/ield on
quantum geometry Alfaro  et~al.\ specify general conditions for
the semi-classical state.
The idea is to work with a~class of
states for geometry and matter that satisfy the following
conditions:
\begin{enumerate}\itemsep=0pt
\item The state is ``peaked'' on f\/lat and continuous geometry when probed
on length scales larger than a~characteristic scale $L$, $L\gg\ell_{\rm P}$.
\item On length scales larger than the characteristic length the state is ``peaked'' on the classical f\/ield.
\item The expectation values of operators are assumed to be well-def\/ined and geometric corrections
to the expectations values may be expanded in powers of the ratio of
the physical length scales, $L$ and $\ell_{\rm P}$.
\end{enumerate}
The authors dub these states ``would-be semi-classical states''.
States peaked on f\/lat geometry and a~f\/lat connection are expected
for semiclassical or coherent states that model f\/lat space.
Lacking
the quantum Hamiltonian constraint for the gravitational f\/ield and
thus also for the associated semi-classical states, the work of
Alfaro  et~al.\ is necessarily only a~forerunner of the
detailed analysis of semi-classical states.
See~\cite{ST1, ST2} for
further work on semiclassical states and dispersion relations.
To
parameterize the scaling of the expectation value of the
gravitational connection in the semiclassical state the authors
introduce a~parameter $\Upsilon$ that gives the scaling of the
expectation value of the geometric connection in the semi-classical
state $\ket{W \phi}$
\begin{gather*}
\bra{W \phi}A^i_a\ket{W \phi}\sim\frac{1}{L}\big({\ell_{\rm P}}{L}\big)^\Upsilon\delta_a^i,
\end{gather*}
where $\phi$ are the matter f\/ields.
The determination of the scaling
is a~bit of a mystery.
Alfaro  et~al.\ propose two values for
$L$: The ``mobile scale'' where $L=1/p$, and the ``universal'' value
where~$L$ is a~f\/ixed constant,
$p$~is the magnitude of the
3-momentum of the particles under consideration.
We will see in the
next section that matching the modif\/ications to the ef\/fective f\/ield
theory suggests a~universal value $L\gg\ell_{\rm P}$ and $\Upsilon
\ge0$, so we will use the universal value.
It is not surprising
that Lorentz-violating (LV) terms arise when the spatial distance $L$ is
introduced.

Expanding the quantum Hamiltonian on the semi-classical states
Alfaro   et~al.\ f\/ind that particle dispersion relations are
modif\/ied.
Retaining leading order terms in $p/\smp$, the scaling
with~$(L\smp)$ and next to leading order terms in $\smp$, but
dropping all dimension~3 and~4 modif\/ications for the present, the
modif\/ications are, for fermions,
\begin{gather}
\label{neutdisp}
E^{2}_{\pm}\simeq\big[1+2\kappa_1\left(L\smp\right)^{-\Upsilon-1}\big]p^{2}+m^{2}
\pm\frac{\kappa_9m^2}{\smp}p
\mp\frac{\kappa_7}{2L\smp^2}\left(L\smp\right)^{-\Upsilon}p^{3}-\frac{\kappa_3}{\smp^2}
p^4,
\end{gather}
where $p$ is the magnitude of the 3-momentum and the dimensionless
$\kappa_i$ parameters are expected to be $O(1)$ (and are unrelated to the Planck scale~$\smp$.
The labels are for
the two helicity eigenstates.
These modif\/ications are derived from
equation (117) of~\cite{AM-TUfermions}, retaining the original
notation, apart from the Planck mass $\smp$.

Performing the same expansion for photons Alfaro  et~al.\ f\/ind
that the semi-classical states lead to modif\/ications of the
dispersion relations, at leading order in $k/\smp$ and scaling $(L
\smp)$
\begin{gather}
\label{photdisp}
\omega_\pm^{2}\simeq k^{2}\big[1+2\theta_7(L\smp)^{-2-2\Upsilon}\big]\pm\frac{4\theta_8}{\smp}
\big[1+2\theta_7(L\smp)^{-2-2\Upsilon}\big]k^{3}+\left(2\theta_8-4\theta_3\right)\frac{k^4}{\smp^2},
\end{gather}
where the $\theta_i$ parameters are dimensionless and are expected
to be $O(1)$.
The leading order term is the same
polarization-dependent modif\/ication as proposed in Gambini and
Pullin~\cite{GP}.
In the more recent work \cite{ST1, ST2} the structure
of the modif\/ication of the dispersion relations was verif\/ied but,
intriguingly, the corrections do not necessarily scale with an
integer power of~$\smp$.

As is clear in the derivation these modif\/ied dispersion
relations (MDR) manifestly break LLI and so are models of LQG with a~preferred frame.
The ef\/fects are
suppressed by the Planck scale, so any $O(1)$ constraints on
the parameters are limits placed on Planck-scale ef\/fects.
These constraints, without a~complete dynamical framework that establishes the
conservation, or deformation, of energy and momentum, must come from
purely kinematic tests.

Interestingly, as we will see in Section~\ref{LV}, Alfaro  et~al.\ found the modif\/ications to the dispersion relations corresponding to the dimension 5 and the CPT-even dimension 6 LV operators in the ef\/fective f\/ield theory
framework.
Of course given the limitations of the model they did not derive the complete particle dynamics of the EFT framework.

It was suggested in~\cite{KP} that dif\/ferent
choices for the canonical variables for the ${\rm U}(1)$ f\/ield theory could
remove the Lorentz violating terms.
However Alfaro  et~al.\
pointed out that this is inconsistent; the only allowed canonical
pairs in LQG are those that have the correct semi-classical limit
and are obtained by canonical transformation~\cite{ARM-TU}.

Finally, we must emphasize that these derivations depend critically on assumptions about the semi-classical weave state, the source of the local Lorentz symmetry violations.

\subsubsection{Quantum f\/ield theory in curved space from LQG}\label{QFT-LQG-MDR}

Sahlmann and Thiemann studied
dispersion relations in a~framework of QFT on curved space from
basic LQG principles by heavily making use of approximations~\cite{ST1, ST2}.
In the
f\/irst step QFTs on discrete space are constructed on an essentially
kinematic level.
Rather than taking the total Hamiltonian constraint of
gravity and matter, the matter Hamiltonians of gauge, bosonic and
fermionic f\/ields are treated as observables, dependent on geometric
variables of the background.
Then the gravitational f\/ield is assumed
to be in a~coherent state, where expectation values for f\/ield
variables yield the classical values and the quantum uncertainties
are minimal.

The Hilbert space of matter states ${\cal H}^{\rm Fock}_{\rm
m}(m)$ depends on the state $g$ of the geometry.
The vacuum
state $\Omega_{\rm m}(g)$ is the ground state of some
(geometry-dependent) matter Hamiltonian operator $\hat{H}_{\rm
m}(g)$.
In QG $m$ becomes an operator, and so $\Omega_{\rm
m}(g)$ becomes a~``vacuum operator'', i.e.\
a function of the
matter degrees of freedom with values in ${\cal L}({\cal H}^{\rm
kin}_{\rm geom})\otimes{\cal H}^{\rm kin}_{\rm m}$.
$\cal L$ is
the space of linear operators on a~background independent Hilbert
space of kinematic states of geometry, ${\cal H}^{\rm kin}_{\rm
m}$ is a~kinematic matter Hilbert space.
From this vacuum
operator a~vacuum state may be constructed in principle as
expectation value in a~state of quantum geometry, which is peaked at
classical f\/lat space.

As a~technical detail and interesting twist, the construction of
annihilation and creation ope\-rators involves fractional powers of
the Laplacian on a~background metric.
In~\cite{ST1, ST2} these operators
are constructed but the spectra required to calculate fractional
powers are not known.
To circumvent this problem the
expectation values of the Laplacian in a~coherent state, mimicking
f\/lat space, are calculated f\/irst, then fractional powers are taken.
In
coherent states this approximation coincides with exact calculations
in zeroth order in~$\hbar$.
The coherent state employed is modeled
by spin networks with an irregular 6-valent lattice.
Creation and
annihilation operators, an approximate vacuum state, and approximate
Fock states are constructed from the matter Hamiltonians in the
sense of the described approximation.
The inf\/luence of geometry is
included in an ef\/fective matter Hamiltonian, the matrix elements of
which are def\/ined in the following way
\begin{gather*}
\big\langle\psi_{\rm m},\hat{H}^{\rm ef\/f}_{\rm m}(g)\,\psi'_{\rm m}\big\rangle_{{\cal H}_{\rm m}^{\rm kin}}
:=\big\langle\psi_{\rm grav}(g)
\otimes
\psi_{\rm m},\hat{H}\psi_{\rm grav}(g)
\otimes
\psi'_{\rm m}\big\rangle_{{\cal H}_{\rm grav}^{\rm kin}
\otimes
{\cal H}_{\rm m}^{\rm kin}}.
\end{gather*}
Here the general building principle of a~matter-geometry
Hamiltonian, related to a~graph $\Gamma$ is
\begin{gather*}
\hat H_\Gamma=\sum_{v,l;v',l'}\hat M_l(v)^{\dag}\,\hat G\big(v,l;v',l'\big)\,\hat M_{l'}\big(v'\big),
\end{gather*}
where $\hat M$ a~matter operator with some discrete (collective,
matter and geometry) label $l$, $v$ is a~SNW vertex and $\hat G$ is
an operator of geometry.

Recently work in the context of cosmological models has also
found hints of a~modif\/ied dispersion relation~\cite{DLT}.
Working in the
context of a~quantized Bianchi~I model with a scalar f\/ield, the authors found, when
taking back-reaction into account, that the scalar f\/ield modes
propagate on a~wave number-dependent metric and the
dispersion relation is modif\/ied~\cite{DLT}.

%\looseness=1
The discreteness and irregularity of the underlying lattice breaks
translation and rotation symmetry.
There are no exact plane wave
solutions for matter f\/ields, only in the long distance limit the
irregularities average out and so for long wavelengths~-- compared to
the lattice spacing~-- there are at least approximate plane waves.
In
this limit the matter Hamiltonians simplify suf\/f\/iciently, so
that the sketched program becomes feasible and yields an
energy-momentum dispersion relation for low energies, which carries
the imprints of both discreteness and f\/luctuations.

With these possible modif\/ications to particle dispersion relations,
the obvious next step is to explore the ef\/fects that arise from
these Lorentz violating modif\/ications.
Early studies
\cite{limits,JLM} used particle kinematics phenomenology, modif\/ied
dispersion relations plus energy-momentum conservation.
We now
know, see e.g.~\cite{JLMrev}, that constraints require a~full
dynamical framework for the f\/ields.
The most obvious, and certainly
most developed framework is ef\/fective f\/ield theory.
We review the
physical ef\/fects of cubic modif\/ications to the dispersion relations
in ef\/fective f\/ield theory in Section~\ref{LV}.
We brief\/ly discuss
alternate frameworks and higher dimension modif\/ications in Sections~\ref{alternate} and~\ref{beyondD5}.

\subsubsection{MDR from Hamilton--Jacobi theory}
\label{smolin-frame}

In~\cite{LD} by Smolin the occurrence of corrections to particle
kinematics in the low-energy limit of QG is made plausible in a~very
general way.
This derivation is based on the quantum f\/luctuations around
classical GR in the connection representation.
The dynamics is formulated in the Hamilton--Jacobi
theory with the aid of the action functional $S[A]$.
Canonical
conjugate momenta are given by
\begin{gather*}
{E^a}_i(x)=\frac{1}{\rho}\frac{\delta S[A]}{\delta{A_a}^i(x)},
\end{gather*}
with $\rho$ being a~constant with dimension (length)$^{2}$.
The solutions to
the dynamics form a~trajectory $({{A^0}_a}^i(t),{{E^0}^a}_i(t))$
with some parameter $t$.
The parameter can be chosen to be
proportional to the Hamilton--Jacobi functional.
This functional, in
turn, can be written as an integral over a~density $\cal S$ on the
spatial manifold, or some local coordinate neighborhood,
\begin{gather*}
S[A]=\int_\Sigma{\cal S}[A].
\end{gather*}
Therefore, on the classical space-time there exists a~time $T$
proportional to ${\cal S}[A]$, def\/ining a~sli\-cing.
The slicing
constructed in this way is determined by the classical solution.
When connections depart slightly from the classical trajectory the
slicing f\/luctuates.
So variations of functions on
conf\/iguration space, evaluated at the classical trajectory, can be
related to variations on space-time, expressed as
\begin{gather}\label{dT}
\frac{\rm d}{{\rm d}T}=\mu \frac{\delta}{\delta{\cal S}[A]},
\end{gather}
where $\mu$ is a~constant of dimension (length)$^{-1}$.
The solution ${{E^0}^a}_i(t)$ def\/ines a~matrix
\begin{gather*}
g=-{\rm d}T^2+\sum_ie^0_i\otimes e^0_i
\end{gather*}
with the orthonormal inverse triad $e^0_i$ related to ${{E^0}^a}_i$.

In the neighbourhood of a~classical trajectory the connection $A$ can
be formulated through a~dependence on $\cal S$ and quantities ${a_a}^i$
\cite{LSCS}, so that
\begin{gather*}
\frac{\delta}{\delta{A_a}^i(x)}=\frac{1}{M} {{E^0}^a}_i \frac{\delta}{\delta\cal
S}+\frac{\delta}{\delta{a_a}^i}.
\end{gather*}
The f\/irst term can be understood as variation in the internal time
coordinate, ${a_a}^i$ contains the gravitational degrees of freedom.

Going over to quantum theory we construct the operator
\begin{gather*}
\hat{E^a}_i(x)=-\hbar\rho\frac{\delta}{\delta{A_a}^i(x)}
\end{gather*}
and the semiclassical state functional
\begin{gather*}
\Psi_0[A]=e^{i\frac{S[A]}{\hbar}}\qquad\text{with}\qquad\hat{E^a}_i \Psi_0[A]={{E^0}^a}_i \Psi_0[A].
\end{gather*}

To study semiclassical QG ef\/fects on the propagation of a~matter
f\/ield $\phi$, Smolin takes quantum states in the Born--Oppenheimer form
$\Psi[A,\phi]=\Psi_0[A] \chi[A,\phi]$.
In the neighborhood of the classical trajectory
\begin{gather*}
\chi[A,\phi]=\chi[{\cal S},{a_a}^i,\phi]
\end{gather*}
and the action of $\hat{E^a}_i$ on such functions is
\begin{gather*}
\hat{E^a}_i(x) \chi[A,\phi]=-i\hbar\rho\frac{\delta\chi[A,\phi]}
{\delta{A_a}^i(x)}=\left({{E^0}^a}_i \frac{i\hbar\rho}{M}\frac{\delta}{\delta{\cal
S}(x)}-i\hbar\rho \frac{\delta}{\delta{a_a}^i(x)}\right)\chi[{\cal
S},{a_a}^i,\phi].
\end{gather*}
By~\eqref{dT} we have
\begin{gather*}
\frac{i\hbar\rho}{M}\,\frac{\delta}{\delta{\cal
S}(x)}=\frac{i\hbar\rho}{M\mu} \frac{\rm d}{{\rm d}T},
\end{gather*}
where $\frac{\hbar\rho}{M\mu}$ has dimension of length or time in
natural units.
The only length scale in the problem (leaving aside $\Lambda$) being the
Planck length, we may write
\begin{gather*}
\frac{\hbar\rho}{M\mu}=\alpha\ell_{\rm P}\equiv\frac{\alpha}{\smp},
\end{gather*}
where the constant $\alpha$ is determined later on in~\cite{LD}.

On the classical trajectory $\chi[{\cal
S},{a_a}^i,\phi]=\chi\big[T,{a_a}^i,\phi\big]$.
At the semiclassical
level we can neg\-lect~$\delta/\delta{a_a}^i$, which describes
couplings of matter to gravitons.
Finally we have
\begin{gather*}
\hat{E^a}_i \Psi[A,\phi]=\Psi_0[A] {{E^0}^a}_i\left(1-i\alpha
\ell_{\rm P} \frac{\rm d}{{\rm d}T}\right)\chi\big[T,{a_a}^i,\phi\big].
\end{gather*}

Now consider a~semiclassical state of def\/inite frequency with
respect to the time $T$,
\begin{gather*}
\chi\big[T,{a_a}^i,\phi\big]=e^{-i\omega T}\chi_\omega\big[{a_a}^i,\phi\big].
\end{gather*}
The action of the triad operator on such a~function is
\begin{gather*}
\hat{E^a}_i(x) \Psi[A,\phi]=\Psi_0[A] {{E^0}^a}_i\,(1-\alpha
\ell_{\rm P}\omega) \chi_\omega\big[{a_a}^i,\phi\big].
\end{gather*}

So the classical solution ${{E^0}^a}_i(x,T)$ ef\/fectively goes over
into
\begin{gather*}
{{E^0}^a}_i(x,T,\omega)={{E^0}^a}_i(x,T)(1-\alpha\ell_{\rm
P}\omega),
\end{gather*}
which implies an energy-dependent spatial metric,
\begin{gather*}
g\rightarrow g(\omega)=-{\rm d}T\otimes{\rm d}T+\sum_ie_i^0\otimes
e^0_i(1-\alpha\ell_{\rm P}\omega).
\end{gather*}
This may be interpreted as a~``rainbow metric''~\cite{rain}.
Assuming
that the corresponding contravariant metric in momentum space to
be the inverse spatial metric, we arrive at a~universal modif\/ication of dispersion
relations,
\begin{gather*}
m^2=-g(\omega)^{\mu\nu} k_\mu k_\nu=\omega^2-\frac{k_i^2}{1-\alpha
\ell_{\rm P}\omega}.
\end{gather*}
In view of the universality of the ef\/fect~-- the independence of
the form of matter in discussion~-- and the absence of any (explicit) preferred frame vector
f\/ield, which could distinguish a~preferred reference frame, it is
argued in~\cite{LD} that the proper framework for these MDR is not Lorentz
invariance breaking but a~deformation with a helicity-independent,
energy-dependent speed of photons.
We shall come back to this point in Section~\ref{ncg}.

In~\cite{LD2} a~more general formulation is presented, which does
not rely on connection representation and so is not restricted to
LQG, but rather makes Lorentz invariance deformation plausible for a
wider class of QG theories.
This attribute is shared with the models described
in the next section.

\subsection{Broken Poincar\'e symmetries: Finsler geometry} \label{sec:finsler}

As we recalled in Section~\ref{MDRmodels}, a~common idea behind QG phenomenology is that a semi-classical spacetime should leave some imprint on the propagation of light, in particular through a modif\/ied dispersion relation.
A natural way to encode this idea is to approximate a semi-classical
spacetime by a~medium whose properties are to be determined by the
specif\/ic underlying QG theory.
As is known in solid state
physics, the description of light propagation in a~special medium (see, for
example, references in~\cite{Skakala:2008jp}) is conveniently expressed with
\textit{Finsler metrics}, which are a~generalization of the
notion of Lorentzian/Riemannian metrics.
From this perspective, it
seems then quite natural to explore Finsler geometries as a
candidate to describe ef\/fectively QG semi-classical ef\/fects~\cite{Girelli:2006fw}.
The mathematics behind Finsler geometries are
not yet as solid as in the Riemannian or Lorentzian geometry cases.
For example the notions of signature and curvature of a~Finsler
metric are still active topics of discussion among the specialists.
Finsler geometry provides however a~nice framework to develop
new mathematics and QG phenomenology.
In fact it also
provides ways to theoretically constrain the possible QG
phenomenological proposals~\cite{Ratzel:2011zz,Schuller:2011nh}.

Let us recall the construction.
The propagation of the
electromagnetic f\/ield $u^A=(E^i,B^j)$ (Capital Latin letters designate a~pair of spatial indices.), in this
medium/semi-classical spacetime is described by some ef\/fective
Maxwell equations.
We assume here that for simplicity, spacetime
is $\R^4$.
Following~\cite{Ratzel:2011zz}, we will make the assumption that these ef\/fective equations are still
\textit{linear partial differential equations}
\begin{gather*}%\label{effective eq}
D^{\smp}_{AB}u^B(x)=\sum_{n=1}^N Q_{AB}^{{\alpha_1}\cdots {\alpha_n}}(\smp)\partial_{\alpha_1}\cdots \partial_{\alpha_n}u^B(x)=0.
\end{gather*}
The spacetime indices $\alpha_i$ run from $0$ to $3$.
We note the presence of the scale $\smp$ which encodes the QG ef\/fects.
The standard Maxwell equations are recovered when $\smp\dr0$
\begin{gather}\label{electro}
D^{\smp=0}_{AB}u^B=
\begin{pmatrix}-\delta_{ik}\partial_0&\epsilon_{ijl}\partial_j\\\epsilon_{ijk}\partial_j&\delta_{il}\partial_0
\end{pmatrix} \begin{pmatrix}E_k\\B_l\end{pmatrix} =0
\quad\Leftrightarrow\quad
\begin{cases}
 -\partial_0E_i+\epsilon_{ijl}\partial_j B_l=0, \\ \epsilon_{ijk}\partial_j E_k+\partial_0B_i=0.
\end{cases}
\end{gather}
Assuming there are no
\textit{non-linear} ef\/fects means that the coef\/f\/icients
$Q_{AB}^{\alpha_1\cdots \alpha_n}(\smp)$ do not depend on the f\/ields $u^A$ (but they could depend on the position $x$).
This is not such a~strong restriction since the existing proposals
of QG modif\/ied Maxwell equations such as Gambini--Pullin's~\cite{GP} are of this type (see Section~\ref{MDR}).
To
solve these equations, it is common to use the short wave-length
approximation, i.e.\ the Eikonal approximation.
This means that we consider the particle approximation of the electromagnetic f\/ield.
Skipping the details found in~\cite{Ratzel:2011zz}, this approximation leads to the modif\/ied
dispersion relations or mass-shell constraints in terms of momentum~$p$
\begin{gather*}
\cM(x,\smp,p)=\det\big(Q^{{\alpha_1}\cdots {\alpha_n}}(\smp)p_{\alpha_1}\cdots p_{\alpha_n}\big)=0.
\end{gather*}
This is the \textit{eikonal equation},
which can be seen as the \textit{covariant dispersion relation}.
The
reader can quickly check that keeping only the usual Maxwell
equation, leads to a~standard dispersion
relation\footnote{$\eta^{\mu\nu}$ is the Minkowski metric expressed
in the cotangent bundle.}
$\cM(x,\smp=0,p)=\eta^{\mu\nu}p_\mu p_\nu=p^2=0$.
The particle approximation gives rise to an action,
expressed in phase space or the cotangent space $T^*\R^4$, of the
type
\begin{gather*}
\text{\ss}_{\rm mod}=\int{\rm d}\tau \big(\dot x^\mu p_\mu-\lambda(\cM(x,\smp, p))\big).
\end{gather*}
If one wants to introduce the
concept of energy $E$ and spatial momentum $\vec p$, we need to
introduce an observer frame\footnote{Note that we invert the
notation of the frames with respect to~\cite{Ratzel:2011zz},
between~$T\R^4$ and~$T^*\R^4$.}.
However, we need the
frame $e$ expressed in the cotangent bundle.
We recall that the
Gauss map $G:T^*\R^4\dr T\R^4$ in the massless case (or the
Legendre transform in the massive case) allows to jump from a
description in the cotangent bundle to the tangent bundle.
Indeed,
from the variations of the action, we obtain $\dot x$ in terms of
$p$.
The Gauss map (and the Legendre transform) inverts this
expression.
Explicitly, this map allows to express momentum (i.e.\
a covector) as a~function of vectors\footnote{Hence the dif\/ference
between massless and massive cases.}
\begin{gather*}
\dot x^\mu=\lambda\frac{\partial\cM(x,\smp,p)}{\partial p_\mu}
\ \longrightarrow \
{p_\mu}={f_{\smp}}(x,\dot x^\mu,\lambda).
\end{gather*}
Let us postpone to later the
discussion on the existence of such a~map and let us assume it
exists and it is invertible.
The observer is encoded by a~curve in the spacetime manifold $\R^4$ and its tangent vector $v^\mu$
def\/ines the time part ${\epsilon_0}^\mu$ of the frame ${\epsilon_\alpha}^\mu$.
The other components ${\epsilon_i}^\mu$, $i=1,2,3$, are determined such that they span the tangent plane.
Now we can def\/ine the dual frame ${e^\alpha}_\mu$ in the cotangent bundle using the inverse Gauss map
(or inverse Legendre transform in the massive case).
We def\/ine ${e^0}_\mu=G\mone({\epsilon_0}_\mu)$ and the rest of the frame ${e^i}_\mu$, $i=1,2,3$,
is def\/ined as spanning the rest of the cotangent plane, so that ${e^\alpha}_\mu$ def\/ines a~frame in the cotangent plane.
To have a~physical notion of energy $E$ and 3-momentum, we need to project the momentum $p_\mu$ on $e$.
This point is actually often forgotten in the literature.
The modif\/ied dispersion relation expressed in terms of energy and 3-momentum is then recovered
from $\cM(x,\smp,Ee^0+p_i e^i)=0$.

We emphasize that in this approach the notion of energy or 3-momentum
is def\/ined in terms of a~frame in the cotangent bundle, def\/ined
itself in terms of an observer frame and the (inverse) Gauss map.
Contrary to the usual approach in QG phenomenology, where one
def\/ines arbitrarily the notion of energy, without specifying the
notion of frame observer or having a~good control on it.

Performing the Gauss map at the level of the action, we obtain its
expression in the tangent bundle,
\begin{gather*}
\text{\ss}=\int{\rm d}\tau\,\lambda
\cM^{\sharp}(x,\dot x,\smp).
\end{gather*}
$\cM^{\sharp}(x,\dot x)$ is a~homogeneous
function of $\dot x$, of degree dif\/ferent than $\cM$, in general.
In
the standard electromagnetic case on $\R^4$, given in terms of~\eqref{electro}, we recover
$\cM^{\sharp}(x,\dot x)=\sqrt{\dot x^\mu\eta_{\mu\nu}\dot x^\nu}$.
In this special case, the Minkowski metric is recovered with
\begin{gather*}
\eta_{\mu\nu}=\frac{\partial^2}{\partial\dot
x^\mu\dot x^\nu}\big(\cM^{\sharp}(x,\dot x)\big)^2=\frac{1}{2}
\frac{\partial^2}{\partial\dot x^\mu\dot x^\nu}\left(\dot x^\mu
\eta_{\mu\nu}\dot x^\nu\right).
\end{gather*}
If the polynomial
$\cM^{\sharp}(x,\dot x)$ is more general, one obtains a~metric
\begin{gather*}
g_{\mu\nu}(x,\dot x)=\frac{\partial^2}{\partial\dot x^\mu\dot x^\nu}
\big(\cM^{\sharp}(x,\dot x)\big)^2,
\end{gather*}
which depends both on \textit{the position $x$ and the vector $\dot x$}.
This type of metric is called \textit{Finsler metric}.
Riemann was actually aware of this possible extension (i.e.\
a~metric which depends also on vectors) of the metric, but it is only in 1917 that Finsler explored this generalization for his PhD thesis.
This generalized notion of metric can be seen as the rigorous implementation of the notion ``rainbow metric''~\cite{rain}.

The notion of a~Lorentzian Finsler metric is still a matter of
discussion in the Finsler community.
The most recent proposals that
have been developed concurrently with the approach recalled above
can be found in~\cite{wolfrath, Ratzel:2011zz}.

The reader could argue that it might not be possible to identify a~well def\/ined (invertible) Gauss map or Legendre transform in general.
However if few natural assumptions are added, \cite{Ratzel:2011zz} pointed out that the Gauss map and the Legendre map are well def\/ined.
Let us summarize these assumptions.
(Clearly they can be discussed and one could explore the consequences of removing them.)
\begin{itemize}\itemsep=0mm
\item
The f\/irst assumption is one that we made at the beginning: we assumed that the equations encoding the ef\/fective dynamics are \textit{linear} partial dif\/ferential equations.

\item We take as a~very reasonable (or conservative) assumption that if we know the initial conditions,
we can predict the propagation of the plane-waves at any later time
in the semi-classical spacetime.
Not having this assumption would
make prediction dif\/f\/icult.
On the other
hand, one can argue that in the full QG regime, the notion of
causality could disappear.
We assume here that at the
semi-classical level we do not see such ef\/fects.

\item
Finally, we expect that the notion of energy $E$ def\/ined using the observer frame should be positive for any observer.
\end{itemize}

\looseness=-1
It is quite striking that these three physical assumptions
(linearity, predictability, energy positivity) can be translated
into very dif\/ferent types of mathematics, such as algebraic
geometry and convex analysis.
With these assumptions,
the Legendre transform (massive case) and the Gauss map (massless case) are well-def\/ined and
invertible.
Taken together these assumptions put strong
constraints on the possible shape of the covariant dispersion
relation $\cM(x,\smp,k)$.
\cite{Ratzel:2011zz}~showed that
some well-known QG motivated modif\/ied dispersion relations (such as
the Gambini--Pullin's~\cite{GP}) actually do not satisfy
some of the above assumptions.
This means that these MDR cannot be understood in this setting and it is quite unlikely that they can be physically interpreted in the context of Lorentz symmetry breaking.
For the full details we refer to the
\cite{Ratzel:2011zz}, the presentation of necessary mathematics
would go beyond the scope of this review.

To conclude this quick overview of the Finsler framework, we hope we have conveyed to the reader that there exist more general metric structures than the Lorentzian ones, which seem to be natural candidates to encode the semi-classical QG ef\/fects.
This mathematical framework has been introduced fairly recently in the QG phenomenology
framework~\cite{Girelli:2006fw}.
Even more recently, \cite{Ratzel:2011zz}~has shown that this framework can be very much constrained
from mathematical arguments so that we do not have to explore blindly in every direction what we can do.
There is a~nice mathematical
framework to guide us, which awaits to be further developed.
Finally, (some of) the Lorentz symmetries are broken in the Finsler geometry framework.
There is no known way to accommodate for spacetime symmetries consistent with the scales present in the modif\/ied dispersion relation.
This is consistent with the analogy of a~medium, which will in general contain some preferred direction and/or scale.

\subsection{Non-linear realization of Poincar\'e symmetries}\label{sec:deformed} \label{sec:nonlinear}

Besides the obvious way to introduce an invariant scale by choosing
a preferred reference frame and so sacrif\/icing Lorentz invariance,
it is possible to keep the relativity principle intact in
presence of a~second invariant quantity, a Planck scale $\smp$ in addition to the speed of
light.
One does so by modifying both the transformation
laws from one inertial system to another, and the law of
energy-momentum conservation in such a~way that a MDR become compatible with observer independence.
This can be achieved through a~non-linear realization of
(some of) the Poincar\'e symmetries~\cite{Magueijo:2001cr, Magueijo:2002am}.
This is a~f\/irst attempt to describe deformed symmetries since
in this framework, the realization of the full Poincar\'e symmetries in spacetime is not clear.%\looseness=1

In the papers~\cite{Magueijo:2001cr, Magueijo:2002am}, Magueijo and Smolin
do not make any specif\/ic assumption regarding the structure
of momentum space.
It can be f\/lat or curved, it is left open.
To f\/ix
the notations, def\/ine f\/irst the momentum $\pi$ and the inf\/initesimal
boost which is realized as $J_{0i}=\pi_0\partial_{\pi_i}-\pi_i\partial_{\pi_0}$.
It acts linearly on $\pi$
\begin{gather*}
J_{0i}\act \pi_j= \delta_{ij}\pi_0, \qquad J_{0i}\act \pi_0= - \pi_i.
\end{gather*} We introduce then a~map $U_{\smp}:\R^4\dr\R^4$ so that one
can def\/ine the non-linear realization $K_{0i}$ of the boosts
$J_{0i}$
\begin{gather}\label{non-linear boost}
K_{0i}=U_{\smp}{J_{0i}}
U_{\smp} \mone.
\end{gather} The choice of $U_{\smp}$ is such that the
non-linear boost $K_{0i}$
still satisf\/ies
the Lorentz algebra
\begin{gather}\label{lorentz non-linear}
\com{K_i,K_j}=
\epsilon_{ij}^k J_k,\qquad\com{J_i,K_j}=\epsilon_{ij}^k K_k,\qquad
\com{J_i,J_j}=\epsilon_{ij}^k J_k,
\end{gather} where $J_i$ encodes the
inf\/initesimal rotations.
An example is given by
\begin{gather*}
U_{\smp}[\pi_0]=e^{\frac{\pi_0}{\smp}\pi^\mu\partial_{\pi^\mu}},
\qquad\text{so that}\quad
U_{\smp}[\pi_0](\pi_\mu)=\frac{\pi_\mu}{1-\frac{\pi_0}{\smp}}.
\end{gather*}
Note that this map is not
unitary.
In this example, we have a~maximum energy $\smp$ as one can
check by applying a~boost on $\pi_\mu$.
This specif\/ic non-linear
realization leaves invariant the modif\/ied dispersion relation
\begin{gather*}
\mm(\smp, \pi) = \frac{\pi^2}{(1-\frac{\pi_0}{\smp})^2}.
\end{gather*}
Clearly other choices of non-linear realizations can be performed.
They are simply constrained to satisfy~\eqref{lorentz non-linear}
and can be chosen to implement a~maximum energy or a maximum 3d
momentum.

Another way to present this proposal is to consider a~momentum
$p_\mu=U_{\smp}(\pi_\mu)$ as the measured ``physical'' momentum.
The meaning of~\eqref{non-linear boost} is that f\/irst
we go to the linear auxiliary variable~$\pi$, perform the boost
transformation, then deform back to the physical variable~$p$.
The
physical meaning of the variable~$\pi$ is then not very clear.

The case of constructing multiparticles has been discussed in this
context.
Once again, the idea is to use the linear momentum $\pi$ to
induce the sum of the physical momenta $p$.
A preliminary proposal
was
\begin{gather}\label{bad-sum}
\big(p^{(1)}\oplus p^{(2)}\big)
\equiv
U_{\smp}\left(U\mone_{\smp}\big(p^{(1)}\big)+U\mone_{\smp}\big(p^{(2)}\big)\right).
\end{gather} This
proposal however suf\/fered from an immediate drawback: the ``soccer
ball problem''.
If it makes sense that a~fundamental particle has a
momentum bounded by the QG scale, large systems (such as a~soccer
ball) have doubtlessly a~momentum larger than the QG scale.
For
instance a~f\/lying mosquito has a momentum bigger than the QG scale.
The sum in~\eqref{bad-sum} is such that the total momentum of two
particles will still be bounded by the QG scale.
The solution to
this problem is to have a~rescaling of the QG scale so that in the
large limit, we can deal with systems which have larger momentum
than~$\smp$.
Magueijo and Smolin therefore proposed to deal with a
sum
\begin{gather*}%\label{good-sum}
\big(p^{(1)}\oplus p^{(2)}\big)
\equiv
U_{2\smp}\left(U\mone_{\smp}\big(p^{(1)}\big)+U\mone_{\smp}\big(p^{(2)}\big)\right).
\end{gather*}
This is a~rescaling of the Planck scale implemented by hand.
One should f\/ind a justif\/ication for such rescaling by considering a more complete model.
Dif\/ferent arguments have been proposed in~\cite{Girelli:2004ue, Magueijo:2001cr}.
We note, \textit{en passant}, that this addition of momenta is commutative,
since it is based on the standard commutative addition $\pi_1+\pi_2$.
This approach is therefore not equivalent to the non-commutative approach, which we will discuss later.

To reconstruct spacetime, Magueijo and Smolin propose to see the
coordinates as the in\-f\/i\-ni\-te\-si\-mal translations on momentum space.
In
the specif\/ic example they chose, the coordinates are commutative for
the Poisson bracket
\begin{gather*}x_0=t=1-
\left(1-\frac{\pi_0}{\smp}\right)\left(\left(1-\frac{\pi_0}{\smp}\right)\partial_{\pi_0}-
\frac{\pi_i}{\smp}\partial_{\pi_i}\right),\qquad x_i=
\left(1-\frac{\pi_0}{\smp}\right) \partial_{\pi_i}.
\end{gather*} They are
however functions over the full phase space and not only in
conf\/iguration space.
Their physical meaning is not very clear.
The
def\/inition of the translations in spacetime is not given explicitly.
For a~recent attempt to formulate a particle-dependent,
``non-universal DSR'', see~\cite{pd}.

\subsection{Modif\/ied reference frame}\label{sec:ref}
This approach tries to provide a~physical meaning to the dif\/ferent momenta $p$, $\pi$
introduced in the previous subsection.
In~\cite{Liberati:2004ju,Aloisio:2005qt,Aloisio:2006nd},
the authors recall that to measure the momentum of a~particle $\pi_\mu$, we actually use a reference frame
${e_a}^\mu$, $a=0,\dots,3$.
It is the local inertial frame which can be constructed at every point of the
spacetime manifold from the metric $g_{\mu\nu}$ through
$
g_{\mu\nu}= {e_a}^\mu {e_b}^\nu \eta^{ab}.
$
The outcome of the measurement is then
$
p_a= {e_a}^\mu \pi_\mu.
$
In standard Minkowski spacetime, we have ${e_a}^\mu\sim
{\delta_a}^\mu$ so that $\pi_a\sim p_a$
Two reference frames
${e_a}^\mu$, ${\overline{e}_a}^\mu$ are related by a~Lorentz
transformation ${\Lambda_b}^a$ if
${\overline{e}_b}^\mu={\Lambda_b}^a{e}_a^\mu$.
This transformation induces the
standard linear realization of the Lorentz transformation on $p_a$
\begin{gather}\label{changeframe}
p_a={e_a}^\mu\pi_\mu \dr
\overline{p}_b={\overline{e}_b}^\mu\pi_\mu={\Lambda_b}^a
{e}_a^\mu \pi_\mu= {\Lambda_b}^a p_a.
\end{gather}
The f\/irst idea~--
proposed in~\cite{Liberati:2004ju}~-- is to argue that due to some
QG ef\/fects, the measured momentum $p$ will be a~non-linear function
of the components ${e_a}^\mu\pi_\mu$
\begin{gather}\label{frame1}
p_b= f_b({e_a}^\mu \pi_\mu, \smp).
\end{gather}
Changing frame by performing a~Lorentz transformation on $e$ as in
\eqref{changeframe} will clearly lead to a~non-linear realization $\tilde\Lambda$ of
the Lorentz transformation on $p$.

When constructing explicitly some models to implement this idea, the
actual relationship between momentum and frame has been modif\/ied in
\cite{Aloisio:2005qt, Aloisio:2006nd}.
Instead of~\eqref{frame1},
the measured momentum should be def\/ined with respect to an ef\/fective
tetrad
$
{E^\mu}_\alpha(e,\pi,\smp).
$
In this sense, this is a~picture close to the Finsler geometry
approach (cf.\ Section~\ref{sec:finsler}) or Smolin's derivation from LQG (cf.\ Section~\ref{smolin-frame}), since the frame is now
momentum dependent.
An important assumption is that the map
relating a~trivial frame $e$ with the ef\/fective frame $E$
\begin{gather}
{e^\mu}_\alpha \xrightarrow{\uu_{\smp}} {E^\mu}_\alpha(e, \pi,
\smp)
\label{umap}
\end{gather}
should be \textit{reversible}, at least in some approximation.
When the f\/luctuations
have appropriate symmetries, the map $\uu_{\smp}$ takes the simple form ${E^\mu}_\alpha=F(e,\pi,\smp){e^\mu}_\alpha$,
and $F(e,\pi,\smp)\rightarrow1$ when $\smp\rightarrow\infty$.
For
example, the Magueijo--Smolin dispersion
relation~\cite{Magueijo:2001cr,Magueijo:2002am}
\begin{gather*}
\frac{p_0^2-p^2}{1-\frac{p_0}{\smp}}=m^2
\end{gather*} can be expressed as
$p_\alpha=F(e,\pi,\smp){e^\mu}_\alpha\pi_\mu$
with
\begin{gather*}
F(e,\pi,\kappa)=\frac{1}{{1-\frac{\pi_\mu{e^\mu}_0}{\smp}}}.
\end{gather*}
Lorentz transformations act linearly on the tetrad f\/ield $e_\alpha$,
so their action $\tilde\Lambda$ on the ef\/fective tetrad $E_\alpha$ is specif\/ied by the
following commutative diagram
\begin{gather*}%\label{classicaltransf}
\begin{array}{@{}lcl}
{e^\mu}_\alpha&\stackrel{\Lambda}
{\longrightarrow}&{\overline e^\mu}_\alpha={\Lambda^\beta}_\alpha{e^\mu}_\beta\\
\downarrow\uu_{\smp}&&\downarrow\uu_{\smp}\\
{E^\mu}_\alpha&\stackrel{\tilde\Lambda}
{\longrightarrow}&{\overline E^\mu}_\alpha
\end{array}
\end{gather*}
This induces a~non-linear transformation of the
measured momentum $p$,
\begin{gather*} p_\alpha=\pi_\mu{ E^\mu}_\alpha
\rightarrow p'_\alpha=\pi_\mu{\overline E^\mu}_\alpha =\pi_\mu
\uu_{\smp}\left(\Lambda \cdot \uu_{\smp}^{-1} (E)\right).
%\label{eq:nlt}
\end{gather*} Dif\/ferent proposals argue that QG f\/luctuations
lead to an ef\/fective frame ${E^\mu}_\alpha(e,\pi,\smp)$
\cite{Aloisio:2006nd,Girelli:2006sc}. We have recalled above in
Section~\ref{smolin-frame} Smolin's argument in this sense. Another
motivation comes from models from quantum information theory~\cite{Girelli:2007xn}. From the QG perspective it is natural to
consider that frames should be quantized. Some physical aspects of
the use of quantum reference frame can be explored using f\/inite-dimensional systems, such as spin systems. For example, we can use
three quantum spins $\vec J_{a}$, $a=1,2,3$, as 3d reference frame
and look at the projection of another quantum spin $\vec S$ in this
frame. The corresponding observable is then
\begin{gather*}\mathfrak{S}_{a}=
\vec S \cdot \vec J_{a} = S_i {J^i}_{a}.
\end{gather*} We can take by analogy
with the QG semi-classical limit a~semi-classical frame, that is we
consider the reference frame $\vec J_a$ given in terms of coherent
states $|\psi\rangle$.
We have therefore the semi-classical
reference $\langle\psi|\vec J_{a}|\psi\rangle\equiv\langle\vec
J_{a}\rangle$. We consider the spin $\vec S$ projected in this
semi-classical frame
\begin{gather*}\tilde{\mathfrak{S}}_{a}=\vec S\cdot
\langle \vec J_{a} \rangle.
\end{gather*} A priori the frame $\langle\vec
J_{a}\rangle$ is independent of the system $\vec S$.
However it was
shown~\cite{Poulin2006} that consecutively measuring the observable
$\mathfrak{S}_{a}$ leads to a~kickback of the system on the frame,
which state then becomes dependent on the system $\vec S$.
Hence
consecutive measurements induce the map
\begin{gather}\label{umap2}
\langle
\vec J_{a}\rangle\,\dr\,\langle\vec J_{a}\rangle(\vec S),
\qquad a=1,2,3.
\end{gather} This is the analogue of the deformation
\eqref{umap}.
It was shown that this map can be decomposed as a
rotation together with some decoherence~\cite{Poulin2006}.
The
decoherence part is non invertible, so the map~\eqref{umap2} is not
properly reversible.
It is reversible only in the approximation
where the decoherence can be neglected.
For further details we refer to~\cite{Girelli:2007xn, Girelli:1900zz}.

\subsection{Non-commutative space-time}\label{sec:dsr} \label{ncg}

Non-commutative spaces can be understood as geometries where coordinates become operators.
In this sense we move away from the phase space structure we discussed earlier and use
the quantum setting.
Non-commutative geometries have been mostly introduced by mathematicians
and are therefore very well def\/ined mathematically.
In general one uses the algebraic
concepts of Hopf algebra and so on.
In order to restrict the amount of material,
we shall present only a~pedestrian overview of the topic and refer the more curious reader to the relevant references.

Historically, one of the f\/irst examples of non-commutative space is
due to Snyder~\cite{Sn} who tried to incorporate the Planck length in
a covariant way, i.e.\ without breaking the Lorentz symmetries.
The idea is simple and follows similar philosophy as
in the LQG case.
Space coordinates can be ``discretized'' if they are operators
with a~discrete spectrum.
Snyder used the subspace generated by the inf\/initesimal de Sitter
boosts $J_{4\mu}$, of the Lie algebra $\so(4,1)$ generated by the
elements $J_{AB}$ to encode spacetime.
The coordinates are
\begin{gather*}
X_\mu= \frac{1}{\smp} J_{4\mu},
\qquad\text{such that}\quad
[X_\mu, X_\nu]=\frac{1}{\smp^2} J_\mn \in \so(3,1).
\end{gather*}
The spatial coordinates
$X_i=\frac{1}{\smp}J_{4i}$ are represented as (inf\/initesimal)
rotations and therefore have a~discrete spectrum.
The time
coordinates has instead a~continuum spectrum.
From the def\/inition
of the Lie algebra $\so(4,1)$, there is a~natural action of the
inf\/initesimal Lorentz transformations on $X_\mu$,
\begin{gather*} [J_\mn,
X_{\alpha}] = [J_\mn, \frac{1}{\smp} J_{4\alpha}]= \eta_{\mu\alpha}
X_{\nu}- \eta_{\nu\alpha} X_{\mu}.
\end{gather*} Hence it is possible to
have a~discrete structure with Lorentz symmetry implemented.
By assumption, momentum space is not f\/lat but curved: it is
de Sitter space instead of the standard hyperplane $\R^4$.
Snyder picked a~choice of coordinates on the
de Sitter space such that Lorentz symmetries are implemented.
Considering the embedding of de Sitter in $\R^5\ni\pi^A$ given by
$\pi^A\eta_{AB}\pi^B=-\smp^2$, Snyder picked
\begin{gather*}p^\mu=
\frac{\pi^\mu}{\pi^4},\qquad
\pi^4=\frac{1}{\sqrt{1-\frac{p^2}{\smp^2}}}.
\end{gather*} The dispersion
relation would be the classic one, since the Lorentz symmetries are
the usual ones,
\begin{gather*} p^2=m^2.
\end{gather*} Note that his choice of
coordinates is actually very close to the one picked up in special
relativity.
Indeed in this context, we have the space of 3d speeds
which becomes the hyperboloid
\begin{gather*} V^\mu\eta_{\mn}V^\nu=c^2,
\end{gather*}
where the natural choice of coordinates is
\begin{gather*}v^i=
\frac{V^i}{V^0},\qquad V^0=\frac{1}{\sqrt{1-\frac{v^2}{c^2}}}.
\end{gather*}
Snyder did not discuss the addition of momenta.
However since
momentum space is now curved it is clear that the addition has to be
non-trivial.
We can take inspiration from the addition of velocities
in special relativity: it is given in terms of a~product which is
non-commutative and non-associative.
The space of velocities is then
not a~group but instead a K-loop or a~gyro-group~\cite{ungar}.

A similar addition can be def\/ined for the Snyder case~\cite{Girelli:2010zw}.
The addition of momenta will then be
non-commutative and non-associative.
This means that the notion of
translations becomes extremely non-trivial.
We shall come back on
this structure when recalling the recent results of ``relative
locality''~\cite{RL}.
For dif\/ferent approaches to deal with Snyder's
spacetime we refer to~\cite{Girelli:2010wi} and the references
therein.

Snyder's spacetime is related to the Doplicher--Fredenhagen--Roberts (DFR) spacetime,
which was constructed independently of Snyder's spacetime, using dif\/ferent tools~\cite{Doplicher:1994tu},{\samepage
\begin{gather*}
\com{X_\mu,X_\nu}=\frac{1}{\smp}Q_{\mu\nu},\qquad
\com{Q_{\mu\nu},X_\alpha}=0=\com{Q_{\mu\nu},Q_{\alpha\beta}}.
%\label{DFR-bracket}
\end{gather*}
The operators $Q_{\mu\nu}$ can be seen as (commutative) coordinates on some extra dimensional space.}

This space can be seen as an abelianized version of Snyder spacetime
\cite{Carlson:2002wj, Girelli:2010wi}.
Starting from Snyder's
commutation relations (i.e.\ the algebra $\so(4,1)$), consider
$Q_{\mu\nu}=\frac{1}{\smp'}J_{\mu\nu}$ and the limit
$\smp,
\smp'\dr\infty$ with $\frac{\smp'}{\smp^2}=\frac{1}{\tilde\smp}$
f\/ixed. It is not complicated to see that Snyder's algebra gives an
algebra isomorphic to the DFR algebra in
this limit.

Both Snyder's and DFR's non-commutative spaces can also be
understood as non-com\-mu\-ta\-ti\-ve spaces embedded in a~bigger
non-commutative space of the Lie algebra type~\cite{Girelli:2010wi}.
Once again in this case there is no deformation of the Lorentz
symmetries, the modif\/ied dispersion relation is again the standard
one. However there are non trivial uncertainty relations which
implement a~notion of minimum ``area''~\cite{Doplicher:1994tu},
\begin{gather*}
\cop x_0(\cop x_1+\cop x_2+\cop x_3)\geq\smp^{-2},\qquad\cop x_1
\cop x_2 + \cop x_2\cop x_3 + \cop x_3 \cop x_1 \geq \smp^{-2}.
\end{gather*}

The famous Moyal spacetime
\begin{gather}\label{moyal}
\com{X_\mu,X_\nu}=\frac{i}{\kappa^2}\theta_{\mu\nu}\one,\qquad\com{X_\mu,\one}=0,
\end{gather} is as an example of the DFR spacetime, as amy abe seen by projecting
$Q_{\mu\nu}$ onto a~specif\/ic eigenspace
\cite{Filk:1996dm}.
The Moyal spacetime is the most
studied of the non-commutative spacetimes.
There exists a~huge
literature on the physics in this spacetime, we refer only to
\cite{Hinchliffe:2002km, Balachandran:2005eb}, where the relevant literature can be found.

The tensor $\theta_\mn$ is a~tensor made of c-numbers.
It is
invariant under translations and Lorentz transformations (unlike the
$Q_\mn$ in the DFR space which transform under Lorentz
transformations).
It is not complicated to check that~\eqref{moyal}
transforms covariantly under an inf\/initesimal translation.
However
the case of the (inf\/initesimal) Lorentz transformations is more tricky since
\begin{gather*}
J_\mn\act\com{X_\alpha,X_\beta}=
i\left(\eta_{\nu\alpha}\theta_{\mu\beta} - \eta_{\mu\alpha}\theta_{\nu\beta}
 + \eta_{\nu\beta}\theta_{\alpha\mu} - \eta_{\mu\beta}\theta_{\alpha\nu}\right)\one
\neq\frac{1}{\kappa^2}J_\mn\act\left(i\theta_{\alpha\beta}
\one\right)=0,
\end{gather*}
where we have used $J_\mn\act X_\alpha=\eta_{\nu\alpha}X_\mu-\eta_{\mu\alpha}X_\nu$ and the Leibniz law
\begin{gather*}
J_\mn\act\left(X_\alpha X_\beta\right)=\left(J_\mn\act
X_\alpha\right)X_\beta+X_\alpha\left(J_\mn\act
X_\beta\right).
\end{gather*} One can then say that the Lorentz symmetries
are broken since the non-commutativity is not consistent with the
change of frame.
Most often in the study of Moyal spacetime, this
is the chosen perspective.
One can also look for a~deformation of
the action of these Lorentz symmetries, to make them compatible with
the non-commutative structure~\cite{Oeckl:2000eg}.
This deformation
can be understood as a~modif\/ication of the Leibniz law
\begin{gather*}
J_\mn\act(X_\alpha X_\beta)=
(J_\mn\act X_\alpha)X_\beta+X_\alpha(J_\mn\act X_\beta)
\nonumber\\
\phantom{J_\mn\act(X_\alpha X_\beta)=}
-\frac12\theta^{\rho\sigma}
\big(((\eta_{\rho\mu}\partial_\nu-\eta_{\rho\nu}\partial_\mu)\act X_\alpha)\partial_\sigma\act X_\beta
\nonumber\\
\phantom{J_\mn\act(X_\alpha X_\beta)=}
+\partial_\rho\act X_\alpha((\eta_{\sigma\mu}\partial_\nu-\eta_{\sigma\nu}\partial_\mu)\act X_\beta)\big).
\end{gather*}
It can be extended easily to arbitrary product of functions.
With this new Leibniz law, the non-commutativity~\eqref{moyal} will be consistent with the Lorentz symmetries as one can check.

The proper way to encode this modif\/ication is to use algebraic structures such as \textit{quantum groups}~\cite{majid}.
Indeed, the modif\/ied Leibniz law comes from a~non-trivial coproduct structure.
There exists therefore a~deformation of the Poincar\'e group that is the symmetry group of Moyal's spacetime and as such this spacetime can be seen as a f\/lat non-commutative spacetime.

There is no deformation of the action of the Lorentz transformations on the momenta so that it is the usual mass-shell relation that one considers.
Furthermore since the translations are not modif\/ied either, there is no modif\/ication of the addition of momenta.
We refer to~\cite{Hinchliffe:2002km, Balachandran:2005eb} and references therein for some discussions of the phenomenology of this space.

An important example of non-commutative space for QG phenomenology is the $\kappa$-Minkowski spacetime.
We also mention his cousin the $\kappa'$-Minkowski spacetime.
Their non-commutative structures are def\/ined, respectively, as
\begin{gather}
\com{X_0,X_i}=\frac{1}{\kappa}X_i,\qquad\com{X_i,X_j}=0,\qquad i,j=1,2,3,\label{kappamink}\\
\com{X_1,X_\mu}=\frac{1}{\kappa}X_\mu,\qquad\com{X_\mu,X_\nu}=0,\qquad\mu,\nu=0,2,3.\nonumber
\end{gather}
In the $\kappa$-Minkowski case, it is the time coordinate that is not
commutative with the space coordinates, whereas in the $\kappa'$-Minkowski case,
it is a~space coordinate (here $X_1$, but clearly other cases can be
considered) that is non-commutative with the others.
They encode the most rigorous models of deformed special relativity\footnote{Unfortunately the name of DSR has been given to many dif\/ferent approaches.} (DSR).

These non-commutative spacetimes provide examples where both the Lorentz transformations and the translations are deformed.
Let us focus on the case of translations in $\kappa$-Minkowski.
The commutator cannot be covariant under an inf\/initesimal translation if we use the usual Leibniz law.
Indeed, the commutator is invariant if we use the standard Leibniz law
\begin{gather*}
\partial_\mu\act(X_\alpha X_\beta)=\eta_{\alpha\mu}X_\beta+\eta_{\beta\mu}X_\alpha
\quad\Longrightarrow\quad
\partial_\mu\act[X_\alpha,X_\beta]=0.
\end{gather*}
But due to~\eqref{kappamink}, and $\partial_\mu\act X_\alpha=\eta_{\mu\alpha}$, we see we get for example
\begin{gather*}
\partial_j\act[X_0,X_i]=0=\eta_{ij},
\end{gather*} which is a~contradiction.
We can then argue either that the
translation symmetries are broken or that we can modify the Leibniz
law in order to have an action of the translations compatible with
the non-commutative structure.
One can check that the modif\/ied
Leibniz law on the spatial derivatives (the time derivative
satisf\/ies the usual Leibniz law)
\begin{gather*}
\partial_i\act(X_\alpha X_\beta)
=(\partial_i\act X_\alpha)X_\beta
+\big(e^{\frac{\partial_0}{\smp}}\act X_\alpha\big)
(\partial_i\act X_\beta),
\nonumber\\
e^{\frac{\partial_0}{\smp}}\act X_\alpha=
\begin{cases}
X_\alpha,       & \alpha\neq0,\\
X_0+1/\smp,\quad& \alpha=0
\end{cases}
\end{gather*} does the job.
As
in the Moyal case, there is a~deformation of the Poincar\'e group
which encodes such a~modif\/ication.
This quantum group is called the
$\kappa$-\textit{Poincar\'e group}~\cite{kappa,Majid:1994cy}.
When one identif\/ies the translations with momenta,
the alter-ego of the modif\/ied Leibniz law is a~modif\/ied addition law
of momenta
\begin{gather}\label{mod add}
(p\oplus q)_0=p_0+q_0,\qquad(p\oplus q)_i=
p_i+ e^{p_0/\smp}q_i,
\end{gather} which comes from the non-trivial
coproduct of the $\kappa$-Poincar\'e quantum group.
Just as in the
Snyder case, the momentum space in the $\kappa$-Minkowski case is
given by the de Sitter space.
However in this case, the de Sitter
space $dS\sim\SO(4,1)/\SO(3,1)$ is equipped with a~group product,
so that the momentum addition~\eqref{mod add} is non-commutative but
associative.
The group structure can be obtained by factorizing the
group $\SO(4,1)=G.\SO(3,1)$, where $G$ is the group (actually two
copies of the group called $\AN_3$) encoding momentum space.
This
implies that there is an action of the Lorentz group on momentum
space $G$ but also a~back action
of $G$ on
$\SO(3,1)$~\cite{majid, Majid:1994cy}.
Putting all together, we get
a non-linear realization of the Lorentz symmetries on momentum space
($N_i$, $R_i$ are respectively the boosts and the rotations)
\begin{gather*}
[R_i,p_j]=\epsilon_{ij}^lp_l,
\qquad
[R_i,p_0]=0,\nonumber\\
[N_i,p_j]=\delta_{ij}\left(\sinh\frac{p_0}{\smp}-\frac{{\bf p}^2}{2\smp^2}e^{p_0/\smp}\right),
\qquad
[N_i,p_0]=p_i e^{p_0/\smp}.%\label{kappa poincare algegbra}
\end{gather*} Just like for the translations, the compatibility of the
non-commutative structure~\eqref{kappamink} with Lorentz
transformations implies a~deformation of the Leibniz law for the
action of the Lorentz transformations~\cite{Majid:1994cy}.
This is
also inherited from the $\kappa$-Poincar\'e group.
This non-commutative
space was one of the f\/irst frameworks used to discuss the
non-trivial propagation of gamma-rays~\cite{majid-amelino}.
We refer to
\cite{AmelinoCamelia:2003ex,HHMdsr,KowalskiGlikman:2006vx,
AmelinoCamelia:2010pd} for a~discussion of the phenomenology of this
space.

We have presented a~set of non-commutative spaces and discussed the realization of the Poincar\'e symmetries.
It is interesting to ask whether all these spaces can be classif\/ied.
To our knowledge, there are two types of possible classif\/ications:
\begin{itemize}\itemsep=0mm
\item The f\/irst one is to look at all the possible deformations of the Poincar\'e
group using quantum group techniques, i.e.\ algebraic techniques.
If
one identif\/ies the momentum operator with the translation, this
provides a~set of dif\/ferent momentum spaces, from which we can
determine by duality the dual spacetime.
Therefore determining the
deformations of the Poincar\'e group specif\/ies momentum spaces and
the relevant non-commutative spaces.
In 4d (as well as in 3d), this
classif\/ication was performed~\cite{Podles:1994tb} and a~set of 21
deformations have been identif\/ied.
The Moyal spacetime and the
$\kappa$- and $\kappa'$-Minkowski spaces\footnote{Historically, the
$\kappa$-Poincar\'e deformation was discovered through two dif\/ferent
approaches: the In\"on\"u--Wigner contraction of the quantum group
$\SO_q(4,1)$~\cite{kappa} and the bicrossproduct construction
\cite{Majid:1994cy}.} are of course among them.
However, Snyder's
spacetime is not among them since it is related to a~non-associative
deformation.
This classif\/ication is hence missing some (at least
historically) interesting spaces.

Note that among this classif\/ication, there is a~number of momentum spaces that appear as Lie
groups.
It would be interesting to check whether the full set of 4d Lie groups appear in this classif\/ication.

Finally, we emphasize, that a~priori all these 21 deformations are
equally valid candidates for a~non-commutative description
of QG semi-classical \textit{flat} spacetime.

\item The second approach is geometric in nature but it has not been performed in detail yet.
It consists
in classifying all the possible 4d smooth loops (i.e.\ manifold
equipped with a~product, a~unit and an inverse) that carry an action of the Lorentz group.
This would provide
a classif\/ication of possible momentum spaces, and one would
need to check if one can f\/ind some action of the Lorentz group on
them to get the relevant Poincar\'e group deformation.

Dif\/ferent works have pointed out that classifying the loops can be
seen as a~classif\/ication of the possible connections on some given
manifold (especially in the homogenous case).
For example on the de Sitter space as momentum space, one
can have a~group structure $G=AN_3$ which would give rise
to $\kappa$-Minkowski spacetime, or a~K-loop structure, giving rise to
Snyder's spacetime~\cite{Girelli:2010zw, RL, Frei}.
These dif\/ferent choices amount to dif\/ferent
types of connections on the de Sitter space.

We should note however that with this geometric classif\/ication, we would get the non-associative spaces but
we would miss momentum spaces of the quantum group or of the Moyal type.
Hence the algebraic and geometric approaches are complementary.

\end{itemize}

The Moyal and $\kappa$-Minkowski spacetimes have attracted most
attention in the QG phenomeno\-logy/non-commutative communities.
Among these two, the Moyal spacetime is the best understood,
probably because the non-commutativity is in fact of a~simpler
nature (a twist) than $\kappa$-Minkowski.
Historically Moyal spacetime
was also identif\/ied before $\kappa$-Minkowski.
In the LQG community,
the $\kappa$-Minkowski case is the most popular example, though
$\kappa'$-Minkowski also appears in 3d QG~\cite{Meusburger:2008dc}.

Non-commutative structures appear in the LQG context through the failure
of area operators acting on intersecting surfaces to commute~\cite{Ashtekar:1998ak}.
Currently, there is no derivation of non-commutative ``coordinates''
in 4d LQG. We shall come back in Section~\ref{sec:derqftqg} on
a~derivation of non-commutative f\/ield theory for matter using the group f\/ield theory approach.

\subsection{Relative locality}
This recent development in the construction of ef\/fective theories
proposes a~tentative interpretation of non-commutative coordinates
at the classical level.
The basic claim of ``Relative Locality'' (RL) is that we live in phase space, not in space-time.
In this
view, rather than a~global space-time, there are only
energy-momentum dependent cotangent spaces (interpreted as
spacetime) of the \textit{curved} momentum space $\cM$, a~concept
f\/irst proposed in~\cite{rain}.
RL attempts to describe ``classical,
non-gravitational quantum gravity ef\/fects'', i.e.\
remnants of QG,
when gravity and quantum theory are switched of\/f by going to the
limits $\hbar\rightarrow0$ and $G\rightarrow0$.
In this limit the
Planck length goes to zero as well, whereas the Planck mass $\smp$
is f\/inite. In this way an invariant energy scale is obtained, but
not an invariant length. The scale $\smp$ can be introduced by
considering a~homogeneous momentum space with (constant) curvature
$\smp$.
In fact even more general curved momentum spaces can be
introduced.
If momentum space is not homogeneous of curvature
$\smp$, but of a~more general type, the scale $\smp$ will still
appear in the modif\/ied dispersion relation or the non-trivial
addition in order to have dimensional meaningful quantities.
Quite
strikingly the structures associated to momentum space~-- dispersion
relation, addition of momenta~-- can be related to geometric
structures on momentum space.
Indeed, it can be shown that the
metric on momentum space encodes the dispersion relation, whereas
the sum of momenta is encoded through a~connection (which does not
have to be either metric compatible or torsion free or f\/lat).

To be more explicit, consider an event.
Following Einstein an event can be seen as the inter\-section of dif\/ferent worldlines.
From a~quantum f\/ield theory perspective, an event can be seen as a~vertex in a Feynman diagram,
with a given number of legs.
Each leg can be seen as a~particle labelled by $J$ and momentum $p_\mu^J$.
In case of a general addition $\oplus$, we can write the total momentum at this vertex as
\begin{gather}\label{56}
P_\mu^{\rm tot}=\big(p^1\oplus\cdots \oplus p^N\big)_\mu
\approx
\sum_Ip_\mu^I+\frac{1}{\smp}
\sum_{I<J}\Gamma^{\alpha\beta}_\mu p_\alpha^I p_\beta^J+\cdots,
\end{gather}
with the connection\footnote{Note in this approach the notation of covectors
and vectors are inverse to the notation we are using in general in this review,
since the basic manifold is momentum space.
Hence a~covector will have index up,
whereas a~vector will have the index down.} $\Gamma^{bc}_a$ on momentum space.

A non-zero torsion is equivalent to the non-commutativity of the addition,
whereas non-associativity of the addition is equivalent to a~connection with
non-zero curvature~\cite{RL, Frei}.
Hence, a~Lie algebra type non-commutative
space which has therefore a~momentum addition (in general) non-commutative but
associative will correspond~-- in this RL context~-- to a~choice of a~f\/lat connection
with torsion.
In particular, the $\smp$-Minkowski spacetime illustrates this case~\cite{mercati}.

The particle $J$ has momentum $p^J\in\cM$ and position $x_J\in
T^*_{p^J}\cM$.
At the intersection of the worldlines when
particles interact, we expect conservation of momenta, that is
$P_\mu^{\rm tot}=0$.
This means that this event (the vertex) sits in
the cotangent space $T^*_{p=0}\cM$ at the origin of momentum space
and has coordinates $z^\mu$.
To get the vertex interaction in $T^*_{0}\cM$, we therefore need to parallel transport
the (covector) coordinates $x_J$ to $T_0\cM$
\begin{gather*}
{{\big(\tau_J^{-1}\big)}^\mu}_\nu x_J^\nu=z^\mu,
\end{gather*} where ${\tau_J^\mu}_\nu$ encodes the parallel transport
associated to the particle $J$.
In~\cite{ACFKGS1}, the authors showed
that an action for $N$ particles with the constraint
implementing momentum conserva\-tion~$P_\mu^{\rm tot}$ lead to such a~construction.
There
is a~unique coordinate $z^\mu$ for the interaction vertex and in
particular (cf.~\eqref{56})
\begin{gather*}
{\tau_J^\mu}_\nu=\frac{\partial
P^{\rm tot}_\nu}{\partial p^J_\mu}.
\end{gather*} The interaction coordinates do
not Poisson commute,
\begin{gather}\label{zcom}
\{z^\mu,z^\nu\}=
\frac{1}{\smp}{T_\sigma}^{\mu\nu}z^\sigma+\frac{1}{\smp^2}
{R_\sigma}^{\mu\nu\rho}p_\rho z^\sigma+\cdots,
\end{gather}
${T_\sigma}^{\mu\nu}$ and ${R_\sigma}^{\mu\nu\rho}$
are respectively the torsion and the Riemann tensor for the connection $\Gamma$
associated to the momenta addition~\eqref{56}.
These coordinates are therefore
interesting candidates for the meaning of
the quantum operators encoded in the non-commutative geometry as
discussed in Section~\ref{ncg}.
Indeed, if we take the case of a
connection with constant torsion but zero curvature,~\eqref{zcom}
becomes the classical analogue of a~Lie algebra type non-commutative
space.
In particular one can retrieve the $\kappa$-Minkowski classical
case~\cite{mercati}.

The transition to the reference frame of a~distant observer is
carried out simply by a~translation with the inf\/initesimal form
\begin{gather*}
x^\mu_J\dr x^\mu_J+\delta_\epsilon x^\mu_J,
\end{gather*}
where
\begin{gather*}%\label{57}
\delta_\epsilon x_J^\mu=\epsilon^\nu\{x_J^\mu,P_\nu^{\rm tot}\}
\approx b^\mu+\frac{1}{\smp} \epsilon^\nu\sum_{I>J}\Gamma_\nu^{\mu\beta} p_\beta^I.
\end{gather*}
In consequence, for dif\/ferent particles $\delta x_I^\mu$ is
dif\/ferent, so that the translated endpoints of the worldlines do not
meet at the vertex and the interaction appears non-local for distant
observers.
In this way locality becomes relative to a~certain
extent, as pointed out in~\cite{UJ,LeeA,GACMM,LeeB,SH,SH2}.

In~\cite{RL} the emission of a~low-energy and a high-energy photon
and their absorption by a~distant detector as a model for radiation
from gamma-ray-bursts is discussed in detail.
In RL, the speed of
light is an invariant, but the trajectories of photons at dif\/ferent
energies with their origin and end points lie in dif\/ferent copies of
space-time.
To compare them and to calculate a~possible time delay
between the photons, one must parallel transport the corresponding
cotangent spaces into one.
Note that, for example for an absorption
event, the endpoint of the detector's worldline before the
absorption, the origin of its worldline after the absorption and the
endpoint of the photon's worldline do not coincide in general.
These
non-localities at the absorption or emission events are relative,
depending on the observer's reference frame, but the resulting time
delay in f\/irst order,
\begin{gather*}%\label{58}
\Delta T=-\frac{1}{2} TEN^{+++},
\end{gather*}
is an observer-independent invariant.
$T$ is the running time of the
high-energy photon in the detector's frame (at rest with the source
in the model), $E$ is the photon's energy in this frame and
$N^{+++}$ denotes the component of the non-metricity tensor
$(N_{abc}=\nabla_a g_{bc})$ of the connection along the photon
direction.
In the RL framework current observations
\cite{FERMI,Abdo} can be interpreted as implying a~bound on
non-metricity
\begin{gather*}%\label{59}
|N^{+++}|\leq\frac{1}{0.6 \smp}.
\end{gather*}

A second ef\/fect, derived from the same model is dual gravitational
lensing: Two photons with proportional momenta need not propagate
into the same directions.
When the connection of the curved momentum
space has torsion, a~rotation angle of
\begin{gather*}%\label{60}
\Delta\theta=\frac{E_1+E_2}{2} |T^a|
\end{gather*}
is predicted, where the vector $T^a$ arises by projection of the
torsion tensor into the direction of the photons' momenta and $E_i$
are the photons'energies. Ef\/fects of curvature show up in the
approximation quadratic in $E/\smp$~\cite{Frei}.
The framework
of $\kappa$-Ponicar\'{e} algebras is an example with non-metric
connection, zero curvature and non-zero torsion.
Further experiments
that may measure or bound the geometry of momentum space at order
$\smp^{-1}$ include tests of the linearity of momentum
conservation using ultracold atoms~\cite{Arz} and the development of
air showers produced by cosmic rays~\cite{Anton}.

\subsection{Generalized uncertainty principle}

This approach is connected with the strong gravity regime rather
than with the amplif\/ication of ``low'' energy ef\/fects.
Nevertheless,
it has formal similarity with DSR theories.
But the concept of the generalized uncertainty principle (GUP) has a
physically compelling basis: QG ef\/fects should occur at densities
comparable with the Planck density, not in the presence of large,
extended masses.
For a~modeling of gravity-caused modif\/ications of
scattering processes at extreme energy, see~\cite{GU}.
GUP is tied
to the center-of-momentum (c.o.m.) energy of two or more particles,
concentrated in a~small region in an interaction process, or the
energy of one particle in relation to some matter background, like
the CMB. When in a~scattering process the c.o.m.
energy is high
enough, so that in the scattering region the energy density comes
close to the Planck density, there is signif\/icant space-time
curvature and gravity is non-negligible.
The gravitational inf\/luence
is described by a~local energy dependence of the metric.
At this
point GUP introduces a~split between the {\em momentum} and the {\em
wave vector} of a~particle.
In the case of two scalar particles
scattering in the c.o.m.
system the asymptotic momenta are~$p^\mu$
and~$-p^\mu$, related linearly to the wave vectors
$k^\mu=p^\mu/\hbar$ in the asymptotic region.
The curvature caused
by the energy density is described by a~dependence of the metric on
the wave vector, $g_{\mu\nu}=g_{\mu\nu}(k)$, leading to a~modif\/ied
dispersion relation
\begin{gather}\label{k}
k^\mu g_{\mu\nu}(k) k^\nu=m^2,
\end{gather}
where $m$ is a~mass parameter.
Relation~\eqref{k} containing
higher-order terms in $k$, $k^\mu$ is not a~Lorentz vector and will
not transform according to standard f\/lat-space Lorentz
transformations.
Provided we do not assume graviton production, the
asymptotic momenta are conserved, but in the interaction region the
relation between $p^\mu$ and $k^\mu$ becomes nonlinear.

Formally, the asymptotic momenta, which are acted upon linearly by
the Lorentz group, play a~role analogous to that of the
pseudo-variables in DSR. But, whereas in DSR the latter ones are
mere auxiliary quantities, here they have a~clear, distinct physical
meaning.
The nonlinear variables $k^\mu$, on the other hand, play a
more or less auxiliary role, in contrast to the nonlinear ``physical
variables'' in DSR.

So far, this is nothing more than an ef\/fective description of
gravity, when it plays a~role in high energy particle interaction.
The place where it is encoded in the ef\/fective theory is the form of
the function $k(p)$, or its inverse, respectively.
This function
could, in principle, encode Newtonian gravity, GR, or QG. The input
from QG that is made here is the existence of a~minimal length,
$1/\smp$.
General conditions on the functional
dependence of $k$ on $p$ are given in~\cite{GU}:
\begin{enumerate}\itemsep=0pt
\item For energies much smaller than $\smp$ the usual linear
relation is found.
\item For large energies, $k$ goes asymptotically to $\smp$, not to inf\/inity.
\item The function is invertible, i.e.\
it is monotonically
increasing.
\end{enumerate}
With these conditions satisf\/ied, $k$ remains bounded when $p$ grows
arbitrarily and the (ef\/fective) wavelength $\lambda=2\pi/|k|$ does
not decrease below the invariant length.
Theories of this type have
been examined in various contexts as to their analytical structure
and phenomenological consequences~\cite{KGM,MM1,MM2,ACam}.

Recalling the quantum mechanical relation $p=\hbar k$, an
energy-dependent relation between momentum and wave vector can be
formulated as an energy dependence of Planck's constant,
$p^a=\hbar(p) k^a$, thus introducing an additional quantum
uncertainty.
Such a~modif\/ication of quantum mechanics was suggested
for the f\/irst time by Heisenberg~\cite{Hei}.
The physical idea
is that a~suf\/f\/iciently high energy particle, released in the
interaction at a~detecting process, curves and disturbs space-time
so that an additional position uncertainty arises, which enhances the
quantum mechanical one.
In this way the accuracy of position
measurement is bounded from below by an invariant length scale.

To formulate GUP in terms of commutation relations, we postulate the
canonical relation for the wave vector with a~coordinate~$x$ and
derive the relations between coordinates and momenta,
\begin{gather*}%\label{61}
[x^a,k_b]=i\hbar\delta^a_b,\qquad
[x^a,p_b]=i\hbar\frac{\partial p_b}{\partial k_a}.
\end{gather*}
This results in the generalized uncertainty relations
\begin{gather*}%\label{62}
\Delta x^a\Delta
p_b\geq\frac{\hbar}{2}\left|\left\langle\frac{\partial p_b}{\partial
k_a}\right\rangle\right|.
\end{gather*}

Comparison with DSR-type theories shows modif\/ied dispersion
relations as a~common feature, in this sense they are almost two
sides of the same coin~\cite{self}.
The interpretation of MDRs, however,
is quite controversial.
DSR deals with particle propagation in f\/lat
space, GUP deals with QG ef\/fects in regions with strong gravity.
Both approaches could be compared with each other and with the angle
operator~\cite{angle_obs} by calculating corrections to scattering
cross sections.
A~comparison of DSR versus GUP was
made in~\cite{remarx}, resulting in an opposite inf\/luence of DSR and
GUP on scattering cross sections.

\subsection{QG decoherence}
\label{alternate}

Another very general approach, without close relation to a~specif\/ic
QG theory, is decohe\-ren\-ce~\cite{deco}.
This framework considers
quantum theoretical f\/luctuations of the Minkowski metric, so the
proper time of particles f\/luctuates at a~time scale $\lambda T_{\rm
P}$, a~few orders above the Planck time.
In matter wave
interferometry, described in~\cite{deco}, atoms act as clocks with
a very high frequency.
When in such an experiment a~matter wave is
split into two components and recombined, space-time f\/luctuations
are expected to cause decoherence, a~process, where ingoing pure
states become mixed by the dissipative action of the f\/luctuating
background.
\cite{deco} presents one realization and potential
observable consequences, which predict a~factor $\lambda$ of the
order $10^3$.
Other realizations can be found, for example, in
references in~\cite{deco} and in~\cite{kelley}, where impacts on
neutrino physics are considered.
For an early contribution see~\cite{quant}.

\section{Quantum f\/ield theory frameworks}\label{section4}

In this section we review Lorentz symmetry violating EFTs, associated constraints, a~possible ``combinatoric lever arm'' in scattering experiments, and non-commutative f\/ield theory.

\subsection{Constraints on Lorentz violation with ef\/fective f\/ield theory}
\label{LV}

A successful area of quantum gravity phenomenology in recent years
is in the area of Lorentz symmetry breaking.
There is a
well-developed framework, many new ef\/fects, and very strong
constraints on these ef\/fects, even below the
Planck scale.
However we do not know whether violations of local
Lorentz invariance occur in LQG~-- in fact there
are strong arguments suggesting LLI should be preserved as discussed in Section~\ref{LLILQG}.
The now-extensive body of work provides a~f\/irm
foundation for future derivations, constraints and observational
searches.

When local Lorentz symmetry is broken a~cascade of physical ef\/fects
appear.
Whether the LV is via symmetry breaking or by an additional
background f\/ield, the ef\/fects are studied by adding the possible
terms with the new Lorentz violating vector and tensor f\/ields.
To
organize these ef\/fects and the associated constraints, the terms in
the particle Lagrangian are characterized by their mass dimension.

Without ties to a~specif\/ic fundamental theory, the minimal
Standard Model Extension provides a~framework of organizing
renormalizable higher dimension terms in the standard mo\-del~\cite{CK}.
There is an extensive literature on these ef\/fects and
the associated constraints.
For more on these constraints see
\cite{K,B}.
See~\cite{smetables} for data tables on the standard
model extension parameters, updated annually.

There is a~wealth of ef\/fects that arise from LV. Some of these include
\begin{itemize}\itemsep=0pt
\item sidereal variation in signals as the Earth moves relative to the preferred frame or directions.
\item new processes such as photon decay, photon splitting, and vacuum Cerenkov radiation
\item shifting of thresholds of allowed processes in LLI physics such as the GZK threshold
\item kinematic ef\/fects arising from modif\/ications to dispersion relations that accumulate over cosmological distances or over a~large number of particles
\end{itemize}
Work on dimension 5 and 6 operators is more recent and this section
will focus on these.
For more on work in LV prior to the developments in LQG see the reviews
\cite{JLMrev,JLMrev2,Mrev,LMrev}.
The primary reference to consult for dimension~5 LV QED is~\cite{JLMrev}.

Ef\/fective Field Theory (EFT) is the framework for describing much of
the standard model of particle physics.
According to the Wilsonian
view of renormalization, at the relatively low energies (as compared
to the Planck scale in the case of quantum gravity) explored by
accelerators, the dominant interactions in the Lagrangian/action are
necessarily the relevant terms at these energies.
This means that
when introducing new ef\/fects it is natural to use the EFT framework
to model the physics.
In this section the EFT framework is used to
explore the possible LV dimension 5 operators in QED, yielding cubic
modif\/ications to dispersion relations.

Including a~time-like background four vector f\/ield $u^\mu$ in QED Myers and
Pospelov found three LV dimension 5 operators that were (i)
quadratic in the same f\/ield, (ii) gauge invariant, and (iii)
irreducible under use of the equations of motion or a~total
derivative~\cite{MP}.
Due to the background vector f\/ield $u^\mu$
there is a~preferred frame, usually assumed to be the frame in which
the cosmic microwave background is isotropic. The resulting
operators (up to factors of 2 used here to simplify the dispersion
relations) are
\begin{gather*}
\cL_{LV}=\frac{\xi}{2\smp}u^\mu F_{\mu\nu}\,u\cdot\partial\big(u_\sigma\tilde{F}^{\sigma\nu}\big)+
\frac{1}{\smp}u^\mu\bar{\psi}\gamma_\mu\left(\zeta_1+\zeta_2\gamma_5\right)(u\cdot\partial)^2\psi.
\end{gather*}
Recall that since $\smp=M_{\rm P}$, the parameters $\xi,\zeta_i$ are all dimensionless.
These terms
violate CPT symmetry.
The resulting modif\/ied dispersion
relations include cubic modif\/ications compared to the LLI case.
For photons,
\begin{gather}
\label{photonMDR}
\omega_{\pm}^2=p_\gamma^2\pm\frac{\xi}{\smp}p_\gamma^3,
\end{gather}
where the signs indicate left and right polarization and the modif\/ication depends on the magnitude of the momentum.
While, for fermions,
\begin{gather}
\label{fermionMDR}
E_\pm^2=p^2+m^2+\eta_{\pm}\frac{p^3}{\smp}
\end{gather}
in which $\eta_{\pm}=2(\zeta_1\pm\zeta_2)$ and the signs are the positive and negative helicity
states.
The analysis of the free particle states is in~\cite{JLMrev} where it is shown that the positron helicity states
have a~relative sign compared to the electron; there are only two
fermion parameters $\eta_\pm$.

When the sign of the modif\/ication is negative (positive) the energy decreases (increases) relative to the LLI theory.
Thus the curve $E(p)$ f\/lattens (steepens) out at high energies.
This strongly af\/fects the process rate, thresholds, and the partitioning of momentum.

Comparing these dispersion relations to those found with the heuristic LQG computations
in Section~\ref{GP}, \eqref{neutdisp}~and~\eqref{photdisp} we see that the dimension 5 CPT-odd terms arose in
that calculation, giving similar modif\/ications in the dispersion
relations when
\begin{gather*}
\eta_{\pm}=\mp\frac{\kappa_7}{2}\left(\frac{1}{L\smp}\right)^{1+\Upsilon}\qquad \text{and}\qquad \xi\simeq4\theta_8.
\end{gather*}
(Recall that $L$ is the characteristic scale of the semi-classical state, above which the geometry is f\/lat.)
The photon MDR is essentially identical.
The fermion
MDR has only one helicity parameter and has an additional
suppression due to the scaling $L\smp$.
Further work on LQG
coherent states would clarify this scaling and illuminate the
additional scaling arising in the preliminary calculation, perhaps
even show an associated custodial symmetry.
The dimension 6 terms
are identical to the EFT framework.
However, none of these
operators considered have been derived unambiguously from the
framework of LQG.

The EFT analysis has been extended in a~variety of ways.
Hadrons~\cite{MTML} and even heavy nuc\-lei~\cite{SMS} were included, the framework has been folded into the Standard Model Extension~\cite{KM}.
Additionally, the analysis was generalized to arbitrary 4-vectors is~\cite{GGACM}.
In the original work of Myers and Pospelov the 4-vector $u^\mu$
only had a~non-vanishing time component in the preferred frame, chosen due to the stringent constraints on spatial anisotropy set by clock comparison and spin-polarized matter experiments.
Given the remarkable bounds on the parameters in the Myers--Pospelov model discussed below, the authors of~\cite{GGACM} work with a~model of anisotropic media and show that the bounds in the Myers--Pospelov model may be weakened when analyzed in the more general, anisotropic model.

In the context
of LV EFT a~variety of new phenomena occur~\cite{JLM,JLMrev}
\begin{itemize}\itemsep=0pt
\item New processes, forbidden in the usual LLI theory, can occur.
\item Thresholds for processes in the LLI theory can shift.
\item Upper thresholds can occur; momenta can be high enough so that processes turn of\/f.
\end{itemize}

The modif\/ications become important when the mass term is comparable to the modif\/ication.
Thus for dimension 5 operators the cubic corrections become important when
$p_{\rm crit}\approx(m^{2}M_{\rm P})^{1/3}$.
Since the ef\/fects arise at high momentum many calculations are done for $m\ll p\ll\smp$,
which allows for considerable simplif\/ication.
We'll use $\simeq$ to denote results in this ``high momentum'' limit.

It is worth keeping in mind that this framework excludes a~wider
class of theories, for instance those that contain violations of
local energy-momentum conservation.
This wider arena was brief\/ly
reviewed in Section~\ref{alternate}.
Nevertheless at some energy
scale the new theories must match known results and f\/it within the
EFT framework so at lower energies the framework is an excellent
approximation.
The EFT framework also has the advantage that clear
physical predictions can be made such as in the cases of particle
process thresholds, where the rates of new particle processes
determine the thresholds~\cite{JLMrev}.

\subsubsection{Physical ef\/fects giving current constraints}
\label{eftconstraints}

To give a~f\/lavor of the nature of the constraints in the next
subsections we sketch the derivations of the current tightest
constraints on dimension 5 LV QED. There are many other ef\/fects but
in the interest of reviewing those that give both a~sense of the
calculations and the strongest constraints, we focus on vacuum
birefringence and photon decay.
The f\/ield is comprehensively
reviewed in~\cite{JLMrev}, to which the reader should turn for
details.
A recent update is~\cite{LMrev}.

The f\/irst phenomenon,
arising from birefringence, is purely kinematical.
The remaining
constraints on dimension 5 LV QED are dynamical in the sense that
the dynamics of EFT is employed to derive the constraints.
The
ef\/fects usually involve an analysis of process thresholds.

Threshold constraints require answers to two questions, ``Is the
process allowed?'' and ``Does it occur?'' A typical process
involves a~decay of an unstable particle into two particles.
In the
EFT framework we have usual f\/ield theory tools at our disposal so we
can compute the rate of decay using the familiar expression from
f\/ield theory.
Denoting the outgoing momenta~$p'$ and~$p''$ and
helicity with $s$ and the matrix element by $M(p,s,p',s',p'',s'')$
the rate is
\begin{gather}
\Gamma(p,s,s',s'')=
\int\frac{1}{8(2\pi)^2E_{{\bf p}s}}\frac{d^3p'd^3p''}{E_{{\bf p}'s'}E_{{\bf p}''s''}}
\left|M(p,s,p',s',p'',s'')\right|^2
\nonumber\\
\phantom{\Gamma(p,s,s',s'')=}
\times\delta^{(3)}({\bf p-p'-p''}) \delta(E_{{\bf p}s}
-E_{{\bf p}'s'}-E_{{\bf p}''s''}).\label{rate}
\end{gather}
Because of the modif\/ications in the dispersion relations the nature of the integration dif\/fers
from the simple textbook case.
The threshold for the process is derived by determining the momentum at which the matrix element
is non-vanishing and the momentum space volume is suf\/f\/iciently large to ensure that the process is rapid.
Due to the modif\/ications in the dispersion relations the momentum space volume dif\/fers signif\/icantly
from the corresponding LLI calculation.

\subsubsection{Kinematic constraints arising from birefringence}\label{biref}

From the form of the dispersion relation
\eqref{photonMDR} it is clear that left and right circularly~polarized photons travel at dif\/ferent speeds.
Linear polarized high
energy radiation will be rotated through and energy-dependent angle,
depolarizing the radiation.
After a~distance $d$ the polarization
vector of a~linearly polarized plane wave with momentum $k$ rotates
as~\cite{JLM}
\begin{gather}
\label{polarization}
\theta(d)=\frac{\omega_+-\omega_-}{2}d\simeq\xi\frac{k^2d}{2\smp}.
\end{gather}
Given the f\/inite bandwidth of a~detector, $k_1<E<k_2$, the
constraint on the parameter $\xi$ can be derived from the
observation of polarization in the relevant bandwidth; if the LV
term was large enough there would be no measured net polarization.
If some polarization is observed then the angle
of rotation across the bandwidth must be less than~$\pi/2$
\cite{JLMS,GK}. The simple detection of a~polarized signal yields
the constraint, from~\eqref{polarization}, of
\begin{gather*}
\xi\lesssim\frac{\pi\smp}{(k_2^2-k_1^2)d}.
\end{gather*}
For a~ref\/inement of this argument, and other approaches relying on knowledge of the source, see \cite[Section~IV.B]{LMrev}.
The current best constraint on $|\xi|$ is discussed in Section~\ref{currentconstraints}.
An intriguing {\em change} in polarization during a~GRB was recently reported~\cite{Yonetoku}.

\subsubsection{Dynamical constraints arising from photon stability}
With LV photons can decay via pair production, $\gamma\rightarrow e^+e^-$.
The threshold for photon decay is determined as described above, by investigating the rate.
Then given the observed stability of high energy photons, constraints can be placed on the new LV terms in the dispersion relations.

Photon decay generally involves all three LV parameters
$\xi$, $\eta_+$, and $\eta_-$.
However as argued in~\cite{JLMrev}, we
can obtain constraints on the pairs of parameters $(\eta_-,\xi)$ or
$(\eta_+,\xi)$ by considering the case in which the electron and
positron have opposite helicity.
We present the calculation for
$\eta_+\equiv\eta$ and will see that the~$\eta_-$ case is easily
obtained from this one.
The LV terms are~$\eta_+p_-^3$ for the
electron and $-\eta_+p_+^3$ for the positron.
In the threshold
conf\/iguration the outgoing momenta are parallel and angular momentum
is not conserved.
But, above the threshold, the outgoing momenta
can deviate from the parallel conf\/iguration and, with the additional
transverse momentum, the process will conserve angular momentum.

For slight angular deviations from the threshold conf\/iguration of
the outgoing momenta the matrix element does not vanish and is
proportional to the perpendicular momentum of the outgoing particles~\cite{JLMrev}.
The volume of the region of momentum space where
photon decay occurs is determined by energy conservation and the
boundary of the region occurs when the perpendicular momentum
vanishes.
We denote the photon momentum $k$, electron momentum~$p_-$, positron momentum~$p_+$, and the helicity parameter $\eta=\eta_+<0$.
Thus, from $\omega=E_++E_-$ and equa\-tions~\eqref{photonMDR} and~\eqref{fermionMDR} we see that
\begin{gather*}
k\sqrt{1\pm\xi\frac{k}{\smp}}=p_+\sqrt{1+\frac{m^2}{p_+^2}-\eta\frac{p_+}{\smp}}+
p_-\sqrt{1+\frac{m^2}{p_-^2}+\eta\frac{p_-}{\smp}}
\end{gather*}
or, using conservation of momentum and the high momentum limit, the expression of energy conservation becomes
\begin{gather}\label{econs}
\pm\xi\frac{k^2}{\smp}\simeq\big(m^2+p_\perp^2\big)
\left(\frac{1}{p_+}+\frac{1}{p_-}\right)+\frac{\eta}{\smp}\big(p_-^2-p_+^2\big),
\end{gather}
where $p_\perp$ is outgoing particle transverse momentum.
The LV terms in the
dispersion relation raise or lower the particle's energy as a
function of the momentum.
Because of the f\/lattening out of the
energy at high momentum for negative~$\eta$, the outgoing energy is
reduced if one particle carries more momentum than the other.
The
threshold is thus determined by an asymmetric momentum partition~\cite{JLM}.
Partitioning via an asymmetry parameter $\Delta>0$ we
have, with $p_-=(k/2)(1-\Delta)$ and $p_+=(k/2)(1+\Delta)$,
\begin{gather}\label{EconsDelta}
\pm\xi\frac{k^2}{\smp}=\frac{4}{k}\big(m^2+p_\perp^2\big)\frac{1}{1-\Delta^2}-\frac{\eta k^2}{\smp}\Delta.
\end{gather}
This relation is enforced in the rate of equation~\eqref{rate} by
the energy-conserving delta-function.
Carrying out the integrations
over the momenta, the integral for the rate is reduced to a~single
integration over the longitudinal momentum, when the perpendicular
momentum is determined by energy conservation.
Then
\begin{gather*}
p_\perp^2=\pm\xi\frac{k^3}{4\smp}\big(1-\Delta^2\big)-m^2+\frac{\eta k^3}{4\smp}\Delta\big(1-\Delta^2\big).
\end{gather*}
Given the strong constraints on $\xi$ via vacuum birefringence (see Section~\ref{biref}),
$\xi\simeq0$, we proceed with vanishing $\xi$.
The volume in momentum space for photon decay opens up as the perpendicular momentum increases from zero.
The threshold is then determined by $p_\perp=0$, or
\begin{gather*}
-\frac{4m^2\smp}{\eta k^3}+\Delta\big(1-\Delta^2\big)=0.
\end{gather*}
A simple optimization shows that the asymmetry is $\Delta=1/\sqrt{3}$ and the outgoing momenta are
$(k/2)(1\pm1/\sqrt{3})$ yielding a~threshold of
\begin{gather*}
k_{th}=\left(\frac{6\sqrt{3}m^2\smp}{\eta}\right)^{1/3},
\end{gather*}
as determined from~\eqref{econs} with $\xi=0$.
The rate increases
rapidly above this threshold~\cite{JLMrev} so observation of photons
up to this energy places limits on the size of the parameter $\eta\equiv\eta_+$.
Thus, when we consider the 80~TeV photons
observed by HEGRA~\cite{aharonian} we have $|\eta_+|<0.05$. In
general high energy photon observations place limits on one
helicity. The more general case with non-vanishing $\xi$ may be
derived from~\eqref{EconsDelta}, with the complete allowed region in
$(\eta,\xi)$ parameter space determined numerically.

It is interesting to note that, contrary to what one might expect
from early work~\cite{JLM,limits}, the threshold for the process is
not determined by minimizing the outgoing energy, which is found
from the solution of $(4m^2\smp/\eta k^3)\Delta-(1-\Delta^2)^2=0$.
Instead we start by asking, does the process
occur? The rate~\eqref{rate} then shows that the threshold is
determined by the opening up of momentum space due to non-vanishing~$p_\perp$.

There are a~wide variety of other processes that can contribute to
the dimension 5 LV QED constraints.
In addition to the ones
discussed above, there is vacuum Cherenkov $(\e^\pm\rightarrow
e^\pm\gamma)$, helicity decay $(\e^\pm\rightarrow e^\mp\gamma)$,
fermion pair production $(e^\pm\rightarrow e^{\pm}e^{\pm}e^\mp)$,
photon splitting $(\gamma\rightarrow n\gamma)$, and the shifting of
the photon absorption $\gamma\gamma\rightarrow e^+e^-$ threshold.
For more on these the reader should consult~\cite{JLMrev,LMrev,JLMrev2}.
In addition the model has been
generalized to include processes including dimension 6 CPT even
operators in QED and for hadrons~\cite{MTML}, and for nuclei~\cite{SMS}.

\subsubsection{Neutrino physics}

Planck-scale modif\/ications of neutrino physics are usually
considered in the framework of LLI violating ef\/fective f\/ield theory.
It is well-known that neutrinos in a~def\/inite f\/lavor state
$|\nu_\alpha\rangle$, $\alpha=e, \mu, \tau$, are superpositions of
neutrinos in def\/inite mass states, $|\nu_i\rangle$, $i=1, 2, 3$,
\begin{gather*}
|\nu_\alpha\rangle=\sum_iU_{\alpha i}^*\,|\nu_i\rangle
\end{gather*}
with the unitary mixing matrix $U_{\alpha i}$.
On the other hand,
neutrinos in a~def\/inite mass state are superpositions of f\/lavor
states,
\begin{gather*}
|\nu_i\rangle=\sum_\beta U_{i\beta}\,|\nu_\beta\rangle.
\end{gather*}
In~\cite{JC} neutrino oscillation in a~simplif\/ied two-f\/lavor model
is considered.
The typical oscillation length $L$, i.e.\
the
distance, after which a~neutrino, moving approximately at the speed
of light in special relativity, is likely to have changed its
f\/lavor, is derived in a~simple, straight-forward calculation.
It
depends on the neutrino's energy,
\begin{gather*}
L(E,m)=\frac{2\pi}{\Delta p}\approx\frac{4\pi E}{\Delta m^2},
\end{gather*}
with $\Delta m^2=m_2^2-m_1^2$ and $\Delta p=p_2-p_1$.
The
approximate equality is based on $p_i=\sqrt{E^2-m_i^2}\simeq E-\frac{m_i^2}{2E}$ for each $i=1, 2$.

Planck scale modif\/ications arise when we assume modif\/ied dispersion
relations of the type (leaving aside possible birefringence ef\/fects)
\begin{gather}\label{ndis}
E^2=p^2+m^2+\sum_n\eta_i^{(n)} \frac{p^n}{M_{\rm P}^{n-2}}
\end{gather}
with coef\/f\/icients $\eta_i$, possibly dif\/ferent for each mass state.
Such MDRs would imply f\/lavor oscillations even for neutrinos with
negligible mass, as long as the coef\/f\/icients $\eta_i^{(n)}$ dif\/fer
from each other, i.e.\
when Planck-scale ef\/fects imply dif\/ferent
corrections for dif\/ferent mass states.

For the sake of simplicity, here we present only the version
suggested in~\cite{JC}, based on the (exact) alternative MDR
\begin{gather*}
p^2+m^2=E^2\left[1-\frac{(E-m)^2}{M_{\rm P}^2}\right],
\end{gather*}
which avoids oscillations of neutrinos with zero rest mass.
With
this, the modif\/ied oscillation length becomes
\begin{gather*}%\label{Em}
L'(E,m)=\frac{2\pi}{\frac{1}{2E} \Delta m^2-\frac{E^2}{M_{\rm
P}^2} \Delta m},
\end{gather*}
with the mass dif\/ference $\Delta m=m_2-m_1$.
Neutrino detectors like
IceCube may be sensitive to oscillation length corrections of
atmospheric neutrinos~\cite{MMGLS}.
For cosmogenic neutrinos, which
have oscillated many times on their way to earth, a~detection of such
ef\/fects would be very dif\/f\/icult.

In ef\/fective f\/ield theory with higher mass dimension terms, coupled
to a~LLI breaking four-vector and exact energy and momentum
conservation, neutrinos are favorite objects for thres\-hold
calculations.
Neutrino splitting, $\nu\rightarrow\nu\nu\bar\nu$ for
example, is forbidden in SR, but could be made possible by LV
ef\/fects.
Generally, for processes derived from the MDR~\eqref{ndis}
and energy-momentum conservation one obtains a~threshold energy
$E_{\rm th}\sim(m^2M_{\rm P}^{n-2})^\frac{1}{n}$.
With $\eta_i$
assumed to be of order unity, the small neutrino mass leads to a
threshold energy of $\sim20$~TeV for $\nu$ splitting.
If this
process takes place, it must result in a~cutof\/f in the energy
spectrum of cosmic neutrinos.
Taking the decay length for neutrinos
above the threshold energy into account, for $\eta_i\sim1$ in
\cite{MMGLS} an estimate of $10^{18}{-}10^{19}$~eV for the cutof\/f is
given, when the neutrinos come from a~distance of the order of Mpc.
Observation of higher neutrino energies would lower the upper bounds
to the parameters $\eta_i^{(n)}$.

\subsubsection{Current bounds from astrophysical observation: summary}
\label{currentconstraints}

\begin{figure}[t]\centering

\begin{tabular}{@{}cc@{}}
    \includegraphics[angle=90,scale=0.43]{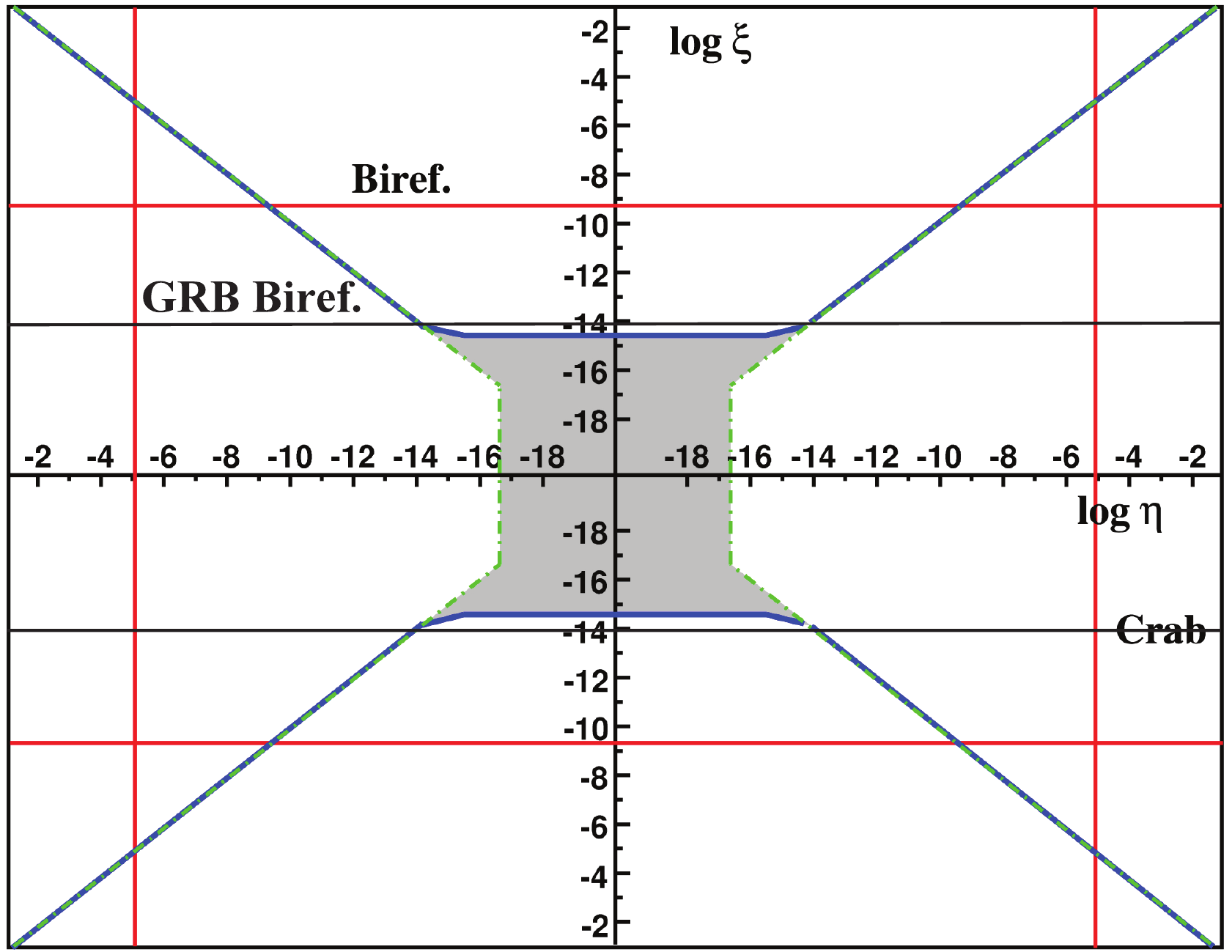} & \includegraphics[angle=90,scale=0.43]{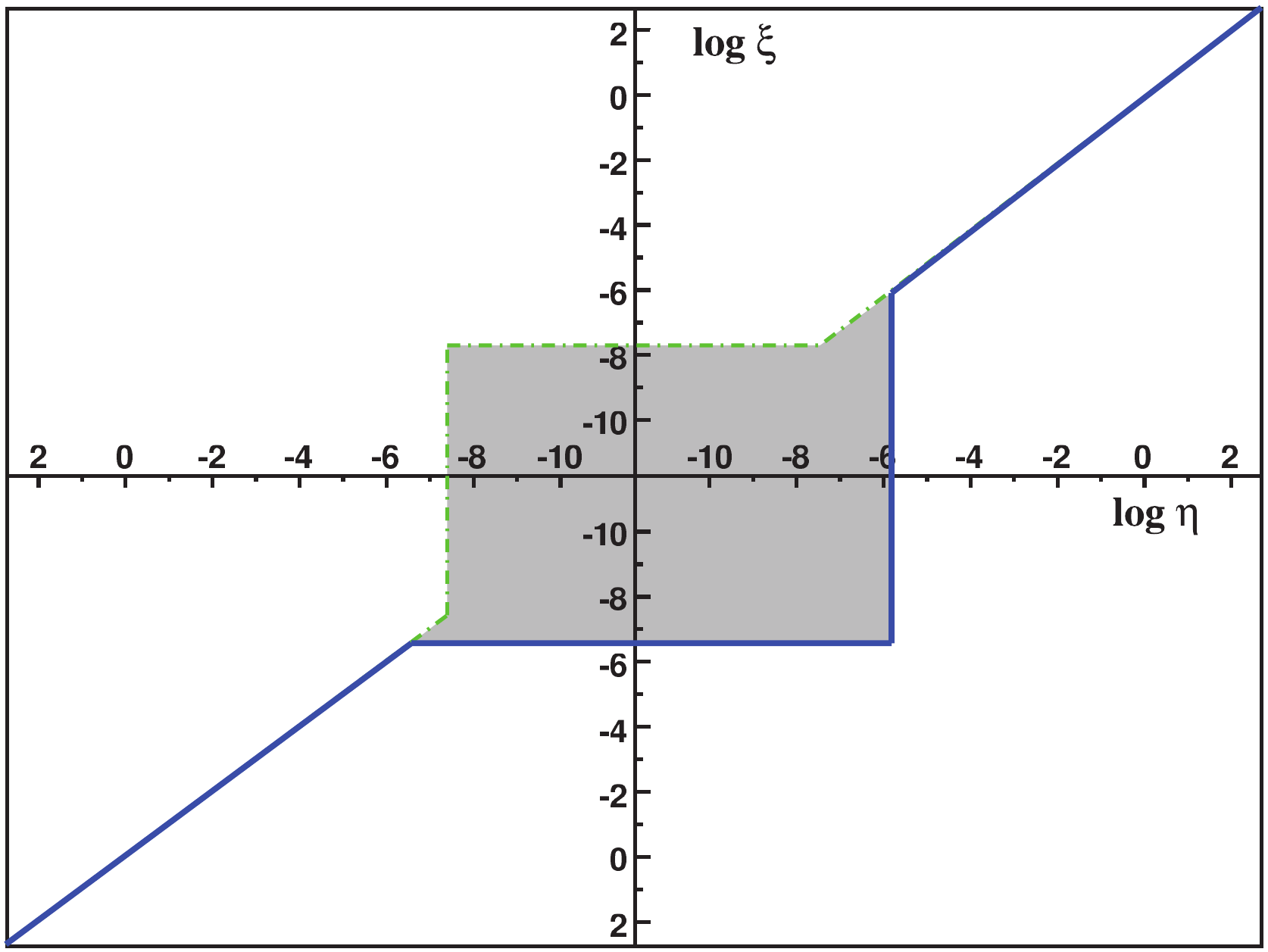} \\
    (a) & (b)
\end{tabular}

\caption{\label{astroconstraint}
The constraints on dimension 5~(a) and dimension 6~(b) QED parameters.
The log-log plots show
the allowed region (dark grey) and the constraints.
The current
constraints on dimension 5 parameters are shown in red and grey.
The grey horizontal lines are due to constraints due to the lack of
birefringence (see Section~\ref{biref}) from GRB photons~\cite{LGBCF,SGRB}.
The red vertical lines are due to the analysis of
the Crab spectrum~\cite{MLCK}.
The blue solid lines show limits that
would arise from an upper threshold on pair production at $k_{\rm th}
\simeq10^{20}$ eV. The green dash-dot lines show the constraints
that would arise from a~$\gamma$-decay threshold of $k_{\rm th}\simeq
10^{19}$ eV. The dimension 6 constraints (b) have the same color
coding.
The GRB birefringence constraint is not relevant to the
dimension 6 case since the operator is parity even, contrary to what
is reported in~\cite{LGBCF}.
Original plots courtesy of Stefano
Liberati.
See~\cite{LMrev} for more detail.}
\end{figure}

The current best constraints on the dimension 5 LV QED parameters
arise from $\gamma$-ray bursts and from an extensive analysis of the
spectrum from the Crab Nebula~\cite{MLCK}.
Using the vacuum
birefringence ef\/fect and recent observations of the $\gamma$-ray
burst GRB041219a, the dimension~5 photon parameter $\xi$ has been
constrained $|\xi|\lesssim10^{-14}$~\cite{LGBCF,SGRB}.
At 95
conf\/idence level $|\eta_\pm|<10^{-5}$ from the analysis of the
spectrum from the Crab nebula~\cite{MLCKU}.
Since the terms are
already suppressed by the Planck scale, terms arising from dimension
5 operators are suf\/f\/iciently constrained as to appear ruled out.
The
current constraints, and some prospects for further improvements are
illustrated in Fig.~\ref{astroconstraint}.

Other astrophysical arenas have been suggested for testing these
theories including neutron stars,~\cite{camacho,GM}.
But it is
clear from these studies that in the EFT framework the ef\/fects are
too small to be observationally accessible~\cite{GM}\footnote{The
result does not immediately extend to the DSR context
\cite{GM,dsr_chandra}.
In the DSR context there are no new
thresholds and the usual special relativity thresholds are only
slightly modif\/ied~\cite{HHMdsr}.}.

Considering that the Planck scale is already factored into the
modif\/ications, the severity of the constraints is impressive.
These
results from astrophysics inform the development
of quantum gravity.
However, there is another reason to suspect
that these modif\/ications do not occur in the EFT framework, the
naturalness problem.

\subsubsection{The naturalness problem}

Within the EFT framework a~naturalness problem suggests that the LV
operators are ruled out.
The argument is based on renormalization:
The higher dimension LV operators generate lower dimension operators
through radiative corrections and the low-energy ef\/fective theory
will contain all terms consistent with the symmetries of the
microscopic theory.
The usual power-counting arguments that give
divergences also determines the natural size of the Planck-scale
suppressed terms.
Generically these appear at dimension 4 or less,
with no Planck scale suppression.
One can show~\cite{CPSUV,CPS2}
that the operators generated by radiative corrections produce
ef\/fects that are incompatible with current particle data.
Thus the
LV parameters would have to be unnaturally f\/ine-tuned to cloak the
LV.

\looseness=-1
Within the EFT framework this naturalness problem is signif\/icant.
However if there is a~custodial symmetry then the radiative terms
will not appear at lower dimension.
Even a~subgroup of the symmetry
manifest in the macroscopic, continuum approximation is suf\/f\/icient.
A good example of this is in Euclidean lattice theory where the
discrete rotation sub-group on a~hyper-cubic lattice, which becomes
the full rotation group as the lattice spacing goes to zero, is
enough to prevent the generation of the lower dimensional operators.
Other symmetries existing in the vast range of scales from the
``low'' TeV scale to the high Planck scale might protect the theory
from these terms.
This is the case for a~model with supersymmetry;
the symmetry-preserving operators f\/irst appear at dimension~5~\cite{NP}.
This solves the naturalness problem in that the lower
dimensional operators are protected by a~custodial symmetry.
But,
of course, supersymmetry is not a~low-energy symmetry.
When
supersymmetry is softly broken with explicit symmetry breaking terms
at a~scale $M_{\rm SUSYB}$, the lower dimension terms return, albeit
additionally suppressed by factors of $M_{\rm SUSYB}/\smp$ raised to
some power so, again, the naturalness problem returns at dimension 5
and the parameters would have to be f\/ine-tuned.
At dimension~6
however, the additional suppression is suf\/f\/icient to evade current
limits, provided the supersymmetry breaking scale $M_{\rm SUSYB}<100$
TeV~\cite{BNP}.

\looseness=-1
Another possibility was raised recently. Gambini~et~al.~\cite{GRP1} study a~Euclidean lattice model, with distinct spatial,
$a$, and temporal, $(1+a\mu)a$ lattice scales.
The authors f\/ind
that at the one loop order the leading contribution to ef\/fective
dimension 4 operators is $a\mu$.
If one takes the stand that the
lattice spacing $a$ should remain small but f\/inite then the lower
dimension term in the self-energy is suppressed by the additional
factor $\mu$ and thus could evade the current bounds on this ef\/fect.
Nonetheless within the EFT framework of LV this does not solve the
naturalness problem since the theory would have to be f\/ine-tuned
through the scale $\mu$.
See~\cite{P} for further comments.

Incidentally Gambini~et~al.\ show that in the 4D ``isotropic
scaling'' case in which the temporal lattice scaling is also~$a$,
that lower dimensional corrections do not arise, since the
``isotropy'' symmetry protects the theory from such corrections.
The
authors argue in Section~IV.B of~\cite{GRP2} that in a~generally
covariant theory the propagators should be constructed with
reference matter f\/ields, physical ``rods and clocks'' rather than the
f\/lat background used in EFT. Since the distribution associated to
these physical rods and clocks cannot be inf\/initely narrow, and
should be considerably larger than the Planck scale, they would
provide a~natural cutof\/f for LV ef\/fects without the generation
of corrections at lower dimension.
In the EFT framework, the widths could provide an additional scale
and associated additional suppression, as in the soft breaking of
super symmetry.
It would be interesting to see whether the hint of
just such a~scale and suppression in the modif\/ied dispersion
relations~\eqref{photdisp} could be made more precise.

\subsubsection{Generalizations and prospects for improved constraints}
\label{beyondD5}

Given the naturalness problem and the tight constraints on
dimension 5 LV QED in the EFT framework several generalizations have
been considered.
Building on earlier work dimension 6 CPT-even LV
operators were studied~\cite{M} and constrained~\cite{LMrev}.
In the
pure photon sector in the Standard Model Extension there are bounds
up to dimension 9~\cite{KM}.

Ultra-high energy cosmic rays (UHECR) hold promise for tightening
the current constraints.
One striking aspect of the cosmic ray
spectrum is the expected GZK cutof\/f.
In the LLI theory
cosmologically sourced ultra-high energy protons interact with
cosmic microwave background producing pions and lower energy
protons, neutrinos, and $\gamma$-rays, reducing the energy of the
original photons.
The resulting GZK (so named for Greisen, Zatsepin
and Kuzmin) cutof\/f occurs at $E_{\rm GZK}\simeq5\times10^{19}(\omega_{cmb}/1.3~\text{meV})^{-1}~\text{eV}$,
where
$\omega_{cmb}$ is the energy of the background photon.
LV terms in
the hadronic dispersion relations can shift the threshold for
photo-pion production~\cite{limits,MTML}.
Recently observations in
support of the GZK cutof\/f were announced~\cite{hires,auger1,auger2}
so the LLI theory is favored.
A detailed statistical analysis of
UHECRs severely constrains proton and pion parameters in the
dimension 5 and 6 CPT-even model~\cite{MTML}.

The GZK process provides another mechanism for constraints.
Photo-pion
production from the UHECR protons leads to the production of $\gamma$-ray pairs.
Cosmic ray observatories have placed limits on the fraction of UHE photons in the
cosmic ray spectrum~\cite{auger3,Rphoton} that can in turn place limits on
photon parameters.
With LV, processes can have upper thresholds at high energies
and the existence of an upper threshold for pair production at high energies
increases the photon fraction, yielding a~conf\/lict with current photon fraction
bounds~\cite{GS}.
Likewise a~neutrino f\/lux is created by pions originating from
the same photo-pion production.
The possible tests of LV in neutrino physics
are outlined in~\cite{MMGLS,LMrev}.

Additionally, in an intriguing development in the {\em
non-relativistic} regime Amelino-Camelia et~al.\ argue that
cold-atom-recoil experiments constrain dispersion relations
modif\/ications of the form $mp/\smp$~\cite{ALMT}.

There remains work to be done on the theoretical side, understanding
the status of LLI in LQG and, if verif\/ied, constraining the MDR
parameters, for instance as in equations~\eqref{neutdisp} and~\eqref{photdisp}.
In particular, it would be very interesting to
determine which specif\/ic structures in LQG, together with any
additional assumptions, yield specif\/ic ef\/fects.
As those specif\/ic
ef\/fects are constrained, then the structures and assumptions may be
tested directly.
Some questions for further work include:
\begin{itemize}\itemsep=0pt
\item Since the scaling in terms of $\Upsilon$ holds for dif\/ferent dimension operators,
the clock compa\-ri\-son experiments, for instance, should yield bounds on $\Upsilon$ and
vacuum birefringence for dimension 4 (see~\cite{JLMrev} page 29).
The correlation between
the dimension 3 and 4 results could yield limits on $\Upsilon$.
\item Building on~\cite{ST1, ST2} is there non-integer scaling in the dispersion relations?
What specif\/ic structures in LQG is the scaling tied to?
\item Is there an analogous calculation that can be completed in the spin foam context?
\item Is is not easy to see how such non-integer scaling could occur in EFT, can it be ruled out?
Or, is EFT missing something? (see, e.g.~\cite{GRP1,GRP2}).
\item Can physical measurements and the underlying spatial discreteness of LQG kinematics be
reconciled with the EFT framework?
\end{itemize}

\subsection{EFT Phenomenology without Lorentz violation:\\ a~combinatoric lever arm}

Another model in the EFT framework demonstrated that there may be
``lever arms'' intrinsic to the granularity of space
\cite{angle_obs,angle_loops11}, demonstrating one way in which the
combinatorics of SWN nodes may raise the ef\/fective scale of the
granularity.
The model is based on the angle operator.

The asymmetry in the angular spectrum shifts the distribution of angles
away from the $\sin\theta$ isotropic distribution of polar angles in 3-dimensional
f\/lat space.
The distribution is recovered when the spin through the
three surfaces, ``f\/lux'' or $s_i$, satisf\/ies $1\ll s_i\ll s_3$ for
$i=1,2$.
Fluxes $\vec{s}$ that satisfy these relations are called
``semi-classical''.
These f\/luxes are essentially the areas of the partitions discussed in Section~\ref{angle}.

The model of~\cite{angle_obs} is based on the assumption that
the states of the atom of 3-geometry or intertwiner are equally likely.
In addition to the uniform probability measure, the model assumes that all incident edges to the node are all
spin-$\tfrac{1}{2}$.
The combinatorics of the atom can be solved
analytically for semi-classical f\/luxes.
Given the f\/luxes $\vec{s}$, or areas of the partitions, and the labels
$\vec{j}$ on the intertwiner vertex, or ``intertwiner core'', that connects the three partitions,
the normalized probability
distribution is given $(\vec{n}=2\vec{j})$
\begin{gather*}%\label{ndist}
p_{\vec{s}}(\vec{n})\simeq\prod_{i=1}^3\frac{n_i}{s_i}\exp\left(-\frac{n_i^2}{2s_i}\right).
\end{gather*}

Accessible measurements of the atom include 3-volume, which is approximately
determined by the total area or f\/lux, and angle, determined by the states
$\ket{\vec{n}}$ of the intertwiner core.
The f\/luxes $\vec{s}$ determine a~mixed state,
\begin{gather*}%\label{mixedstate}
\rho_{\vec{s}}=\sum_{\vec{n}}p_{\vec{s}}(\vec{n})P_{\ket{\,\vec{n}}},
\end{gather*}
where $P_{\ket{\vec{n}}}$ is the projector on the orthonormal basis
of the intertwiner core.
The sum is over the admissible 3-tuples of
integers $\vec{n}$ such that $n_i\leq s_i$.
In the discrete case the projector is $P_{\ket{\,\vec{n}}}=\ket{\theta_I}\bra{\theta_I}$,
where $\ket{\theta_I}=\sum_{\vec{n}}c_{\theta_I}(\vec{n})\ket{\vec{n}}$.
The probability of f\/inding the angle eigenvalue
$\theta_I$ in the mixed state $\rho_{\vec{s}}$ is
\begin{gather*}
%\label{discrete_dist}
\text{Prob}(\theta=\theta_I;\rho_{\vec{s}})=\operatorname{tr}\left(\rho_{\vec{s}}P_{\theta_I}\right)
=\sum_{\vec{n}}p_{\vec{s}}(\vec{n})|\braket{n}{\theta_I}|^2\equiv p_{\vec{s}}(\theta).
\end{gather*}
This procedure can be used to calculate the probability in the
continuum approximation~\cite{angle_obs}
\begin{gather}
\label{rawdisttheta}
P_{\vec{s}}(\theta)=\int{\rm d}^3n p_{\vec{s}}(\vec{n})|c_\theta(\vec{n})|^2\delta\left(\theta-\theta(n)\right).
\end{gather}
The integration of equation~\eqref{rawdisttheta} is straightforward
\cite{angle_obs}.
The key step in the calculation is the
identif\/ication of the ``shape parameter'' $\epsilon:=\sqrt{s_1
s_2}/{s_3}$ that measures the asymmetry in the distribution of
angles.
The resulting measure, when expressed in terms of Legendre
polynomials and to $O(\epsilon^3)$, is~\cite{angle_obs}
\begin{gather*}
%\label{moddist}
\rho_\epsilon(\theta)\simeq\sin\theta\left(1-\frac{8}{\pi}P_1(\cos\theta)\epsilon+\frac{3}{2}P_2(\cos\theta)\epsilon^2\right).
\end{gather*}
The af\/fect of the modif\/ied distribution of polar angles is that the
`shape' of space is altered by the combinatorics of the vertex; the
local angular geometry dif\/fers from f\/lat 3-space.
While these
ef\/fects would be in principle observable at any f\/lux, the results
here are valid for semi-classical f\/lux, $1\ll s_j\ll s_3$ for
$j=1,2$.
As mentioned brief\/ly above, the total f\/lux $s=\sum_i s_i$ determines
the 3-volume of the spatial atom and thus an ef\/fective mesoscopic
length scale, $\ell_s=\sqrt{s}/\smp$, greater than the
fundamental discreteness scale of $1/\smp$.
The combinatorics
of the intertwiner provides a~lever arm to lift the fundamental
scale of the quantum geometry up to this larger mesoscopic scale.
So
while the the shape parameter $\epsilon$ is free from the Planck
scale, the ef\/fective length scale, determined by total f\/lux~$s$, is
tied to the discreteness scale of the theory.

If the scale $\ell_s$ of the spatial atom is large enough then the
underlying geometry would be accessible to observations of particle
scattering.
An overly simple example of Bhabha scattering is worked out
in~\cite{angle_obs}.
However, the model needs further development.
In particular for coherent states def\/ined with classical directions lever
arm is too short to raise the scale up to an observable window~\cite{angle_loops11}.
It is an open
question as to whether the long combinatoric lever arm exists for coherent states
built from quantum information intrinsic to the states of the atom of geometry.
Further, the QED vertex should be modeled in detail.

\subsection{Non-commutative f\/ield theory}\label{sec:ncqft}
\subsubsection{Overview}\label{sec:qft-dsr}

As discussed in Section~\ref{ncg}, a~non-commutative geometry can be generated when the coordinates functions~$X_\mu$
are \textit{operators} that do not commute.
To introduce a f\/ield over this non-commutative space, we have to consider a f\/ield over these coordinates operators.
As in Section~\ref{ncg}, we shall focus on f\/lat non-commutative geometries.
In the case of a scalar f\/ield, this can be done rigorously in most of the non-commutative geometries discussed
in Section~\ref{ncg}.

However, instead of using operators and all their machinery, we can simplify the mathematics and use instead
c-numbers $x_\mu$ for the coordinates with a~non-trivial product which encodes the non-commutative structure.
This approach can be understood either as a~specif\/ic representation of the non-commutative algebra
(def\/ined in terms of operators) or as an example of deformation.
At the end, the formalism is equivalent.
The representation is obtained through the Weyl map~$W$, which is such that
\begin{gather*}
W(X_\mu)=x_\mu,\qquad W(X_\mu X_\nu)=x_\mu*x_\nu.
\end{gather*}
We see that the non-trivial operator product is encoded in a~new product denoted $*$.
This is naturally extended to general (analytic) functions.
The deformation can be understood as modifying the point-wise product of functions.
The usual algebra $\cC^\infty(\R^4)$
of commutative functions over spacetime $\R^4$ is equipped with the pointwise product
\begin{gather*}
(f_1.f_2)(x)=f_1(x)f_2(x),\qquad\forall\, f_i\in\cC^\infty\big(\R^4\big).
\end{gather*}
It becomes a~non-commutative algebra if we replace the pointwise product by the $*$-product.
\begin{gather*}
(f_1.f_2)(x)\dr(f_1*f_2)(x).
\end{gather*}
The $*$-product of coordinates functions $x_\mu$ encodes a~similar non-commutative structure as
the operators discussed in Section~\ref{ncg}.
For example, respectively in the Moyal case and the $\kappa$-Minkowski case, we will have
\begin{gather*}
\text{Moyal space:}
\quad
x_\mu*x_\nu-x_\nu*x_\mu=\frac{i}{\smp^2}\theta_{\mu\nu},
\\
\text{$\kappa$-Minkowski space:}
\quad
x_0*x_i-x_i*x_0=\frac{1}{\smp}x_i,\qquad x_j*x_i-x_i*x_j=0.
\end{gather*}
This representation is not unique.
For example for the Moyal spacetime, the most popular representations of the $*$-product
are given by~\cite{Balachandran:2009as}
\begin{gather*}
\text{Moyal:}
\quad
(f_1*_M f_2)(x)
=f_1(x)e^{\frac{i}{2}\theta^{\mn}\overleftarrow{\partial_\mu}\overrightarrow{\partial_\nu}}f_2(x),\\
\text{Voros:}
\quad
(f_1*_V f_2)(x)
=f_1(x)e^{\frac{i}{2}
\left(\theta^{\mn}\overleftarrow{\partial_\mu}\overrightarrow{\partial_\nu}
-\overleftarrow{\partial^\mu}\overrightarrow{\partial_\mu}\right)}f_2(x).
\end{gather*}
This $*$-product construction of a~non-commutative space is analogue to the phase-space formulation
of quantum mechanics introduced by Moyal~\cite{QM-moyal} and Groenewald~\cite{groene}.

With the help of this $*$-product, we can generalize the usual construction
of the scalar f\/ield action with a~$\phi^3$ interaction term for a real scalar f\/ield $\phi$ from
\begin{gather}
\int[d^4x] \left[\left((\partial_\mu\phi).(\partial^\mu\phi)\right)(x)+m^2(\phi.\phi)(x)
+\frac{\lambda}{3!}(\phi.\phi.\phi)(x)\right]
\qquad
\text{to}\nonumber\\
%\downarrow\\
\int[d^4x] \left[\left((\partial_\mu\phi)*(\partial^\mu\phi)\right)(x)+m^2(\phi*\phi)(x)
+\frac{\lambda}{3!}(\phi*\phi*\phi)(x)\right].\label{star action}
\end{gather}
Note that in the case of the DFR spacetime as well as Snyder spacetime interpreted as a~subspace of a larger space, one should also consider the measure on the extra coordinates.
We refer to~\cite{Doplicher:1994tu, Girelli:2010wi} for further details on these cases.

The usual approach in QG phenomenology is to construct a~scalar f\/ield theory as it is the simplest f\/ield theory.
The case of f\/ields with higher spin is more complicated.
Indeed, spin is usually related to a~representation of the Poincar\'e group.
If we deform this symmetry group to accommodate for the non-commutative structure, the representation theory will change and hence what we call spin could change.
Furthermore the relation spin-statistics will be non-trivial since, in the case of a quantum group, the tensor product of representations becomes non-commutative.
See~\cite{Pinzul:2005gx, Balachandran:2005eb} for the Moyal case and \cite{Arzano:2008bt, Young:2007ag}
for the $\kappa$-Minkowski case.
At this stage, it is only in the Moyal case where the notions of spinor~\cite{Borris:2008zr}
and vector f\/ield have been introduced, since the non-commutative structure is a~``mild'' one in terms of deformation (it is a~twist~\cite{majid}).

Since the hope is to measure some semi-classical QG ef\/fects in the propagation of electromagnetic f\/ield, one would like
to construct a~${\rm U}(1)$ gauge theory in one's preferred non-commutative spacetime.
Unfortunately this can be done to our knowledge only in the Moyal case.
Even in this case, things are highly non trivial at the classical level (and of course at the quantum case as well!).
For example, a ${\rm U}(1)$ gauge theory behaves essentially like a non-Abelian gauge theory due to the non commutativity~\cite{SheikhJabbari:1999iw}.
It is dif\/f\/icult to construct non-Abelian gauge theories with simple local groups since the non-commutativity will always destroy the traceless property \cite{Jurco:2000ja}.
There are also constraints on the transformations of the matter f\/ields under gauge transformations \cite{Chaichian:2001mu}.
Since the dispersion relation is not modif\/ied in the Moyal case, we do not expect to see ef\/fects measurable in gamma-ray bursts a priori.

There is no def\/inite construction of an Abelian nor non-Abelian Yang--Mills theory in
either of the $\kappa$-Minkowski, DFR and Snyder cases.
This is an important issue
to address if one intends to make precise predictions for the FERMI experiment.

Once we have def\/ined the scalar f\/ield action in spacetime, we can try to def\/ine it in momentum space, that is we introduce plane-waves and a~Fourier transform.
Majid has introduced in a~general setting the notion of Fourier transform for Hopf algebras~\cite{majid}.
We follow here a more pedestrian presentation.

As we have emphasized in Section~\ref{sec:particle}, momentum space is equipped with a~group structure since we need to add momenta.
Furthermore the pointwise product of plane-waves as functions on spacetime incorporates this momenta addition,
\begin{gather*}
(e_p.e_q)(x)=e^{p\cdot x}e^{iq\cdot x}=e^{i(p+q)\cdot x}=e_{p+q}(x).
\end{gather*}
The generalization to the non-commutative case then uses the $*$-product between plane-waves.
The Moyal case and the $\kappa$-Minkowski case lead to dif\/ferent cases.
\begin{gather*}
\text{Moyal:}
\quad
(e_p*e_q)(x)=e^{i(p+q)\cdot x}e^{\frac{i}{2}p\cdot\theta\cdot q},\\
\text{$\kappa$-Minkowski:}
\quad
(e_p*e_q)(x)=e_{p\oplus q}(x),
\end{gather*}
where we have used the non-trivial sum $p\oplus q$ of~\eqref{mod add} inherited from
the non-Abelian group~$\AN_3$ as discussed in Section~\ref{ncg}.
With this in hand, we def\/ine the Fourier transform and its inverse as
\begin{gather}\label{fourier}
\hphi(p)=\cF(\phi)(p)\equiv
\int\! \big[d^4x\big](\phi*e_p)(x),\qquad\phi(x)
=\cF\mone(\hphi)(x)\equiv
\int \!\big[d^4p\big]\hphi(p)e_{x}(\ominus p),\!\!
\end{gather}
where we have used the relevant measure $[d^4p]$ on momentum space.
For example, $\AN_3$ is isomorphic to half of de Sitter space, hence $[d^4p]$
will be the measure on de Sitter space expressed in the chosen coordinates.
In the case of the inverse Fourier transform $\cF\mone$,
we see the plane-wave as a~function over momentum space and we have to deal with the relevant inverse $\ominus$
of the addition $\oplus$, i.e.\ $\ominus p\oplus p=0=p\oplus(\ominus p)$.
We use the commutative pointwise product for the algebra of functions over the momentum manifold.
Deforming this product as well, i.e.\
to have a~non-commutative momentum space, would mean that we are dealing with a quantum group momentum space.
This case has not yet been studied to our knowledge.

To perform the Fourier transform of the action~\eqref{star action},
it is usually convenient to consider the plane-wave as the eigenfunction of the derivative
$\partial_\mu$.
This requires in general a~careful study of the dif\/ferential calculus
over the non-commutative space as for example in $\kappa$-Minkowski case.
We refer to~\cite{Sitarz:1994rh, Freidel:2007hk} for further details on this.
We assume that the plane wave is the eigenfunction of the derivative so that
\begin{gather*}
\partial_\mu e_p(x)=ip_\mu e_p(x),\qquad\cF(\partial_\mu\phi)(p)=i p_\mu\hphi(p).
\end{gather*}

Note that when dealing with functions over a~group (which we call here momentum space), there is another notion of Fourier transform which is usually used.
For example, in the case of compact groups
(for instance $\SU(2)$ as a 3d Euclidian momentum space),
the Fourier transform one would think to use consists
in decomposing the functions over the group in terms of the matrix elements
of the representations of the group, thanks to the Peter--Weyl theorem.
This is not the Fourier transform we have discussed above in~\eqref{fourier}.
There exists nevertheless a~natural isomorphism between the dif\/ferent types of Fourier transform~\cite{Joung:2008mr}.
To our knowledge the isomorphism has not been studied in detail in the case of non-compact groups for momentum space.

With the Fourier transform~\eqref{fourier} the $\lambda\phi^3$ action becomes, in the Moyal case,
\begin{gather*}
\int[dp][dq]\,\hphi(p)\big(p^2-m^2\big)\hphi(q)\delta(p+q)\\
\qquad
{} -\frac{\lambda}{3!}\int[dp]^3\hphi(p_1)\hphi(p_2)\hphi(p_3)
e^{i\left(p_1\cdot\theta\cdot p_2+(p_1+p_2)\cdot\theta\cdot p_3\right)}\delta(p_1+p_2+p_3),\nonumber
\end{gather*}
and, in the $\kappa$-Minkowski case,
\begin{gather*}
\int[dp][dq]\,\hphi(p)\kk(p)\hphi(q)\delta(p\oplus q)
-\frac{\lambda}{3!}\int[dp]^3\hphi(p_1)\hphi(p_2)\hphi(p_3)\delta(p_1\oplus p_2\oplus p_3).
\end{gather*}
Note that we still have a~conservation of momenta in both cases.
We notice the key dif\/ference: In the Moyal case, the Dirac delta function comes decorated with
a~phase depending on momentum and $\theta$, but the conservation of momenta is obtained through the usual commutative addition of momenta.
In the $\kappa$-Minkowski case, the conservation of momenta is done through
the modif\/ied addition, inherent to the new group structure that we use.
Furthermore we could have a modif\/ied propagator $\kk(p)$~--
related to a modif\/ied dispersion relation~-- in the $\kappa$-Minkowski case.

If we focus on a~general group for momentum space (therefore on a spacetime with Lie algebra type),
we can rewrite the action for a~scalar f\/ield in momentum space just in terms of group elements
\begin{gather}\label{actiongroup}
\int[dg]^2\,\hphi(g_1)\kk(g_1)\hphi(g_2)\delta(g_1g_2)
-\frac{\lambda}{3!}\int[dg]^3\hphi(g_1)\hphi(g_2)\hphi(g_3)\delta(g_1g_2g_3).
\end{gather}
$[dg]$ is the Haar measure on the group of interest.
Note that to be rigorous one should be careful with the dif\/ferent coordinates patches used to cover the group.
We omit this subtlety and refer to~\cite{Joung:2008mr, Freidel:2007yu} for further details.
Equation~\eqref{actiongroup} is nothing but a~sum of convolution products of functions over the group
evaluated at the identity since
\begin{gather*}
(\hpsi\circ\hphi)(g)= \int[dg]^2\,\hphi(g_1)\kk(g_1)\hphi(g_2)\delta\big(g_1g_2g\mone\big)
\qquad\text{with}\quad\hpsi(g_1)
=\hphi(g_1)\kk(g_1),\\
(\hphi\circ\hphi\circ\hphi)(g)= \int[dg]^3\,\hphi(g_1)\hphi(g_2)\hphi(g_3)\delta\big(g_1g_2g_3g\mone\big).
\end{gather*}
The usual commutative scalar f\/ield theory can also be put under this form if we use the Abelian group $\R^4$.
The $\kappa$-Minkowski case appears when the group is non-Abelian and is $\AN_3$.

Deforming the addition of momenta can be seen as another way to deform our theory.
We can start with the standard scalar f\/ield theory def\/ined over momentum space given by $\R^4$
and introduce a~non-trivial addition.
By considering plane-waves with a~non-trivial product, inherited from the deformed momenta addition,
we reconstruct the $*$-product.
This way of proceeding is essentially the dual to the one we have presented here.

The abstract writing of the scalar f\/ield action is useful as it can help to understand how matter can arise from a~spinfoam~\cite{Girelli:2009yz}.
We shall discuss this in the next subsection.

We have discussed the classical def\/inition of the scalar f\/ield action.
Before discussing the quantum case, let us comment on two specif\/ic topics: symmetries and conserved charges.
We have discussed in Section~\ref{ncg} how some non-commutative spaces can be seen as f\/lat non-commutative spaces.
We can therefore expect that the action of the scalar f\/ield will be invariant under the deformed Poincar\'e symmetries~\cite{Chaichian:2004za, Freidel:2007yu}.
A scalar f\/ield will be simply the trivial representation of the deformed Poincar\'e group.
According to the chosen deformation, the main dif\/ference with the usual case will be how the tensor product of f\/ields are transformed.
For example, for a~translation by $\epsilon$ in the $\kappa$-Minkowski case, we have
\begin{gather*}
\hphi(p)\dr e_\epsilon(p)\hphi(p),
\qquad %\nonumber\\
\hphi(p)\otimes\hphi(q)\dr e_{p\oplus q}(\epsilon)\big(\hphi(p)\otimes\hphi(q)\big)
=(e_{p}*e_q)(\epsilon)\big(\hphi(p)\otimes\hphi(q)\big).
\end{gather*}
We notice the appearance of the $*$-product and the modif\/ied momenta addition which encodes
the deformation of the symmetries.
In particular $\hphi(p)\otimes\hphi(q)$ and $\hphi(q)\otimes\hphi(p)$ do not transform
in the same way since $p\oplus q\neq q\oplus p$.
This is another way to see that the tensor product is no longer commutative when using quantum groups.
Nevertheless, it can be shown that the scalar f\/ield action in the dif\/ferent
non-commutative spacetimes is invariant under the relevant deformed symmetries as expected.
In the $\kappa$-Minkowski case, we have a~deformation of the Poincar\'e symmetries
but one can also encounter Lorentz symmetry breaking if one is not careful
about the choice of coordinates patch~\cite{ Freidel:2007yu}.

If we have symmetries, one can expect to have conserved charges following Noether's theorem.
This is indeed true in the non-commutative context.
They have been analyzed for both Moyal~\cite{AmelinoCamelia:2007wk}
and $\kappa$-Minkowski spacetimes~\cite{Freidel:2007hk}.
The analysis relies on the understanding of the dif\/ferential calculus over the non-commutative space.
For example in the $\kappa$-Minkowski case there exist dif\/ferent types
of dif\/ferential calculus \cite{Sitarz:1994rh} which leads then to dif\/ferent notions
of conserved charges \cite{Agostini:2006nc, Freidel:2007hk}.
We refer to the original articles for more detail.

The quantization of non-commutative scalar f\/ield theory can be performed.
The Moyal case has been analyzed in great detail, the other non-commutative geometries much less so.
For a~recent overview of some phenomenology of f\/ield theory
in Moyal spacetime, we refer to~\cite{Balachandran:2010wc}.

As we have alluded few times already, when dealing with a~quantum group, the tensor product
of its representations (i.e.\ here the scalar f\/ield) becomes non-commutative.
If we want to permute representations we have to use a~structure called ``\textit{braiding}'',
which encodes the non-commutativity of the tensor product.
Now, when constructing Feynman diagrams we use Wick's theorem and f\/ields permutations extensively.
Then we have a~choice: either consider the braiding related to the deformation
of the Poincar\'e group, or use a trivial braiding
(i.e.\ the one associated to the usual Poincar\'e group).
In the f\/irst case, this will ensure that the Feynman diagrams
are invariant under the deformed Poincar\'e group.
This is the setting of braided f\/ield theory as developed by Oeckl~\cite{Oeckl:1999zu} and Majid~\cite{majid}.
In the second case, we can encounter symmetry breaking.
In particular non-planar diagrams will often fail to be invariant under the Poincar\'e symmetries.
Quite strikingly, it has been shown that such braiding in the Moyal case means that
the non-commutative scalar f\/ield theory has the same amplitude as the commutative
one~\cite{Fiore:2007vg, Balachandran:2005pn}!
Thus we see that the Moyal deformation is a~``mild'' one.
Quantum gauge theories do feel the non-commutativity since at the classical level already~--
as we have recalled~-- some non-trivial ef\/fects happen.

From this perspective, not considering the braiding in the Moyal case leads
to a~more interesting scalar f\/ield theory, not equivalent to the standard one.
However one has to face a~new problem: the ultraviolet-infrared (UV-IR) mixing.
This problem arises when dealing wit non-planar diagrams which are most sensitive
to the non-commutative structure as we recalled above.
When the external momenta are \textit{not} zero, the space-time non-commutativity regula\-ri\-zes
the ultraviolet divergences, just like one would hope non-commutativity to do.
However when the external momenta are zero, the amplitude of the non-planar diagram diverges again.
Small (external) momenta lead to a high energy divergence~\cite{Minwalla:1999px, Matusis:2000jf}.
This has made the control of the renormalization analysis of the Moyal non-commutative scalar f\/ield theory
quite involved~\cite{Gurau:2008vd, Grosse:2009pa}.

One might wonder if braiding could simplify the amplitudes of a~quantum scalar f\/ield theory
in the $\kappa$-Minkowski case.
A priori, we should not expect to recover the same amplitudes as in
the un-deformed case there one really modif\/ies momentum space by adding curvature.
In particular the measure is dif\/ferent in the f\/lat case and curved case.
This is only a~preliminary argument since the $\kappa$-Minkowski case
faces another issue: the braiding is not completely understood; see~\cite{Young:2008zm} for further comments.
Finally, the IR-UV mixing also arises when considering a scalar f\/ield in $\kappa$-Minkowski~\cite{Grosse:2005iz}.

As a~f\/inal comment, let us recall that non-commutative geometry was introduced
by Snyder~\cite{Sn} as a hope to regularize the divergencies of f\/ield theory.
This hope was not realized in general.
Perhaps the only examples where this holds are DFR space \cite{Bahns:2003vb} and,
when momentum space is a quantum group \cite{Majid:1990gq, RodriguezRomo:1995am}.

\subsubsection{Relating non-commutative f\/ield theory and spinfoam models}\label{sec:derqftqg}
Group f\/ield theories (GFT) are tools to generate spinfoams.
Namely, they are scalar f\/ield theo\-ries with a~non-local interaction term, built on a product of groups.
Upon quantization, using the path integral
formalism, the Feynman diagram amplitudes can be interpreted as spinfoam amplitudes, constructed out of gravity or some topological models.
As we have recalled in Section~\ref{sec:qft-dsr}, even standard f\/ield theories can seen as some kind of group f\/ield theories.
The only dif\/ference is whether the group is Abelian or not, so that the dual space becomes commutative or not.
Realizing this is the f\/irst key to understand how one can recover non-commutative f\/ield theories encoding matter from a~spinfoam model.

GFT built on $\SO(4,1)$, it could probably contain in some ways a~DSR scalar f\/ield theory.
One has to carefully identif\/ies the scalar degrees of
freedom in the spinfoam GFT. One key-dif\/f\/iculty is to identify the DSR propagator since often the spinfoam GFT has a~trivial propagator.

As always, 3d quantum gravity being simpler than 4d quantum gravity models, it provides the ideal framework to illustrate this idea.
We consider therefore Boulatov's GFT which generates the Ponzano--Regge spinfoam amplitude describing Euclidian 3d quantum gravity
\cite{Boulatov:1992vp}.
It is def\/ined in terms of a~real scalar f\/ield $\vphi:\SU(2)^3\rightarrow\R$, which is required to be gauge
invariant under the diagonal right action of
$\SU(2)$,
\begin{gather*}
\vphi(g_1,g_2,g_3)=\vphi(g_1g,g_2g,g_3g),\qquad\forall\,
g\in\SU(2).
\end{gather*}
Boulatov's action involves a~trivial propagator and the tetrahedral interaction
vertex,
\begin{gather}
S_{3d}[\vphi]=\frac12\int[dg]^3\vphi(g_1,g_2,g_3)\vphi(g_3,g_2,g_1)
\nonumber\\
\phantom{S_{3d}[\vphi]=}
-\frac{\lambda}{4!}\int[dg]^6
\vphi(g_1,g_2,g_3)\vphi(g_3,g_4,g_5)\vphi(g_5,g_2,g_6)\phi(g_6,g_4,g_1).\label{boulatov}
\end{gather}
We note that the interaction term is def\/ined over six copies of $\SU(2)$,
the action is therefore very non-local, even in the non-commutative sense.
The equation of motion is
\begin{gather*}
\vphi(g_3,g_2,g_1) =
\frac{\lambda}{3!}\int{\rm d}g_4{\rm d}g_5{\rm d}g_6
\vphi(g_3,g_4,g_5)\vphi(g_5,g_2,g_6)\vphi(g_6,g_4,g_1).
\end{gather*}
In his PhD thesis~\cite{Livine:2003ux}, Livine identif\/ied some solutions of this equation of motion.
\begin{gather*}%\label{sol}
\vphi^{(0)}(g_1,g_2,g_3) = \sqrt{\frac{3!}{\lambda}} \int{\rm d}g\,
\delta(g_1g)F(g_2g)\delta(g_3g),
\qquad F: \ G\rightarrow\R,
\end{gather*}
when $\int[dg]\,F(g)^2=1$ (or $F=0$).

Later on, Fairbairn and Livine realized that the scalar perturbations $\hphi(g)$
around some specif\/ic solutions would actually behave exactly like a~scalar f\/ield theory
with $\SU(2)$ as momentum space~\cite{Fairbairn:2007sv}.
The ef\/fective action for $\hphi$
is constructed using
$\vphi(g_1,g_2,g_3)=\vphi^{(0)}(g_1,g_2,g_3)+\hphi\big(g_1g_3\mone\big)$
and Boulatov's action~\eqref{boulatov}
\begin{gather*}
S_{\text{ef\/f}}[\hphi]=
S_{3d}[\vphi]-S_{3d}\big[\vphi^{(0)}\big]=\int\hphi(g_1)\kk(g_1)\hphi(g_2)\delta(g_1g_2)
\nonumber\\
\phantom{S_{\text{ef\/f}}[\hphi]=}
-\frac\mu{3!}\int[dg]^3\,
\hphi(g_1)\hphi(g_2)\hphi(g_3)\delta(g_1g_2g_3)
-\frac{\lambda}{4!}\int[dg]^4\,\hphi(g_1)\cdots \hphi(g_4)\delta(g_1\cdots g_4),
\end{gather*}
with the kinetic term and the 3-valent coupling given in terms
of $F$
\begin{gather*}
\kk(g) = 1-2\left(\int [dh] F(h)\right)^2-\int [dh] F(h)F(hg), \qquad \text{and} \qquad
\frac\mu{3!} = \sqrt{\frac{\lambda}{3!}} \int [dh] F(h).
\end{gather*}
One can choose $F$ such that $\kk(g)$ becomes the standard propagator $p^2-m^2$ with a~non-zero
mass~\cite{Fairbairn:2007sv}.
We recognize then an action that is very close to the one in~\eqref{actiongroup}.
From this perspective, in 3d, using a GFT to generate spinfoams we can f\/ind some degrees
of freedom which can be interpreted as matter.
Furthermore these matter degrees of freedom have naturally a curved momentum space.
If we perform the Fourier transform \eqref{fourier}, we would recover matter as propagating in
a non-commutative spacetime, of the Lie algebra type.

The 4D extension of this model to recover a~scalar f\/ield in $\kappa$-Minkowski space
from a topo\-lo\-gi\-cal GFT (i.e.\ giving the BF spinfoam amplitudes) was proposed in~\cite{Girelli:2009yz}.
The construction is a bit more involved than in the above example since we have to deal with non-compact groups.
Furthermore momentum space in $\kappa$-Minkowski is the group $\AN_3$
which is not the one used to build the spinfoam model.
One has then to use dif\/ferent tricks to recover this group in the GFT. For further details
we refer to \cite{Girelli:2009yz}.

After these particle and f\/ield theory models we now turn to the LQG formulation of cosmological models,
a~promising observational window for QG phenomenology.
In fact, already in the standard model of cosmology the metric is used in the quantized perturbation variables.

\section{Loop quantum cosmology}\label{section5}

A line of research that renders potentially
observable results is Loop Quantum Cosmology (LQC).
(For readers new to this subject we suggest the recent LQC reviews~\cite{Cobsrev,BCMB,AS,buch}.)
In contrast to the subjects of the foregoing sections in this branch
of QG phenomenology we do not consider amplif\/ications of tiny
ef\/fects in the weak gravitational f\/ield regime, but rather today's
remnants of the strong gravitational regime in the early universe.
Given the observational windows onto the early universe, this
line of work holds promise for accessible hints of fundamental space-time
structure.

We do not have solutions to full LQG that could be restricted to
cosmological models.
So, to model the early universe
and to obtain a~dynamical evolution with
observable consequences, one assumes a~cosmological
background~-- usually highly symmetric, homogeneous or homogeneous and
isotropic models.
With a~scalar inf\/laton f\/ield one can consider
perturbations around the background by means of ef\/fective equations.
From the
ef\/fective equations one can derive estimates for correlation
functions of quantities of scalar and tensorial type, constructed
from perturbations of the connection around the isotropic case and
relevant for the period of inf\/lation.
Finally these
can be compared with the CMB inhomogeneities.

In homogeneous cosmological models the degrees of freedom are
reduced to a~f\/inite number by symmetry reduction
prior to (loop) quantization.
This results in simplif\/ied operators,
and particularly, in a~simplif\/ied constraint algebra, tailored to
the cosmological model under con\-si\-deration.
For such systems we
often know exact or at least numerical solutions.
Although not solutions of
full LQG, but of a~simplif\/ied of\/fspring of LQG, these constructions are
guided by the ef\/fort to be as close as possible to the full theory.

\looseness=1
In the following we illustrate the approach to LQC with the simplest
cosmological model, the Friedmann--Lema\^{\i}tre--Robertson--Walker
(FLRW) model with zero spatial curvature~\cite{AS}.
The
gravitational part of this model is one-dimensional, the only
geometrical dynamical variables are the scale factor $a(t)$ of the
universe and the expansion velocity $\dot{a}(t)$.
The Gauss and the
dif\/feomorphism constraints do not show up explicitly, they are
automatically satisf\/ied, what remains to solve is the Hamiltonian
constraint in form of a~dif\/ference equation.
Discreteness plays a
signif\/icant role only in the very early phase of the universe, in
the ensuing continuous evolution the dif\/ference equation can be
approximated by a~dif\/ferential equation.
The intermediate regime
between these two phases is the domain of quantum corrections to
classical equations.

The metric of the f\/lat FLRW model is usually given in the form
\begin{gather*}
g_{\mu\nu} {\rm d}x^\mu {\rm d}x^\nu=-{\rm d}t^2+a^2(t)\left({\rm
d}x_1^2+{\rm d}x_2^2+{\rm d}x_3^2\right)
\end{gather*}
with a~f\/iducial spatial Euclidian metric.
As the spatial topology of
the model is that of $\R^3$, one has to choose a~f\/iducial
cell $\cal C$ to obtain f\/inite integrals in quantities like the
total Hamiltonian, the symplectic structure, and others.
In comoving
euclidian coordinates the volume of such a~cell is denoted by $V_0$,
the corresponding geometric volume is $V=a^3V_0$.
When introducing
Ashtekar variables, we can, thanks to the symmetry of the model,
choose the homogeneous and isotropic densitized triad and connection
variables
\begin{gather*}
{A_a}^i=V_0^{-1/3}c {\delta_a}^i
\qquad\text{and}\qquad
{E^a}_i=V_0^{1/3}p {\delta^a}_i.
\end{gather*}
In terms of metric variables we have
\begin{gather*}
c=V_0^{1/3}\gamma \dot a
\qquad\mbox{and}\qquad
p=V_0^{2/3}a^2,
\end{gather*}
where $\gamma$ is the Barbero--Immirzi parameter.
The Poisson bracket is independent of the size of the f\/iducial cell,
\begin{gather*}
\{c,p\}=\frac{8\pi G\gamma}{3}.
\end{gather*}
With these variables the gravitational phase space of homogenous and
isotropic models is spanned by one canonical pair and, with a
spatially constant f\/ield $\phi$ and its canonical momentum, we have
f\/inite-dimensional quantum mechanics.
Approaches
of this kind are summarized under the notion of Wheeler--DeWitt (WDW)
theory or ``Geometrodynamics'', see~\cite{WDW}.

In LQC we want to take into account discreteness and so in the
spirit of LQG we construct holonomies from the connection variable.
For this purpose we choose an edge of $\cal C$, whose coordinate
length $V_0^{1/3}$ is multiplied by a~dimensionless parameter
$\mu$.
Like in LQG, where SNW edges carry quanta of area, $\mu$ will
later turn out to be a~measure of area.

The holonomy along such an edge is
\begin{gather}\label{Fhol}
h=\exp\left[\mu c\tau_k\right]=\cos\frac{\mu c}{2} \mathbb{I}
+2\sin\frac{\mu c}{2} \tau_k,
\end{gather}
where $\mathbb{I}$ is the $2\times2$ unit matrix and $\tau_k$ is an $\su(2)$
element(with normalization $(\tau_k)^{2}=-1/4 \,\mathbb{I})$.
Obviously it suf\/f\/ices to take
\begin{gather*}
{\cal N}_\mu(c):=e^\frac{i\mu c}{2}
\end{gather*}
as elementary functions in the connection representation.
The f\/lux
through a~face of $\cal C$ is given directly by $p$.
$\cal N$ and
$p$ make up the holonomy-f\/lux algebra of the f\/lat FLRW model,
promoted to operators, their commutator is
\begin{gather*}
\big[\hat{\cal N}_\mu(c),\hat p\big]=-\frac{8\pi\gamma G\hbar}{3}\frac{\mu}{2} \hat{\cal N}_\mu.
\end{gather*}

Quantum states are represented by almost periodic functions
$\Psi(c)$, i.e.\
countable superpositions of elementary plane wave
functions
\begin{gather*}
\Psi(c)=\sum_n\alpha_n\exp\frac{i\mu_nc}{2},\qquad\alpha_n\in
\C,\qquad\mu_n\in\R.
\end{gather*}
A brief introduction into this formalism, the Bohr compactif\/ication
of the real line, may be found in~\cite[Chapter~28]{T1} and~\cite{buch}.

In the kinematic Hilbert space with the norm
\begin{gather*}
||\Psi||^2=\lim_{D\rightarrow\infty}\frac{1}{2D}\int_{-D}^D\bar\Psi(c)\Psi(c)\,{\rm
d}c=\sum_n|\alpha_n|^2
\end{gather*}
the functions ${\cal N}_\mu$ constructed above form an orthonormal
basis,
\begin{gather*}
\langle{\cal N}_\mu|{\cal N}_{\mu'}\rangle
=\big\langle e^\frac{i\mu c}{2}\big|e^\frac{i\mu'c}{2}\big\rangle
=\delta_{\mu,\mu'}.
\end{gather*}
Note that on the right-hand side there is the Kronecker-$\delta$.
These functions are analogs of the SNW functions in LQG. The actions
of the holonomy and f\/lux operators on a~state function are by
multiplication and derivative, respectively,
\begin{gather*}
\hat{\cal N}_\sigma\Psi(c)=\exp\frac{i\sigma
c}{2} \Psi(c),\qquad\hat
p \Psi(c)=-i\frac{8\pi\gamma\ell_{\rm P}^2}{3}\frac{{\rm
d}\Psi}{{\rm d}c}.
\end{gather*}

It is also possible to go over to the
$p$-representation, which is sometimes more convenient.
Here the
quantum states are functions of $\mu$ and the operators act in the
following way,
\begin{gather}\label{mu}
\hat{\cal N}_\alpha\Psi(\mu)=\Psi(\mu+\alpha),\qquad\hat
p \Psi(\mu)=\frac{4\pi\gamma\ell_{\rm P}^2}{3} \mu \Psi(\mu),
\end{gather}
i.e.\
as shift operators and by multiplication.
Here $\mu$, which was originally introduced as a~dimensionless ratio of lengths in~\cite{AS}, is proportional to area, as a~factor in the eigenvalue of~$\hat p$.

The dynamics of cosmological models will be dealt with in Section~\ref{Latt}.
As the main goal in LQC is to be as close as possible to
the full theory, an
adaption of the full LQG Hamiltonian is more convenient than the
simplif\/ied Hamiltonian constraint resulting from a~symmetry reduced
model.
In this way discreteness enters the dynamics in a~much more
natural way.

Some more general features of LQC, applicable to cosmological models
of dif\/ferent degrees of complexity
summarize the expected LQC corrections.
\begin{itemize}\itemsep=0mm
\item The LQG Hamiltonian constraint contains an inverse volume
expression.
The volume operator has a~zero eigenvalue and therefore
does not have a~densely def\/ined inverse, for the inverse volume an
operator of its own must be constructed.
This is done in such a~way
that for ``large'' volume its eigenvalues go like $V^{-1}$, but for
``small'' volume they do not diverge, but in the limit
$V\rightarrow0$ eventually go to zero.
This construction contains
one parameter, on the value of which it depends, how many Planck
volumes have to be considered as ``large'' or ``small'' in the above
sense, this gives rise to quantum ambiguities.
The well-def\/ined
inverse volume operator is an important ingredient in resolution of
the classical cosmic singularity.
(See~\cite{AS} for more on
singularity resolution.)

\item The classical Hamiltonian constraint contains the
connection, which is not gauge invariant and so has no operator
equivalent in the gauge-invariant Hilbert space.
Like in full LQG,
connection variables are replaced in one or the other way by
corresponding holonomies, as in the example described above.
This
introduces in principle inf\/initely many terms of arbitrary powers of
the connection, leading to corrections in the classical equations.

\item There are quantum back reaction ef\/fects from f\/luctuations, which
occur in any system, when the expectation value of the Hamiltonian
operator is not the classical Hamiltonian function of the
expectation values of its arguments,
$\langle\hat H\rangle(q,p)\neq H\left(\langle\hat q\rangle,\langle\hat p\rangle\right)$.
This is the case for cosmological Hamiltonians.
Back reaction terms
are included into an ef\/fective Hamiltonian in the ef\/fective
Friedmann equations.

\item Ef\/fective Poisson brackets with a~correction parameter $\alpha$,
constructed from correction terms, should be anomaly-free.
Anomaly-freeness means that the constraints remain f\/irst-class,
which is essential for consistency and was shown explicitly in
several cases, but is not established in general in LQC.
\end{itemize}

In the following we consider holonomy corrections and inverse volume
correction in dependence of a~QG length parameter and their
interesting interplay.
Intuitively we can expect that the smaller
the length scale, the smaller the holonomy corrections and the
larger the inverse volume corrections, and vice versa.

\subsection{Holonomy corrections}
As above, we assume the f\/iducial cell $\cal C$ with comoving
coordinate volume $V_0$ and physical volume $V=a^3V_0$ of a
cosmological model partitioned into $N$ elementary building blocks
of volume $v=a^3 \frac{V_0}{N}$.
This gives a~length scale
$L=v^{1/3}$~\cite{BCT}.
Setting $L=\mu V^{1/3}$ with
$\mu$ corresponding to the state functions $|\mu\rangle$ ties $L$ to
the quantum theory.
Here $\mu$ appears f\/irst as a~dimensionless
proportionality factor of (classical) lengths; in quantum theory it
is connected with f\/luxes, see~\eqref{mu}.
A typical QG density
\begin{gather*}%\label{rho}
\rho_{\rm QG}=\frac{3}{8\pi GL^2},
\end{gather*}
becomes $\frac{3}{8\pi}$ times the Planck density $M_{\rm P}/\ell_{\rm P}$, when $L=\ell_{\rm P}$.
Polynomial terms in the
connection in the LQG Hamiltonian constraint operator are replaced
by holonomies~\eqref{Fhol} along an edge, this leads to higher-order
corrections.
Holonomy corrections become large when the exponent~$\mu c$ is of order one.

From the classical Friedmann equation we can express the matter
density in terms of~$L$ and~$c$,
\begin{gather*}%\label{fried}
\rho=\frac{3}{8\pi G}\left(\frac{\dot a}{a}\right)^2=\frac{3(c\mu)^2}{8\pi G\gamma^2L^2}.
\end{gather*}
Thus holonomy corrections are large when
$\rho\approx\gamma^{-2}\rho_{\rm QG}$.
As a~measure for holonomy
corrections we introduce
\begin{gather*}%\label{deltah}
\delta_{\rm hol}=\frac{\rho}{\rho_{\rm QG}}=\frac{8\pi
G}{3} L^2 \rho.
\end{gather*}
This relation implies that we may expect considerable holonomy
corrections in early phases of the universe, when the density is
large.

\subsection{Inverse volume corrections}
Thiemann~\cite{trick} showed that expressions containing the
inverse volume, like~\eqref{145}, which comes from the
Hamiltonian constraint, can be classically expressed in terms of the
Poisson bracket of the connection and the volume,
\begin{gather}\label{145}
\left\{{A_a}^i,V\right\}=\left\{{A_a}^i,\int{\rm d}^3x\,\sqrt{|\det E|}\right\}
=2\pi\gamma G\epsilon^{ijk}\epsilon_{abc}\frac{{E^b}_j{E^c}_k}{\sqrt{|\det E|}}\,{\rm sgn}(\det E).
\end{gather}
So in quantum theory the Poisson bracket can be expressed by a~commutator of
well-def\/ined operators, when the connection is
replaced by the corresponding holonomy.
This construction is at the root of
the cosmological singularity problem.
For holonomies with links of
coordinate length $L/a$ we write
\begin{gather}\label{tr}
\frac{1}{2}\frac{i\hbar L}{a}{\widehat{\big\{{A_a}^i,V_v\big\}}}\dot{e}^a
\sim
\operatorname{tr}\big(\tau^i\hat{h}_{v,e}\big[\hat{h}_{v,e}^{-1},\hat{V_v}\big]\big),
\end{gather}
where $\dot{e}^a$ is the tangent vector to a~link $e$ adjacent to
the node $v$; $h_{v,e}$ is a~holonomy along a link adjacent to $v$;
and $V_v$ is the volume of a~region containing $v$.
There is an
ambiguity in the $\SU(2)$ representation, in which the trace is
taken.
The parameter labeling this ambiguity inf\/luences the scale,
where the transition from the discrete quantum universe to the
continuous classical universe takes place.
It enables us to model
the time scale of inf\/lation~\cite{DIC}.

In the older literature a~f\/ixed
discreteness scale $\mu_0={\rm const}$ with respect to the f\/iducial
metric was employed, which led to problems in the continum limit.
For comparison we present the f\/ixed lattice formulation and
postpone the ref\/ined lattice to the next subsection.
In this case,
a volume eigenstate $|\mu\rangle$ with
\begin{gather*}
\hat V|\mu\rangle=\left(\frac{4\pi\gamma\ell_{\rm
P}^2}{3} |\mu|\right)^{3/2},
\end{gather*}
and for the simplest choice of $j=1/2$ for the
$\SU(2)$-representation in the trace, the inverse volume operator
acts in the following way
\begin{gather*}
\widehat{V^{-1}}|\mu\rangle=\left|\frac{3}{4\pi\gamma\ell_{\rm
P}^2\mu_0}\left(\{V(\mu+\mu_0)\}^{1/3}-\{V(\mu-\mu_0)\}^{1/3}\right)\right|^3.
\end{gather*}
From this, one derives the action of the self-adjoint gravitational
Hamiltonian constraint operator (constructed from curvature terms of
full LQG) on $|\mu\rangle$:
\begin{gather}
\hat H_{\rm g}|\mu\rangle=
\frac{M_{\rm P}}{32\sqrt{3\pi}\gamma^{3/2}\mu_0^3}\nonumber\\
\phantom{\hat H_{\rm g}|\mu\rangle=}
\times
\big\{[R(\mu)+R(\mu+4\mu_0)]|\mu+4\mu_0\rangle+4R(\mu)|\mu\rangle+
[R(\mu-4\mu_0)]|\mu-4\mu_0\rangle\big\}\label{H0}
\end{gather}
with
\begin{gather*}
R(\mu)=\left||\mu+\mu_0|^{3/2}-|\mu-\mu_0|^{3/2}\right|.
\end{gather*}

When the commutator in~\eqref{tr} is expressed in terms of holonomy
and f\/lux operators, its expectation
values in quantum states do not have the classical relationship with
the expectation values of the basic operators.
Classically the f\/lux
through an elementary lattice site is $L^2(a)$, where $L$ is the
length scale introduced in the foregoing subsection, which depends
on the scale factor according to its def\/inition.
In~\cite{BCT} the
f\/lux operator is rewritten in the form $\hat F=\langle\hat
F\rangle+(\hat F-\langle\hat F\rangle)$ and the volume operator as
function of $\hat F$ is expanded in $\hat F-\langle\hat F\rangle$.
With $\langle\hat F\rangle=L^2(a)$ in lowest order one obtains a
correction function $\alpha$ to classical Hamiltonians, depending on
the scale factor, whose expansion for small deviations from the
classical value for large $L(a)$ is
\begin{gather*}%\label{7}
\alpha(a)=1+\alpha_0\delta_{\rm Pl}(a)+\cdots
\end{gather*}
with $\alpha_0$ being a~constant and
\begin{gather*}
\delta_{\rm Pl}:=\left(\frac{\ell_{\rm P}}{L}\right)^4.
\end{gather*}
Volume corrections become large, when $L$ is small of the order of
the Planck length.

In comparison with holonomy corrections we may observe that
$\delta_{\rm Pl}$ is small, when $L\gg\ell_{\rm P}$,
$\rho_{\rm QG}$ becomes small and holonomy corrections become
large.
The relation between these two kinds of corrections can be
better seen in terms of densities,
\begin{gather}\label{15}
\delta_{\rm Pl}
=\left(\frac{8\pi G}{3} \rho_{\rm QG} \ell_{\rm P}\right)^2
=\left(\frac{8\pi}{3}\frac{\rho_{\rm QG}}{\rho_{\rm Pl}}\right)^2
=\left(\frac{8\pi}{3}\frac{\rho}{\rho_{\rm Pl}} \delta_{\rm hol}^{-1}\right)^2.
\end{gather}
Inverse volume corrections depend on the ratio of the QG density to
the Planck density, whereas holonomy corrections depend on the ratio
between the actual density and the QG density.
For small densities
in an expanding universe, holonomy corrections decrease, but~\eqref{15} tells us that $\delta_{\rm Pl}$ cannot simultaneously go
down arbitrarily.
This gives at least a~lower bound for LQC
corrections, from which M.~Bojowald et al.\
in~\cite{BCT} derive
lower bounds to correlation functions of inhomogeneities in the CMB.
Here we do not have only upper bounds for LQC ef\/fects, but an
estimate that gives rather narrow bounds for parameters like~$\alpha_0$.
Here we note that, as the size of the inverse volume
corrections relies on the size of a~f\/iducial cell, in~\cite{AS} it
is argued that they become negligible, when the limit of an inf\/inite
cell is taken, so that it extends over all~$\R^3$.

The correction function $\alpha$ appears also in the ef\/fective
constraint algebra, where the Poisson brackets of two smeared-out
Hamiltonian constraints $H(M)$ and $H(N)$ are modif\/ied in the form
\begin{gather*}%\label{6}
\{H(M),H(N)\}=D\big[\alpha^2q^{ab}(N\partial M-M\partial
N)\big]
\end{gather*} ($D$ is the dif\/feomorphism constraint).
The
ef\/fective algebra is, importantly, f\/irst-class, i.e.\
anomaly-free.
Anomaly freeness was shown for not too large departures from FLRW.

In modeling inf\/lation the inhomogeneities superposed on the FLRW
background f\/ind their way into the classical Friedmann equation and
the equation of motion of the scalar f\/ield in form of holonomy and
inverse-volume corrections.
So they enter into the basis of
inf\/lation models with dif\/ferent sorts of dilaton potentials and
eventually emerge in the perturbation power spectrum of CMB, where
it comes into contact with observations.
The lower bounds of the
predicted LQC corrections are only a~few orders of magnitude away
from the present upper observational bounds~\cite{BCT}, so that we
can hope for an experimental judgement in the not too distant
future.
However, see Section~\ref{lqcphenom} for further discussion on this.

\subsection{Dynamics and lattice ref\/inement}\label{Latt}

In the dynamical evolution of an expanding model universe in LQC,
were it rigorously derived from LQG, we might expect a~steady
creation of new nodes by the Hamiltonian constraint operator,
which keeps the typical link length small.
Without the full LQG
dynamics we cannot see this creation of
links.
What we can do is to try to model a~ref\/inement of the SNW
lattice by hand~\cite{Sa,Sak}.

As mentioned above, the dynamics in LQC is constructed with the aid
of the LQG Hamiltonian constraint, including some kind of matter as
``internal clock'', usually a~scalar f\/ield.
Beside the inverse
volume, in the Hamiltonian the curvature ${F_{ab}}^k$ plays an
important role.
Classically it can be written as limit of holonomies
around a~plaquette in the $(a,b)$ plane, when the area of the
plaquette goes to zero.
However,
due to the discreteness of the spatial geometry, in LQC this limit
does not exist.
The curvature term in the Hamiltonian is expressed
for f\/inite plaquettes, the area of which is chosen to be equal to
the area gap $\Delta A=4\sqrt{3}\pi\gamma\ell_{\rm P}^2$, the
lowest nonzero eigenvalue of the LQG area operator.

The plaquettes introduce a~new length scale in the classical theory,
when we assume that a~face of a f\/iducial cell $\cal C$ is
partitioned into $N$ plaquettes of area $\big(\bar\mu V_0^{1/3}\big)^2$.
The parameter~$\bar\mu$ is distinct from the parameter~$\mu$
which characterizes holonomies. A dynamical length scale, $\bar\mu$~appears in the regularization of the
Hamiltonian constraint operator.

To determine a~relation between $\bar\mu$ and the characteristic
value~$\mu$ of a~state $|\mu\rangle=\Psi(\mu)$ which the
Hamiltonian constraint operator is to act on, we take the area of a
face of the f\/iducial cell
\begin{gather*}
N \Delta A=|p|=\frac{4\pi\gamma\ell_{\rm P}^2}{3} |\mu|
\end{gather*}
and take into account that it is covered by $N$ plaquettes,
\begin{gather*}
N\big(\bar\mu V_0^{1/3}\big)^2=V_0^{2/3}.
\end{gather*}
Eliminating $N$ we f\/ind the discreteness scale in the Hamiltonian
depends on quantum states via
\begin{gather*}
\bar\mu(\mu)=\left(\frac{3\sqrt{3}}{|\mu|}\right)^{1/2}.
\end{gather*}
This means, when the area of a~face of~$\cal C$, or its physical
volume $V$, grow due to a~growing scale factor, the partition of a
f\/iducial cell is ref\/ined.
This is necessary for the size~$\bar\mu$
of the quanta of geometry to remain small, when the universe
expands.

The necessity for this lattice ref\/inement is best seen in
considering the classical, continuous limit. As the scale factor~$a$ becomes large in the course of the dynamical evolution, the
dif\/ference equation stemming from the Hamltonian constraint can be
approximated by a~dif\/ferential equation, the WDW equation, for a
smooth wave function. The wave function oscillates on scales~${\sim}a^{-1}$, and for growing~$a$, this becomes smaller
than the discreteness scale, when the latter were given by a
constant and thus f\/irmly tied to the scale factor.
We refer to~\cite{Sa,Sak,AS} for the details.

For a~f\/ixed lattice scale $\mu_0$, holonomies $\exp(i\mu_0c/2)$ act
as simple shift operators by the constant $\mu_0$ on states
$|\mu\rangle$, the action of $\exp(i\bar\mu c/2)$ is more
complicated, because~$\bar\mu$ is a~function of~$\mu$.
However on
volume eigenstates $|\nu\rangle$ with
\begin{gather}\label{nu}
\nu=\frac{2}{3} \mu^{{3}/{2}}
\end{gather}
and
\begin{gather*}
\hat V|\nu\rangle=\frac{3\nu}{2}\left(\frac{4\pi
\gamma}{3}\right)^{3/2}\ell_{\rm P}^3|\nu\rangle
\end{gather*}
the ref\/ined lattice holonomies act by a~shift and the self-adjoint gravitational Hamiltonian acquires a~form
analogous to~\eqref{H0}
\begin{gather*}
H_{\rm g}|\nu\rangle=
\frac{|\nu|}{64\sqrt{2\pi}\smp\gamma^{3/2}\mu_0^3}
\Bigg[\frac{1}{2}\left\{U(\nu)+U(\nu+4\mu_0)\right\}|\nu+4\mu_0\rangle-2U(\nu)|\nu\rangle
\nonumber\\
\phantom{H_{\rm g}|\nu\rangle=}
+\frac{1}{2}\left\{U(\nu)+U(\nu-4\mu_0)\right\}|\nu-4\mu_0\rangle\Bigg]
\end{gather*}
with
\begin{gather*}
U(\nu)=|\nu+\mu_0|-|\nu-\mu_0|.
\end{gather*}
In summary, lattice ref\/inement fulf\/ills all the following
conditions:
\begin{enumerate}\itemsep=0pt
\item[1)] Independence of the elementary cell chosen in an open
cosmological model to make integration f\/inite.

\item[2)] Inf\/lation becomes ``natural'' in the sense that an inf\/laton mass
$M_{\rm inf}\leq10^2M_{\rm P}$ is suf\/f\/icient, in contrast to much
lower values without lattice ref\/inement.

\item[3)] Factor ordering in the macroscopic WDW equation becomes unique.

\item[4)] The requirement of ``pre-classicality'' is fulf\/illed, i.e.\
quantum
corrections at large scales are avoided.
\end{enumerate}

Consequences for inf\/lation can be seen in~\cite{Sa}.
Here we mention
just a~few facts about modeling inf\/lation with and without lattice
ref\/inement.
We take a~wave function, depending on the volume of the
universe, or $p=V^{2/3}$, respectively, and an inf\/laton f\/ield~$\phi$,
\begin{gather*}%\label{17}
|\psi(p,\phi)\rangle=\sum_\nu\psi_\nu(\phi)\,|\nu\rangle,
\end{gather*}
where $|\nu\rangle$ is an eigenstate of volume, $\nu$ is related
with $\mu$ by~\eqref{nu}.
In the continuous limit, when the
evolution equation of the wave function is approximated by a
dif\/ferential equation, we assume
\begin{gather*}%\label{18}
\psi(p,\phi)=\Upsilon(p)\Phi(\phi).
\end{gather*}
If the dynamics of $\phi$ is approximately driven by the
potential $V(\phi)=V_\phi p^{\delta-3/2}$, where $V_\phi$ is constant
and $\delta=3/2$ in the case of slow-roll, then this leads to an
oscillating function $\Upsilon(p)$ with a~separation of two
successive zeros
\begin{gather*}
\Delta p=\frac{\pi}{\sqrt{\beta V_\phi}} p^{(1-2\delta)/4},
\qquad\text{with}\quad
\beta=\frac{12}{\pi G\hbar^2}.
\end{gather*}
In the continuum limit the wave function must vary slowly on the
discreteness scale $\bar\mu$ of QG (pre-classicality, see~\cite{DIC}).
This can be formulated in the way that the distance
between two zeros in terms of $\mu$, $\Delta\mu=\frac{3}{4\pi\gamma
\ell_{\rm P}^2} \Delta p$ must be at least equal to $4\bar\mu$,
which yields the condition of pre-classicality
\begin{gather*}%\label{19}
\Delta p>16\left(\pi\gamma\right)^{3/2}\ell_{\rm
P}^3p^{-{1/2}},
\end{gather*}
which leads to an upper bound of the inf\/laton potential,
$V(\phi)\leq2.35\cdot10^{-2}\ell_{\rm P}^{-4}$, in contrast to
$V(\phi)\ll10^{-28}\ell_{\rm P}^{-4}$ for f\/ixed lattice.
To be in accordance with the COBE-DMR measurements, the potential
must further satisfy
\begin{gather*}%\label{20}
\frac{[V(\phi)]^{3/2}}{V'(\phi)}\approx5.2\cdot10^{-4}M_{\rm
P}^3.
\end{gather*}
If we choose an inf\/laton potential
$V(\phi)\approx\frac{m^2\phi^2}{2}$ then we obtain, from the least two conditions,
$m\leq10M_{\rm P}$, compared with $m\leq70(e^{-2N_{\rm cl}})M_{\rm P}$ for the f\/ixed lattice.
$N_{\rm cl}$ is the number of $e$-folds.
The condition that a~signif\/icant proportion of the inf\/lationary
regime takes place during the classical era imposes a~condition on
the inf\/laton mass, which is much more natural for the ref\/ined
lattice.

\subsection{Loop Quantum Cosmology: possible observational consequences}
\label{lqcphenom}

With the development of LQC and recent observational missions in cosmology there are rich veins of work to explore in the phenomenology
of the early universe.
There are many approaches to this work.
As
yet there is no consensus on the best route from the discrete
quantum geo\-met\-ry of LQG to observable cosmological predictions.
In this review of LQC phenomenology, which is very brief, we do not attempt a
comprehensive review of the literature but rather provide a~guide to
starting points for further exploration of these veins of work.
Our
scope is further reduced by focusing on observational signatures
that will be accessible in the near future.

The current best observational window on the early universe is the
power spectrum of small angular f\/luctuations in the cosmic microwave
background (CMB) radiation.
In the standard inf\/lationary model
these f\/luctuations are generated by scalar and tensor perturbations.
Thus, the power spectra of the scalar and tensor perturbations are
the key tools for investigations of cosmic background radiation in
electromagnetic and gravitational sectors.
Because of the
dif\/ference in scales at which decoupling occurs, the gravitational
wave background originates at an earlier epoch than the CMB, thus
allowing a~view into the very early universe.
However, observation
of the gravitational background remains a~huge experimental
challenge.

The now-familiar plot of the CMB power spectrum is of
angular correlations of the tempera\-tu\-re-temperature, or ``TT'',
power spectrum.
Polarization modes of the CMB are decomposed into
curl-free electric $E$-modes and gradient-free magnetic $B$-modes.
The
$B$-mode power spectrum arises from two ef\/fects: directly from
primordial gravitational waves (the tensor modes) and indirectly
through lensing from the conversion of $E$-mode to $B$-mode.
Therefore, most intriguingly, it may be possible to study tensor
modes from polarization measurements of the CMB, without directly
observing the gravitational wave background.

As in the case of LV, there are a~number of dif\/ferent perturbation frameworks
under de\-ve\-lop\-ment.
For instance in one class of widely-used
frameworks, the background space-time enjoys LQC modif\/ications
(corrections due to holonomy, inverse volume, or both) and the
perturbations take a~form similar to the perturbations in FLRW
cosmologies, where the linear perturbations of Einstein's equations
are quantized.
Because the background does not follow Einstein's
equations in this framework it is not immediately clear that
perturbation equations are consistent.
There may exist a~consistent
set, but this issue is currently not resolved.

In another framework
(see, e.g.~\cite[Section~VI.C]{AS}), the classical theory is f\/irst
reduced by decomposing the gravitational phase space into
homogeneous and purely inhomogeneous parts (for matter as well as
gravitational variables); the linear perturbations are entirely
within the inhomogeneous phase space.
Then both the background
and the linear perturbations are quantized with LQC techniques.
In related
work of~\cite{AKL,DLT}, a~quantum scalar f\/ield is analyzed on a
quantized Bianchi~I background.
This work provides a~framework for
perturbations on an ef\/fective `dressed' quantum geometry that
may also contain back reaction.

Most current studies are some blend of the traditional framework of
cosmological perturbations and an ef\/fective LQC framework.
The
formulation of these frameworks are currently a~matter of lively
debate; see, for instance, \cite[Section~VI.D]{AS},
\cite[Sections~2.3.4]{BCT}, and~\cite{W-E}.
Nevertheless there is an impressive
body of work in developing the phenomenology of the very early
universe.

\looseness=1
Loopy modif\/ications to the power spectrum have been derived.
LQC
of\/fers at least two modif\/ications to the usual scenario, holonomy
and inverse-volume or inverse-triad corrections as discussed in the last section.
Both these corrections have been
incorporated in models of the background space-time using LQC
methods.
As yet we lack a~comprehensive study of all the LQC ef\/fects
on the scalar and tensor perturbations.
Nonetheless there are many
studies analyzing how specif\/ic models of LQC corrections af\/fect the
power spectra.
We will mention two lines of work, one on af\/fects in the power spectrum of scalar and tensor perturbations and the other on the chirality of tensor perturbations.
This last work is outside the symmetry reduced LQC models and is included here as it concerns
tensor perturbations.

\subsubsection{Scalar and tensor perturbations} In the ef\/fective
Friedmann equation framework scalar perturbations with inverse triad
corrections are discussed in~\cite{BHKS,BHKS2,BHKSS,BC} and with
holonomy corrections in~\cite{WL,W-E,CMBR}.
In \cite{CMBR}
correction terms were introduced without a~gauge choice with the
result that the perturbation equations are anomaly-free.
Tensor
modes in the same framework are discussed in
\cite{BH,CMNS,GCBG,MCGB,BC} with inverse triad corrections and in
\cite{BH,Mcosmo2,Mtensor1,Mtensor2,GB,GB} with holonomy
corrections.

Starting with~\cite{BH} the work in \cite{MCGB,GBCM} develops a
phenomenological model of the tensor modes within a~model bouncing
cosmology with a~single massive scalar f\/ield.
They f\/ind that the
ef\/fects can be modeled with two parameters~\cite{GBCM}, one ``bump
parameter'' is simply related to the inf\/laton mass.
The second
parameter, a~transition wave number, is related in a complicated way
to the critical density and the scalar potential energy-critical
density ratio at the bounce.
The authors f\/ind that the tensor power
spectrum is suppressed in the infra-red regime, agrees with the
standard general relativistic picture in the UV, and has both an
increase in amplitude and damped oscillations at intermediate
scales.
This work suggests that the next generation $B$-mode
experiments could provide a~successful constraints on the model
parameters~\cite{GBCM}.
For more on this approach see
\cite{Mcosmo2,GB,GCBG,MCGB,GBCM,MCBGanomaly}.

The power spectra of scalar and tensor modes, with inverse triad
corrections, are derived in the inf\/lationary scenario with ef\/fective
Friedmann equations in~\cite{CH,BC,BCT}.
In \cite{BC,BCT} the
corrections are parameterized with corrections related to the area
gap, a~parameterization of quantization ambiguities, and how the
number of lattice sites changes in the evolution of the cosmology.
This
leads to an enhancement of power on large angular scales.
The model is compared to WMAP 7 year data as well as other
astronomical surveys in~\cite{BCT}.

Normally in the inf\/lationary scenario, because of the rapid
expansion, the universe is in its vacuum state shortly after the
onset of inf\/lation.
It was pointed out in~\cite{AP,Agullo},
however, that if pre-inf\/lationary physics in LQC led to a~non-vacuum
state then spontaneous generation of quanta would have observational
consequences in terms of non-Gaussianities in the CMB and in the
distribution of galaxies.

The wealth of phenomenological models means that there is guidance
as to many ef\/fects arising from aspects of LQG. The f\/ield has
evolved to the point where these models can be directly compared
with current data.
But there remain many questions on the
derivation of these ef\/fects from a~more fundamental level, both
from LQC and from LQG. For instance, the parameter capturing the
phenomenology of inverse volume corrections depends on the f\/iducial
volume~\cite{BC}.
While the parameter can be f\/ixed by the size of
the Hubble horizon at horizon crossing the
parameterization is debated.
So the status of inverse volume
corrections in the presence of inhomogeneities is a~matter of
current debate (particularly when the spatial topology is
non-compact), see e.g.\
\cite[Section~2.4]{BCT} and~\cite[Section~VI.D]{AS}.

\subsubsection{Chirality of tensor perturbations}

Working with the Ashtekar--Barbero connection formalism and deriving
the tensor perturbations in a~de Sitter background, Magueijo and
collaborators f\/ind that the graviton modes have a~chiral asymmetry
in the vacuum energy and f\/luctuations if the Immirzi parameter has
an imaginary part~\cite{MB,BM1,BM3,BM2}.
This is signif\/icant as the chirality
would leave an imprint on the polarization of the cosmic microwave
background and might be observed with the PLANCK mission.
The asymmetry depends on operator ordering.

\subsection{Phenomenology of black hole evaporation}

The subject of this subsection is closely related to cosmology and
has potentially observable consequences.
In~\cite{Modest}
semiclassical models of Schwarzschild and Reissner--Nordstr\"{o}m
black holes are presented.
They are based on LQG's discreteness of
area and a~resulting repulsive force at extremal densities.
With
these ingredients the space-time metric outside ``heavy'' black holes
(with respect to the Planck mass) is only slightly modif\/ied in
relation to the classical form, but inside the horizon the
singularity is smoothed out and in the limit when the radial
coordinate goes to zero the metric becomes asymptotically Minkowski.
By the introduction of the new coordinate $R=a_0/r$, where $a_0$ is
the LQG-inspired minimal area, the regularized metric is shown to be
self-dual in the sense of T-duality: An observer at
$R\rightarrow\infty$ sees a~black hole with mass ${\sim} m_{\rm
Pl}/m$, when $m$ is the mass of the black hole seen by observers in
the asymptotic f\/lat region $r\rightarrow\infty$.
For ``light'' (=
sub-Planckian) black holes also the outside metric is modif\/ied
considerably.

``Light'' black holes do not evaporate completely, although they
would emit high-energy radiation at an extremely low rate.
They are
supposed to explain two cosmological puzzles: Being practically
stable, ultralight black holes created during the inf\/lation process
could account for dark matter as well as they could be the so far
unknown source for UHECR.

In the sequel~\cite{X,Emiss} it was shown that the discreteness of
area leads to features that distinguish black hole evaporation
spectra based on LQG. They are very distinctly discrete in contrast
to the classical Hawking spectrum and observation, should it become
possible, should be able to distinguish LQG from other underlying QG
theories.

\section{Conclusions}\label{section6}

In this review we have described ways in which LQG, mainly by means of discreteness of the
spatial geometry, may lead to experimentally viable predictions.
We discussed ways in which the discreteness may (or may not) lead to a~large variety of
modif\/ications of special relativity, particle
physics and f\/ield theory in the weak f\/ield limit.
In Sections~\ref{section3}
and~\ref{section4} ef\/fective particle and f\/ield theory frameworks are
presented in some detail.
Where possible we have given current observational bounds on the models.
In these sections, as well as in the LQC section, we have pointed out numerous approaches
and some of the theoretical and experimental open problems.
Many of these are collected below.

Of course QG and QG phenomenology remain open problems.
We lack strong ties between observationally accessible models and LQG. These models have
ans\"{a}tze that are often in striking
contradiction with each other and none has a~clear support from
observations.
Furthermore, should any of the ef\/fective models
presented here be favored by experimental data in the near future,
this will hardly point uniquely at one of the fundamental theories,
or to a~certain version of them\footnote{``In this situation a possible
conclusion is that today is too early to search for a
fundamental QG theory and that one should in the meantime rather
look for an ``old quantum theory of gravity'' in the style of Bohr
and Sommerfeld'', quoting G.~Amelino-Camelia at the workshop
``Experimental Search for QG 2010'', Stockholm.
Such an approach was recently successfully employed in the investigation of the volume operator~\cite{BiaHag}.}.
On the other hand, this f\/ield has seen tremendous progress since the mid-90's when it was tacitly
assumed that there were essentially no experimentally accessible windows into QG.
Quite the contrary, now there are many avenues to explore QG ef\/fects and stringent
bounds have already been placed on ef\/fects originating at the Planck scale.
These developments are an essential f\/irst step toward a~physically viable quantum gravity theory.

Concluding, the subject of LQG phenomenology, and of
QG phenomenology in general, is now far reaching.
We expect that QG
phenomenology will remain a~very active f\/ield and will hopefully bring new perspectives and clarity
on the ad hoc assumptions and models.
Indeed, in spite of its
shortcomings, phenomenology is indispensable for LQG, or any other
quantum theory of gravity, if it is to become a~physical theory.

\appendix

\section{Elements of LQG }
\label{newvariables}

In the f\/irst part of this appendix there is an overview of the
basics of LQG. In the second part we very brief\/ly review the theory's
kinematics.
For more details the reader should consult the recent
brief reviews by Rovelli~\cite{Rrev} and Sahlmann
\cite{SahlmannRev}.
For longer reviews the reader should consult
\cite{LongRevAL,LongRevP,LongRevT1,LongRevT2} and the texts~\cite{T1,R1}.

LQG is a~quantization of GR. Due to the special features of GR it
looks in many points quite dif\/ferent from other quantum f\/ield
theories:
\begin{itemize}\itemsep=0mm
\item LQG takes into account that space and time are not an external
background for physics, but part of physical dynamics.
\item Gravity is self-interacting, but the self-interaction cannot
be treated perturbatively, because the theory is non-renormalizable.
\item Due to general covariance of GR the gauge group is the group
of dif\/feomorphisms, not the Poincar\'{e} group.
\end{itemize}
Whereas non-linearity is shared with other QFTs, the issue of
dynamical space-time is unique for gravity.

For comparison recall that usual QFTs deal with f\/ields on either
Minkowski space or some curved Riemannian space with a~given metric
and the corresponding Levi-Civita connection.
LQG, on the other
hand, is a~QFT on a manifold, a priori without further structure,
and the geometric properties of physical space are realized in form of
dynamical f\/ields.
Concretely, in LQG the basic f\/ield variables
are not metric components, as in GR, but orthogonal bases (triads)
in the tangent space of every point in three-dimensional space and
the connection.

The introduction of triads brings further gauge degrees of freedom
into the game, namely local rotations, i.e.\
elements of the group
$\SO(3)$ (or $\SU(2)$).
Further, due to space-time covariance of GR,
the embedding of the spatial 3-manifold into a~4-manifold and the
choice of a~time coordinate is also a matter of gauge.
In the
canonical formalism this is ref\/lected by the appearance of a~further
gauge generator.

As common in gauge theories, gauge generators are constraints.
In
the present case the Gauss constraint, which has a~formal analogy to
the Gauss law in electrostatics, generates triad rotations.
The theory also has the
dif\/feomorphism constraint and the Hamiltonian constraint, which
generates transitions from one spacelike 3-manifold to another one.
In LQG these constraints are imposed as
operators, which annihilate physical, i.e.\
gauge-invariant, quantum
states.
This leads immediately to a~surprising feature in canonical
QG: The propagation from a~hypersurface to the next one being a
gauge transformation, all physical states are invariant under these
transitions and there is no physical time evolution.
Gauge-invariant
states contain all the history of a~state of the gravitational
f\/ield.
This was pointed out long before the advent of LQG
\cite{WDW}.
Time evolution must be introduced in an operational way in
relation to some suitable kind of matter, which is coupled to
gravity and may be considered as clock.

The problem of non-linearity, coming from gravity's self
interaction, is more a~technical than a conceptual problem.
Non-linearity in interacting QFTs in the standard model is
successfully dealt with by renormalization methods.
The
non-renormalizability of GR appears as a~serious obstacle, but it
can be traced back to the background~-- splitting the metric into a~background, e.g.\
Minkowski
space, and a~(small) f\/ield on it in the form
$g_{ik}=\eta_{ik}+\psi_{ik}$, $|\psi_{ik}|\ll1$ and quantizing ``the
ripples $\psi_{ik}$ on the background'' does not work.
The
conclusion, which is drawn from this in LQG, is that only spatial
geometry as a~whole has a chance to be successfully quantized.

Presently the success and limitations of LQG can be summarized very
brief\/ly in the following way.
Local connection components are not
gauge invariant, they are not even tensor components.
Constructing
gauge-invariant quantities from them is possible by means of closed
contour integrals, so-called Wilson loops.
Their further development
are holonomies and spin networks.
They introduce
non-locality into the theory.
In consequence, metric quantities
(which are not introduced from the beginning as basic variables),
like area and volume, turn out to have discrete spectra with a
fundamental role of the Planck length.
In other words, LQG yields
``quanta of space'' or ``atoms of geometry''.
The existence of a
minimal length provides a~natural ultraviolet cutof\/f for other QFTs.
The main open problem is the non-linearity of the Hamiltonian
constraint.
Thiemann~\cite{trick} succeeded to formulate it as
well-def\/ined operator in several versions, but there remain some
ambiguities.
A more recent approach is the master constraint
programme~\cite{T1}.
The problem of the Hamiltonian constraint
and time evolution have been satisfactorily solved in simplif\/ied
models, mainly in cosmology.

Concerning technicalities, at the very basis of LQG stands a~$(3+1)$
decomposition of space-time and a~canonical formulation of GR in
terms of densitized triad variables ${E^a}_i$ and connection
variables ${A_a}^i:=\Gamma_a^i-\gamma K_a^i$, which are
canonically conjugate, on the spatial 3-manifold\footnote{The index
$a$ is spatial index and $i$ is an $\su(2)$ Lie algebra index.}. The
connection depends on both the spin connection $\Gamma_a^i$ and the
extrinsic curvature $K_a^i$.
The para\-me\-ter~$\gamma$ is the
Barbero--Immirzi parameter, which may be f\/ixed by matching with the
Bekenstein--Hawking black hole entropy formula, see e.g.~\cite{MeissnerBH}.

In the connection representation, the quantum Hilbert space is
spanned by functionals of the connection.
A convenient basis,
invariant under $\SU(2)$ gauge transformations of the triads, is
provided by spin network (SNW) functions, def\/ined with the aid of
graphs $\Gamma$, where to each edge or link $e_I$ a~representation
of $\SU(2)$, corresponding to a~spin~$j$, is associated, the ``color'' of the edge.
SNW functions are constructed from the path ordered
exponential of the connection, the {\em holonomies}
\begin{gather}\label{21}
h_{e_I}(A):={\cal P}\exp\left(-\int_{e_I}\dot
e^a{A_a}^i{}\tau_i^{(j)}\right),
\end{gather}
where $\dot e^a$ is the tangent vector of the edge $e_I$ and
$\tau_i^{(j)}$ the generator of the representation of $\SU(2)$ with
spin $j$.
A SNW function based on a~graph $\Gamma$ with $N$ edges
has the following form
\begin{gather*}%\label{22}
\Psi_\Gamma(A)=\psi(h_{e_1}(A),\ldots,h_{e_N}(A)).
\end{gather*}
The holonomies are connected by intertwiners at the nodes or
vertices in such a~way that $\psi$ is a scalar function.
The
functions $\Psi_\Gamma(A)$, having a~f\/inite number of arguments,
namely the number of edges of $\Gamma$, are called cylindrical
functions.
They may be considered as coordinates on the space of
smooth connections modulo gauge transformations, denoted by ${\cal
A}/{\cal G}$.
The Hilbert space of LQG is the closure of the space
of cylindrical functions on generalized (distributional) connections
modulo gauge transformations with the Ashtekar--Isham--Lewandowski
measure~\cite{AIL1,AIL2,AIL3}, constructed from the Haar measure on~$\SU(2)$.

On this Hilbert space the conf\/iguration variable~${A_a}^i$ would act
as a~multiplication operator, were it well-def\/ined, and the momentum
variable~${E^a}_i$ as a~functional derivative with respect to~${A_a}^i$.
As is common in quantum f\/ield theory, elementary
variables do not enter quantum theory as operators, but as
operator-valued distributions, which have to be regularized by
integrating out with some test functions.
In the case of the
connection, the above def\/ined holonomy opera\-tors~\eqref{21} arise
from integrating~${A_a}^i$ in one dimension, which is natural
for one-forms.
These operators either add a~holonomy along an link
present in a~SNW, or create a new link.

The momentum variable ${E^a}_i$ is a~vector density, which can be
associated with a~two-form $\eta_{abc}{E^a}_i$ with the aid of the
Levi-Civita density $\eta_{abc}$.
So it is natural to smear it out
by integration over a~two-dimensional surface.
Let a surface $\cal S$ be def\/ined by $(\sigma^1,\sigma^2)\rightarrow
x^a(\sigma^1,\sigma^2)$, where $\vec\sigma=(\sigma^1,\sigma^2)$ are
coordinates of $\cal S$ and $x^a$ coordinates in the
three-dimensional space, where $\cal S$ is embedded.
Then
\begin{gather*}%\label{23}
E_i({\cal S}):=-i\hbar\int_{\cal S}{\rm d}^2\sigma n_a(\vec\sigma) \frac{\delta}{\delta{A_a}^i(x(\vec\sigma))}
\end{gather*}
is a~well-def\/ined operator, {\em the flux operator} of the f\/ield
${E^a}_i$ through $\cal S$ with
\begin{gather*}%\label{24}
n_a(\vec\sigma)=\epsilon_{abc}\frac{\partial
x^b(\vec\sigma)}{\partial\sigma^1}\frac{\partial
x^c(\vec\sigma)}{\partial\sigma^2}
\end{gather*}
being the one-form normal to $\cal S$.
The action on a~holonomy of
a link $e$ crossing $\cal S$ is
\begin{gather*}%\label{25}
E_i({\cal S})h_e(A)=\pm i\hbar h_{e_1}(A) \tau_i^{(j)} h_{e_2}(A),
\end{gather*}
where $e_1$ and $e_2$ are two parts of the link $e$, divided by the
intersection point with $\cal S$.
The sign depends on the relative
orientation of $e$ and $\cal S$.
$E_i({\cal S})$ inserts a~generator
at the intersection point; if $e$ and $\cal S$ do not intersect, the
action is zero.
The algebra of these basic operators is called the
{\em holonomy-flux algebra}.

Minkowski space is represented in LQG by a~state of the
gravitational f\/ield, which is not the vacuum state, but a
superposition of excited SNW states.
Discreteness comes in a
natural way from the ``polymeric'' structure of the SNWs, suggesting
the presence of QG ef\/fects even in f\/lat space.

The above formulation is in the ``embedded framework'' of LQG. This
has the advantage of having clear ties to the classical theory but
in the kinematic Hilbert space is non-seperable.
In addition the
state space has physically mysterious continuous moduli that label
equivalence classes of dif\/feomorphism invariant states
\cite{moduli}.
Partly in response to these dif\/f\/iculties an
alternate framework has received increasing attention.
The
combinatorial framework of LQG\footnote{In the ``combinatorial
framework'', the combinatorics refers to an approach to the topology
of space while in the ``combinatorics of the node'' the combinatorics
refers to the combinatorics of $\SU(2)$ representations.} was
introduced by Zapata~\cite{jose,jose2} and recently used as the kinematic
setting for spin foam models in the review~\cite{NL}.
In this
framework the kinematical Hilbert space is separable and is free of
the mysterious moduli.

\subsection*{Acknowledgements}

Thanks to Stefano Liberati, Mathew Mewes, and Jorge Pullin for
correspondence and to Stefano Liberati and Luca Maccione for
permission to use their f\/igures.
We also thank the anonymous reviewers
for their helpful suggestions.
F.H. was supported by the
Ministry of Education of the Czech Republic, contract no.~MSM
0021622409.

\LastPageEnding

\end{document}